\pdfoutput=1
\documentclass[a4paper,11pt]{article}

\usepackage[sort&compress,super]{natbib}
\usepackage[T1]{fontenc}

\usepackage[english]{babel}
\usepackage{amsmath, esint}			%% Equations
\usepackage{amssymb}				%% Additional symbols
\usepackage{graphicx}				%% Main package for figure inclusion
\usepackage{tabularx}                                 %% for tabulars with controllable width
\usepackage{subcaption}				%% allow subfigures
\usepackage{algorithm,algorithmic}		%% Allows [H] option for figures for exact placement
\usepackage{tikz}

\usetikzlibrary{arrows}
\usetikzlibrary{mindmap,trees}
\usepackage{listings}					%% Interface for code typing
\usepackage{color}					%% Allows color generation
\usepackage{xcolor}					%% Allows color generation
\usepackage{mathtools}				%% Declare mathematical operators, e.g. \DeclarePairedDelimiter, e.g.  inner product
\usepackage{lscape}					%% Allow landscape pages
\usepackage{hyperref}					%% Allow URL's and active links within the document
\usepackage{cleveref}					%% Allows referencing subfigures. Must be loaded after hyperref

\DeclarePairedDelimiter{\floor}{\lfloor}{\rfloor}	% Round down

\usepackage{tcolorbox}% http://ctan.org/pkg/tcolorbox
\definecolor{mycolor}{rgb}{0.122, 0.435, 0.698}

\usepackage{lipsum}
\newenvironment{mybox}{\vspace{2mm} \begin{tcolorbox}[colframe=mycolor, left=1pt,right=4pt,top=2pt,bottom=2pt]}{\end{tcolorbox} \vspace{1mm}}

\newcommand{\citeParMetis}{\cite{schloegel+2002, lasalle+2013}}
\newcommand{\citeGMSH}{\cite{geuzaine+2009}}

\newcommand{\citeDune}{\cite{bastian+2008}}

\newcommand{\dune}{\setlength\emergencystretch{3cm}\texttt{DUNE}}
\newcommand{\gmsh}{\setlength\emergencystretch{3cm}\texttt{gmsh}}
\newcommand{\ParMETIS}{\setlength\emergencystretch{3cm}\texttt{ParMETIS}}

\newcommand{\ParaView}{\setlength\emergencystretch{3cm}\texttt{ParaView}}
\newcommand{\visit}{\setlength\emergencystretch{3cm}\texttt{VisIt}}
\newcommand{\hades}{\setlength\emergencystretch{3cm}\texttt{HADES3D}}
\newcommand{\dunegeom}{\setlength\emergencystretch{3cm}\texttt{dune-geometry}}
\newcommand{\dunegrid}{\setlength\emergencystretch{3cm}\texttt{dune-grid}}
\newcommand{\curvgeom}{\setlength\emergencystretch{3cm}\texttt{dune-curvilineargeometry}}
\newcommand{\curvgrid}{\setlength\emergencystretch{3cm}\texttt{dune-curvilineargrid}}
\newcommand{\curvreader}{\setlength\emergencystretch{3cm}\texttt{curvreader}}
\newcommand{\curvwriter}{\setlength\emergencystretch{3cm}\texttt{curvwriter}}

\makeindex

\begin{document}

\lstset{language=C++, breaklines=true}

%\begin{titlepage}

\begin{center}
    
\noindent \textsc{{\Large Dune-CurvilinearGrid}}

\vspace{5mm}

\noindent \textbf{\textsc{{\Large Parallel Dune Grid Manager for\\Unstructured Tetrahedral Curvilinear Meshes}}}
  
\vspace{2mm}
    
{\large
    
\noindent Aleksejs Fomins$^{\mathrm{a,b}}$ and Benedikt Oswald$^{\mathrm{b}}$

  }

\vspace{1mm}

\noindent $^{\mathrm{a}}$ Nanophotonics and Metrology Laboratory (\texttt{nam.epfl.ch}), 
\noindent Ecole Polytechnique F\'ederale de Lausanne (EPFL), Switzerland
  
\vspace{1mm}

\noindent $^{\mathrm{b}}$ LSPR AG, Grubenstrasse 9, CH-8045 Z\"urich, Switzerland.
\noindent Phone +41 43 366 90 74 \\
\noindent
email: \texttt{aleksejs.fomins@epfl.ch} and \texttt{benedikt.oswald@lspr.ch}

\vspace{2mm}

\end{center}

\begin{figure}[H]
    \centering
	\begin{subfigure}[b]{\textwidth} \hspace{4mm} \includegraphics[scale=1.5]{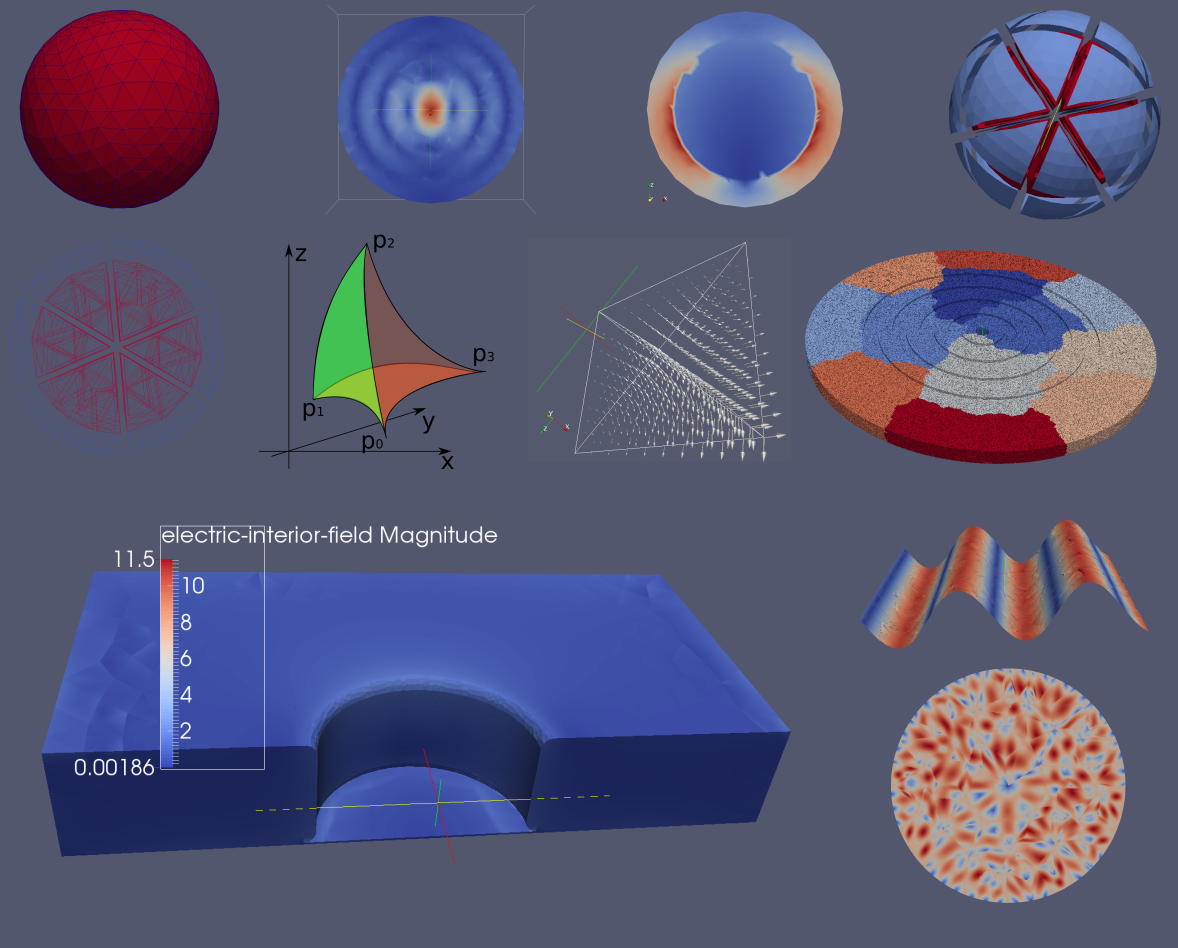} \end{subfigure}
    
% 	\begin{subfigure}[b]{0.3\textwidth} \hspace{4mm} \includegraphics[scale=0.2]{images/32-full} \end{subfigure}
% 	\begin{subfigure}[b]{0.3\textwidth} \hspace{4mm} \includegraphics[scale=0.07]{images/mie-1000-febi-2ndtry}       \end{subfigure}
% 	\begin{subfigure}[b]{0.3\textwidth} \hspace{4mm} \includegraphics[scale=0.08]{images/example_herzian}       \end{subfigure}
% 	\begin{subfigure}[b]{0.3\textwidth} \hspace{4mm} \includegraphics[scale=0.10]{images/gmshreader-vtk-wireframe}       \end{subfigure}
% 	\begin{subfigure}[b]{0.3\textwidth} \hspace{4mm} \includegraphics[scale=0.1]{images/tutorial3-vtk-globalsinusoid}       \end{subfigure}
% 	\begin{subfigure}[b]{0.3\textwidth} \hspace{4mm} \includegraphics[scale=0.15]{images/sphere32discr6ord5}       \end{subfigure}
% 	\begin{subfigure}[b]{0.3\textwidth} \hspace{4mm} \includegraphics[scale=0.08]{images/plane_wave_example}       \end{subfigure}
% 	\begin{subfigure}[b]{0.3\textwidth} \hspace{4mm} \includegraphics[scale=0.15]{images/Edge1_Kind1}       \end{subfigure}
% 	\begin{subfigure}[b]{0.3\textwidth} \hspace{4mm} \includegraphics[scale=0.15]{images/bullseye-core-angle}       \end{subfigure}
% 	\begin{subfigure}[b]{0.3\textwidth} \hspace{4mm} \includegraphics[scale=0.1]{images/cavity-1000-field-at-600nm}       \end{subfigure}
\end{figure}

\pagebreak

%%%%%%%%%%%%%%%%%%%%%%%%%%%%%%%%%%%%%%%%%%%%%%%%%%%%%%%%%%%%%%%%%%%%%%%%
% ACRONYMS
%%%%%%%%%%%%%%%%%%%%%%%%%%%%%%%%%%%%%%%%%%%%%%%%%%%%%%%%%%%%%%%%%%%%%%%%

\noindent
\textbf{List of Acronyms} \\
%\texttt{SNOM/NSOM} - Near-field Scanning Optical Microscopy \\
%\texttt{TERS} - Tip-Enhanced Raman Spectroscopy\\
%\texttt{SERS} - Surface-Enhanced Raman Spectroscopy\\
%\texttt{SPR} - Surface Plasmon Resonance\\
%\texttt{PEF} - Plasmon-Enhanced Fluorescence\\
%\texttt{AFM} - Atomic Force Microscopy\\
%\texttt{SEM} - Scanning Electron Microscopy\\
%\texttt{TEM} - Transverse Electric Magnetic (Mode)\\
\texttt{LSPR} - Localised Surface Plasmon Resonance\\
%
%\texttt{DRA} - Dielectric Resonator Antenna\\
%
\texttt{DUNE} - Distributed Unified Numerical Library\\
\texttt{PDE} - Partial Differential Equation\\
\texttt{DoF} - Degree of Freedom\\
\texttt{HTVFE} - Hierarchical Tangential Vector Finite Elements\\
\texttt{FDTD} - Finite Difference Time Domain \\
\texttt{FEM} - Finite Element Method\\
\texttt{SIE} - Surface Integral Equation (Method)\\
\texttt{CG} - Continuous Galerkin\\
\texttt{DG} - Discontinuous Galerkin\\
\texttt{DGTD} - Discontinuous Galerkin Time Domain\\
\texttt{DGFD} - Discontinuous Galerkin Frequency Domain\\
%\texttt{FETI-DP} - Finite Element Tearing and Interconnect - Dual Primal (Method)\\
%\texttt{DD} - Domain Decomposition\\
\texttt{ABC} - Absorbing Boundary Condition\\
%\texttt{PML} - Perfectly Matched Layer\\
\texttt{BI} - Boundary Integral \\
\texttt{FEBI} - Finite Element Boundary Integral\\
\texttt{MPI} - Message Passing Interface\\
\texttt{POD} - Plain Old Datatype\\
\texttt{RAM} - Random Access Memory\\
\texttt{GPU} - Graphical Processing Unit\\
\texttt{CAD} - Computer-Aided Design\\
\texttt{GSL} - GNU Standard Library\\
\texttt{STL} - Standard Template Library\\
\texttt{GPL} - General Public License\\
\texttt{EPFL} - \'Ecole Polytechnique F\'ed\'erale de Lausanne \\
\texttt{NAM} - Nanophotonics and Metrology (Laboratory, EPFL) \\
\texttt{EDPO} - Ecole Doctorale Photonique (EPFL)\\
%\texttt{PSI} - Paul Scherrer Institut\\
%\texttt{CSCS} - Centro Svizzero di Calcolo Scientifico\\

%%%%%%%%%%%%%%%%%%%%%%%%%%%%%%%%%%%%%%%%%%%%%%%%%%%%%%%%%%%%%%%%%%%%%%%%
% ABSTRACT
%%%%%%%%%%%%%%%%%%%%%%%%%%%%%%%%%%%%%%%%%%%%%%%%%%%%%%%%%%%%%%%%%%%%%%%%

\pagebreak
\vfill

\noindent \textbf{\textsc{ABSTRACT}} - We introduce the \curvgrid{} module. The module provides the self-contained, parallel grid manager \curvgrid{}, as well as the underlying elementary curvilinear geometry module \curvgeom{}. Both modules are developed as extension of the \dune{} \citeDune{} project, and conform to the generic \dunegrid{} and \dunegeom{} interfaces respectively. We expect the reader to be at least briefly familiar with the \dune{} interface to fully benefit from this paper.
\curvgrid{} is a part of the computational framework developed within the doctoral thesis of Aleksejs Fomins. The work is fully funded by and carried out at the technology company LSPR AG. It is motivated by the need for reliable and scalable electromagnetic design of nanooptical devices, achieved by \hades{} family of electromagnetic codes. It is of primary scientific and industrial interest to model full 3D geometric designs as close to the real fabricated structures as possible. Curvilinear geometries improve both the accuracy of modeling smooth material boundaries, and the convergence rate of PDE solutions with increasing basis function order \cite{fahs2011}, reducing the necessary computational effort. Additionally, higher order methods decrease the memory footprint of PDE solvers at the expense of higher operational intensity, which helps in extracting optimal performance from processing power dominated high performance architectures \cite{williams+2009}.
\curvgeom{} is capable of modeling simplex entities (edges, triangles and tetrahedra) up to polynomial order 5 via hard-coded \textit{Lagrange} polynomials, and arbitrary order via analytical procedures. Its most notable features are local-to-global and global-to-local coordinate mappings, symbolic and recursive integration, symbolic polynomial scalars, vectors and matrices (e.g. Jacobians and Integration Elements).
\curvgrid{} uses the \curvgeom{} module to provide the following functionality: fully parallel input of curvilinear meshes in the \gmsh{} \citeGMSH{} mesh format, processing only the corresponding part of the mesh on each available core; mesh partitioning at the reading stage (using \ParMETIS{} \citeParMetis{}); unique global indices for all mesh entities over all processes; Ghost elements associated with the interprocessor boundaries; interprocessor communication of data for shared entities of all codimensions via the standard \dune{} data handle interface. There is also significant support for Boundary Integral (BI) codes, allowing for arbitrary number of interior boundary surfaces, as well as all-to-all dense parallel communication procedures.
The \curvgrid{} grid manager is continuously developed and improved, and so is this documentation. %Among other things, we are working on higher order basis functions for curvilinear grids, non-uniform adaptive h and p-refinement, periodicity. 
For the most recent version of the documentation, as well as the source code, please refer to the following repositories \\

\noindent
\url{http://www.github.com/lspr-ag/dune-curvilineargeometry}\\
\url{http://www.github.com/lspr-ag/dune-curvilineargrid}\\

\noindent
and our website\\

\noindent
\url{http://www.curvilinear-grid.org}

%\end{titlepage}

%%%%%%%%%%%%%%%%%%%%%%%%%%%%%%%%%%%%%%%%%%%%%%%%%%%%%%%%%%%%%%%%%%%%%%%%
% ACKNOWLEDGEMENTS
%%%%%%%%%%%%%%%%%%%%%%%%%%%%%%%%%%%%%%%%%%%%%%%%%%%%%%%%%%%%%%%%%%%%%%%%

\pagebreak
\vspace{15mm}
\noindent \textbf{Acknowledgements} - While the architecture, implementation and down-to-earth programming work for \curvgrid{} grid manager is credited to Aleksejs Fomins and Benedikt Oswald, both authors are pleased to acknowledge the inspiration and support from the wider \dune{} community. We mention names in alphabetical order and sometimes associated with a specific subject. In case we have forgotten to acknowledge an important contributor we kindly ask you to inform us and we will be happy to insert the name immediately. Here we are:
\textit{Peter Bastian}, Professor at University of Heidelberg, Germany for initial suggestion to consider curvilinear tetrahedral grids in order to reduce the computational burden of the complex linear solver;
\textit{Markus Blatt}, Heidelberg, Germany based independent high performance computing and \dune{} contractor, for numerous hints related to the build system and \dune{} architecture;
\textit{Andreas Dedner}, professor, University of Warwick, United Kingdom, for numerous hints related to the \dune{} architecture;
\textit{Christian Engwer}, professor, University of M\"unster, Germany, for fruitful discussions on \dune{} and \curvgrid{} interface;
\textit{Jorrit 'Hippi Joe' Fahlke}, postdoctoral scientist, University of M\"unster, Germany, for numerous hints related to the \dune{} architecture, grid API, grid testing and many other fields;
\textit{Christoph Gr\"uniger}, doctoral student, University of Stuttgart, Germany for support with \dune{} internal implementation;
\textit{Dominic Kempf}, doctoral student, University of Heidelberg, Germany for support w.r.t grid API implementation in \dune{}, especially w.r.t \textit{CMake};
\textit{Robert Kl\"ofkorn}, senior research scientist, IRISI, Norway, for support w.r.t grid API implementation in \dune{};
\textit{Steffen M\"uthing}, postdoctoral scientist, University of Heidelberg, Germany for support with \dune{} internal implementation;
\textit{Martin Nolte}, postdoctoral scientist, University of Freiburg im Breisgau, Germany, for numerous hints related to the \dune{} architecture;
\textit{Oliver Sander}, professor, TU Dresden, Germany, for numerous hints related to the \text{DUNE} architecture, numerical integration and quadrature; \\

\noindent
Also we would like to express special gratitude to Prof. Olivier Martin for supervising the thesis of Aleksejs Fomins, and to Prof. Emeritus. J\"org Waldvogel, who has spent significant amounts of his private time and even visited our company in order to enlighten us about the deep details of numerical integration. We would like to acknowledge enormous support from other software communities, especially Xiaoye Sherry Li for support with SuperLUDist project, George Karypis for support with ParMETIS and Steven G. Johnson for support with his GSL cubature implementation. Finally we would like to express our gratitude to Prof. Torsten Hoefler for advices on newest MPI standards, and Prof. Ralf Hiptmair for fruitful discussions on PDE solvers, including integration. \\

% \noindent
% We are immensely indebted to the above people, and perhaps many others who have contributed with their advices. Community support is the crucial element that makes science possible.

%%%%%%%%%%%%%%%%%%%%%%%%%%%%%%%%%%%%%%%%%%%%%%%%%%%%%%%%%%%%%%%%%%%%%%%%
% BY WHOM DEVELOPMENT WAS SPONSORED
%%%%%%%%%%%%%%%%%%%%%%%%%%%%%%%%%%%%%%%%%%%%%%%%%%%%%%%%%%%%%%%%%%%%%%%%

\vspace{5mm}
{\small
\noindent \textbf{LEGAL NOTICE} - The development of the \curvgrid{} grid manager is sponsored and fully funded by the technology company \textbf{LSPR AG}, Grubenstrasse 9, CH-8045 Z\"urich, Switzerland. \textbf{LSPR AG} exclusively holds all rights associated with \curvgrid{}.
The \curvgrid{} fully parallel grid manager is publicly available via \texttt{Github} and is a free software based on the GPLv2 license. Other licenses can be negotiated through \texttt{curvilinear-grid@lspr.swiss}.
We herewith exclude any liability on the part of \textbf{LSPR AG} since the software is made available as is. Any user uses the software at his own risk and by no means LSPR AG assumes any responsibility for harmful consequences or any other damage caused by the use of the software. The whole project is governed by Swiss Law. We reject any attempt of any other sovereign law to cover what we do.
}

\tableofcontents

\newpage
\section{Introduction}

%%%%%%%%%%%%%%%%%%%%%%%%%%%%%%%%%%%%%%%%%%%%%%%%%%%%%%%%%%%%%%%%%%%%%
% Curvilinear Grid Outline - Section on capabilities of the Grid
%%%%%%%%%%%%%%%%%%%%%%%%%%%%%%%%%%%%%%%%%%%%%%%%%%%%%%%%%%%%%%%%%%%%%

\noindent
Integrating curvilinear geometries into modeling software is an involved multi-level process. It requires meshing software capable of creating accurate higher order elements from analytic designs or experimental data, a curvilinear grid manager able to efficiently manipulate such elements, as well as a PDE solver able to benefit from the curved geometries. The latter is mainly achieved by means of curvilinear basis functions able to accurately describe the associated vector fields (e.g. \textit{div} or \textit{curl}-conforming), adapting to the non-linear "bending of space". Ideas of using curvilinear grids first appeared in the literature in the 1970s \citep{ciarlet+1972, lenoir1986} and have been used in electromagnetic simulations for at least two decades \citep{wang+1993}. \\

\noindent
An impressive example of using a curvilinear $3$-dimensional code together with DG and optimized for parallel GPU processing can be found in the aerodynamics community \cite{Warburton2012}.
Wang et al \cite{wang+2011} demonstrate a 3D curvilinear parallel DGTD (\textit{Time-Domain}) code for solving Maxwell's equations in a homogeneous medium.
Nevertheless, in electromagnetic community curvilinear grids are much less widespread than linear, predominantly used in 2D codes \citep{wang+2011a}.
We believe that the associated challenges are as follows
\begin{itemize}
\item The generation of curvilinear meshes is a challenging process, as naive approaches can result in self-intersecting meshes \citep{toulorge+2013, johnen+2012}. Further, it must be ensured that the generated elements are optimal for the optimal PDE convergence \cite{lenoir1986}.
\item Standard functionality of a grid manager, such as interpolation, local-to-global and global-to-local mappings, integration, calculation of normals and basis functions becomes significantly more difficult in the curvilinear case; additional numerical tools, such as \textit{Lagrange} interpolation, adaptive integration, symbolic polynomial manipulation, and optimization algorithms are needed to provide the desired functionality. 
\item In order to fully benefit from the curvilinear geometries through reducing the total element count, basis functions of order sufficient to resolve the detailed structure of the field are necessary. The widely used CG-based codes require a divergenceless curvilinear basis of flexible order that preserves the field continuity across the element boundary. At the moment of writing authors are not aware of publications presenting such a basis. Fahs\cite{fahs2011} implements a serial 2D and 3D curvilinear DGTD code using polynomially-complete basis, and studies the scaling of the accuracy of electromagnetic benchmarks (\textit{Mie} scattering and layered \textit{Mie} scattering) depending on $p$-refinement. He finds that only in curvilinear geometries increasing the basis function order significantly improves the solution accuracy.
\end{itemize}

\noindent
Until recently, literature presents implementations of curvilinear electromagnetic DGFD codes with several simplifications, limiting the flexibility and detail achievable with moderate computational resources. \\

\begin{figure}
    \centering
	\begin{subfigure}[b]{0.48\textwidth} \hspace{8mm} \includegraphics[scale=0.215]{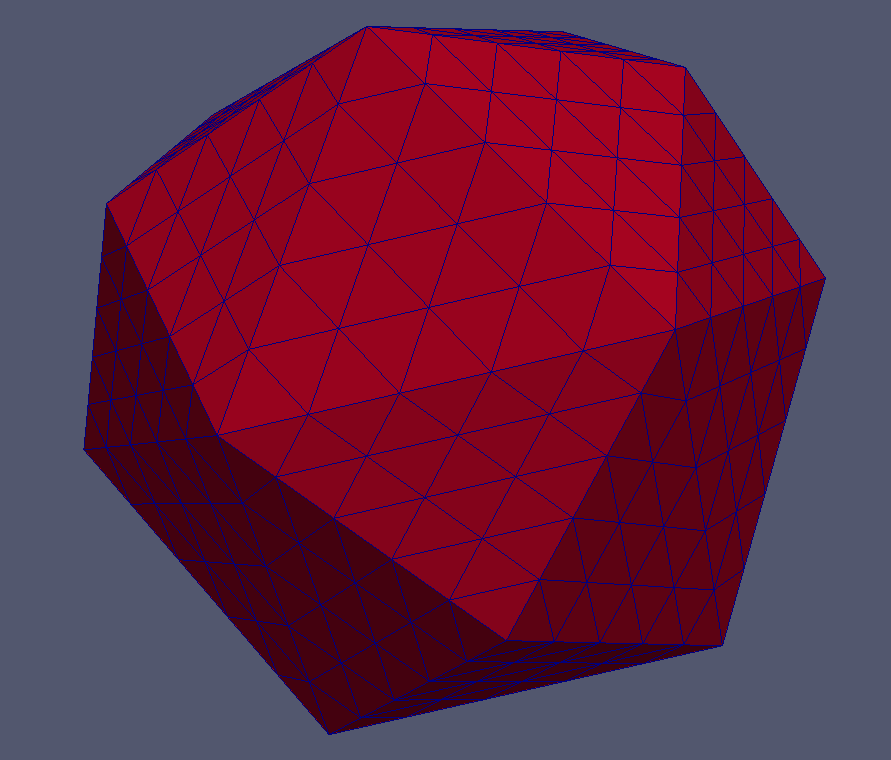} \end{subfigure}
	\begin{subfigure}[b]{0.48\textwidth} \includegraphics[scale=0.2]{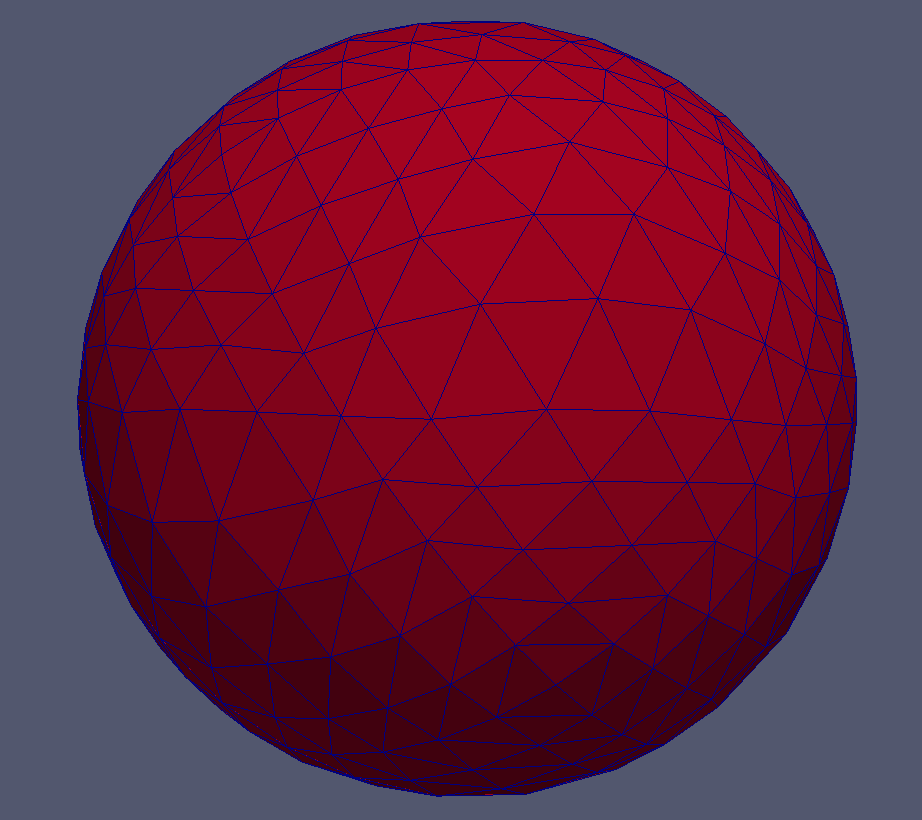} \end{subfigure}
	\captionsetup{width=0.8\textwidth} 
	\caption{Presented is the 32 element tetrahedral mesh of a sphere, using first and fifth order polynomial interpolation. The curvature is represented by virtual refinement of curvilinear tetrahedra via smaller linear tetrahedra. %Visualization software (\ParaView{}, in this case) does not appear to have documented interface for direct curvilinear element output.
	}
	\label{fig:result:spherecurv}
\end{figure}

\noindent
The major objective for a PDE solver optimization is the improvement of accuracy of the PDE solution given a limited computational resource, and curvilinear geometries can offer that. The curvilinear material boundaries decrease or fully eliminate the artificially high jump in the surface derivatives \cref{fig:result:spherecurv}, avoiding the unphysical "corner effects" \cite{volakis1998, jin2014}. Otherwise, the corners have to be smoothened by high \textit{h}-refinement, which leads to unnecessarily high number of \textit{Degrees of Freedom} (DoF). \\

\begin{figure}
    \centering
    \includegraphics[scale=0.2]{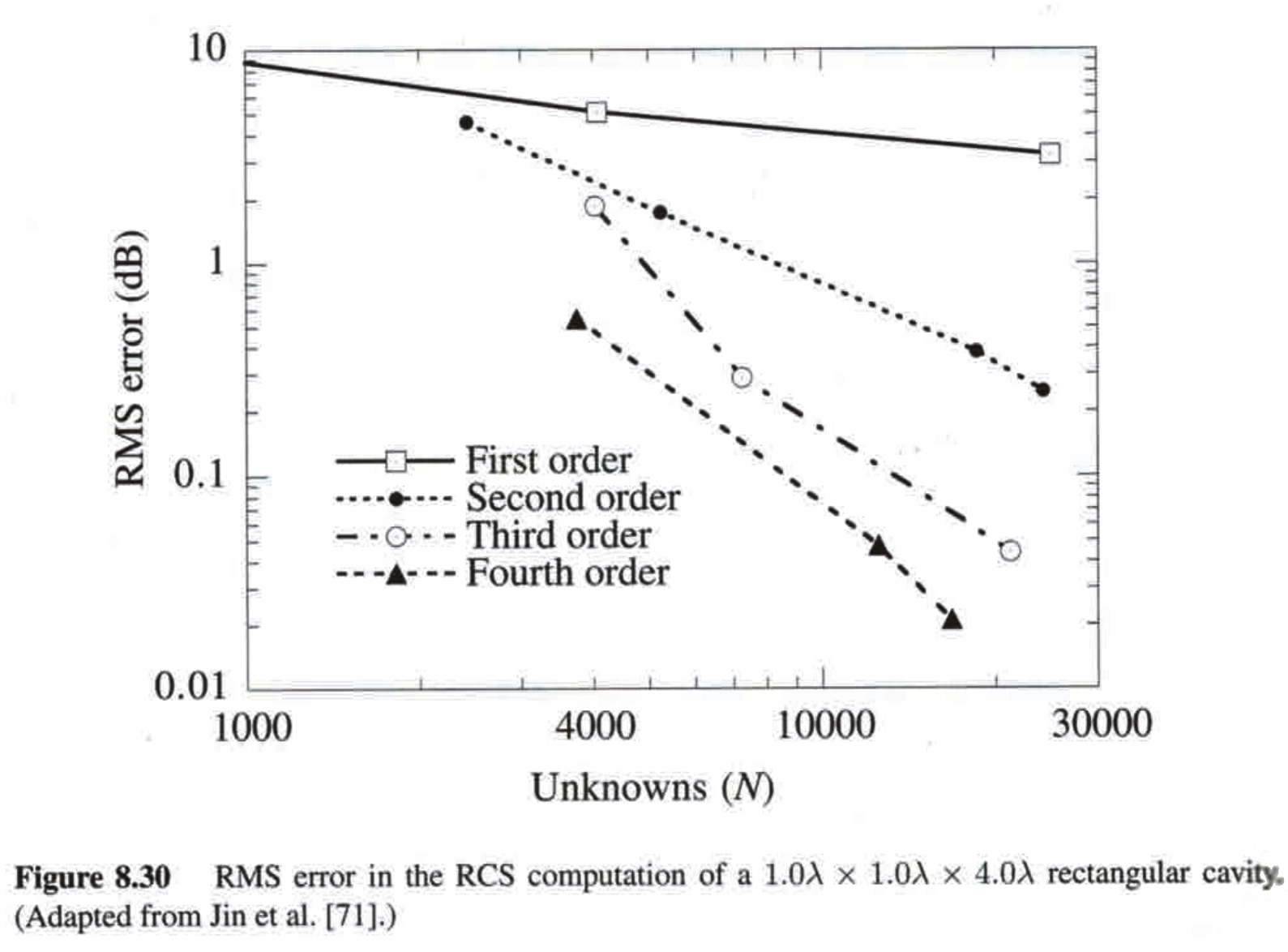}
	\captionsetup{width=0.8\textwidth} 
	\caption{ Jin \cite{jin2014} shows that the improvement of accuracy due to h-refinement improves exponentially with increasing basis order. We thank the author for permission to reproduce this plot. }
	\label{fig:jin:basisconv}
\end{figure}

\begin{figure}
    \includegraphics[scale=0.25]{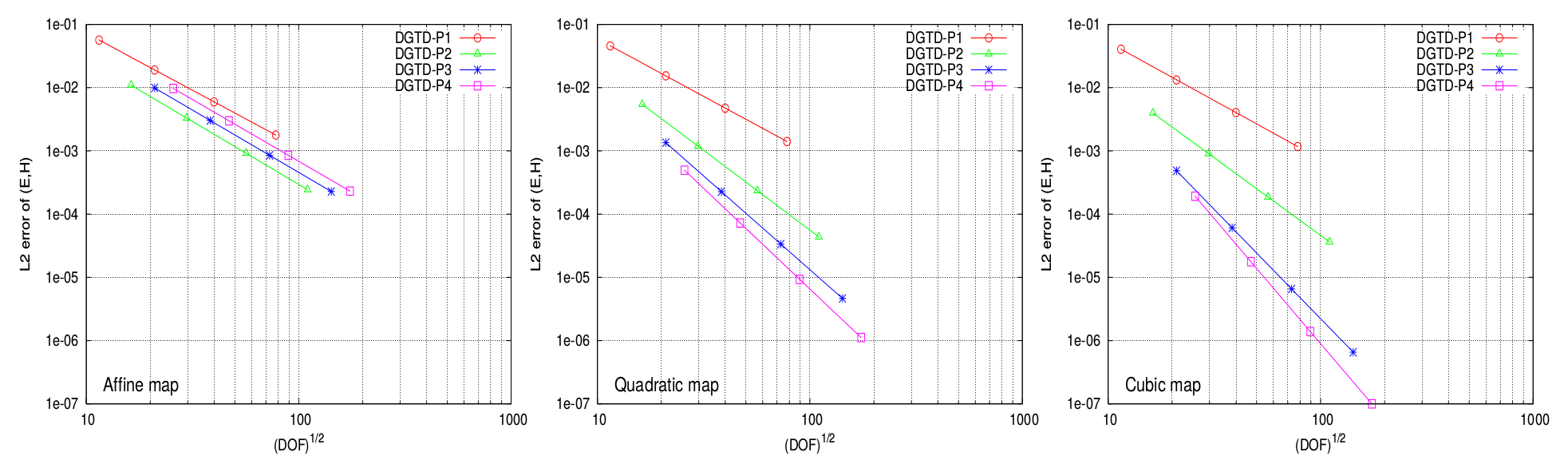}
	\captionsetup{width=0.8\textwidth} 
	\caption{ Fahs \cite{fahs2011} shows that computational accuracy (in terms of $L_2$ norm) improves with increasing basis function order, but it improves faster if the entity interpolation (curvature) order is increased accordingly. We thank the author for permission to reproduce these plots.}
	\label{fig:fahs:curvconv}
\end{figure}

\noindent
Further, the accuracy of a PDE solution improves much faster with increasing basis function order (\textit{p}-refinement) than with increasing element number (\textit{h}-refinement) \cite{jin2014}, \cref{fig:jin:basisconv}. Fahs \cite{fahs2011} shows that, in case of curved material boundaries, this effect can only be exploited if the corresponding element geometries are of sufficiently high order \cref{fig:fahs:curvconv}.

\subsection{Capabilities of CurvilinearGrid}
\label{section-outline-capabilities}

The \curvgrid{} is a self-consistent grid manager supporting 3D tetrahedral curvilinear grids. It depends on the core modules of \dune{} \citeDune{}, as well as an external parallel mesh partition library \ParMETIS \citeParMetis{}. \curvgrid{} also depends on \curvgeom{}, which we developed as a separate \dune{} module. \\

\noindent
\curvgeom{} is capable of interpolating and performing multiple geometric operations over curvilinear simplex entities (edges, triangles and tetrahedra) of orders 1-5 via hard-coded \textit{Lagrange} polynomials, and arbitrary order simplex entities via analytic \textit{Lagrange} interpolation method. \curvgeom{} complies with the standard \dunegeom{} interface, providing methods for local-to-global and global-to-local coordinate mapping, computation of the \textit{Jacobian} matrix, integration element and entity volume. \curvgeom{} has non-cached and cached implementations, where the cached version pre-computes the local-to-global map and its determinant, thus performing considerably faster for integration and mapping tasks. In comparison with the standard \dunegeom{}, \curvgeom{} provides methods to obtain either all interpolatory vertices of an entity or only its corners, as well as the method to obtain the curvilinear order. Additionally, \curvgeom{} provides the methods to obtain the outer normals of subentities of the geometry, and the subentity geometries themselves. Another feature of \curvgeom{} is the symbolic polynomial class and associated differential methods, which allow to obtain analytical expressions for local-to-global map and associated integration element, enabling exact manipulation of geometries of arbitrary order. \curvgeom{} contains its own recursive integration tool, wrapping the quadrature rules provided by dune-geometry. The reason for implementing this functionality is to accurately treat non-polynomial integrands for which the optimal polynomial order required for the desired accuracy is not known. In particular, it happens that curvilinear integration elements are non-polynomial in the general case (see \cref{sec:theory:integration}). The recursive integration scheme is capable to simultaneously integrate multidimensional integrands, such as vectors and matrices. This is highly useful, for example, for integrating outer product matrices. For such matrices the evaluation of all matrix elements at a given coordinate only requires $O(N)$ expensive function evaluations. \curvgeom{} provides a utility for testing curvilinear entities for self-intersection. This is done by sampling the value of integration element across the entity, and ensuring that it never changes sign. \\

\begin{figure}[H]
    \centering
    \includegraphics[scale=0.15]{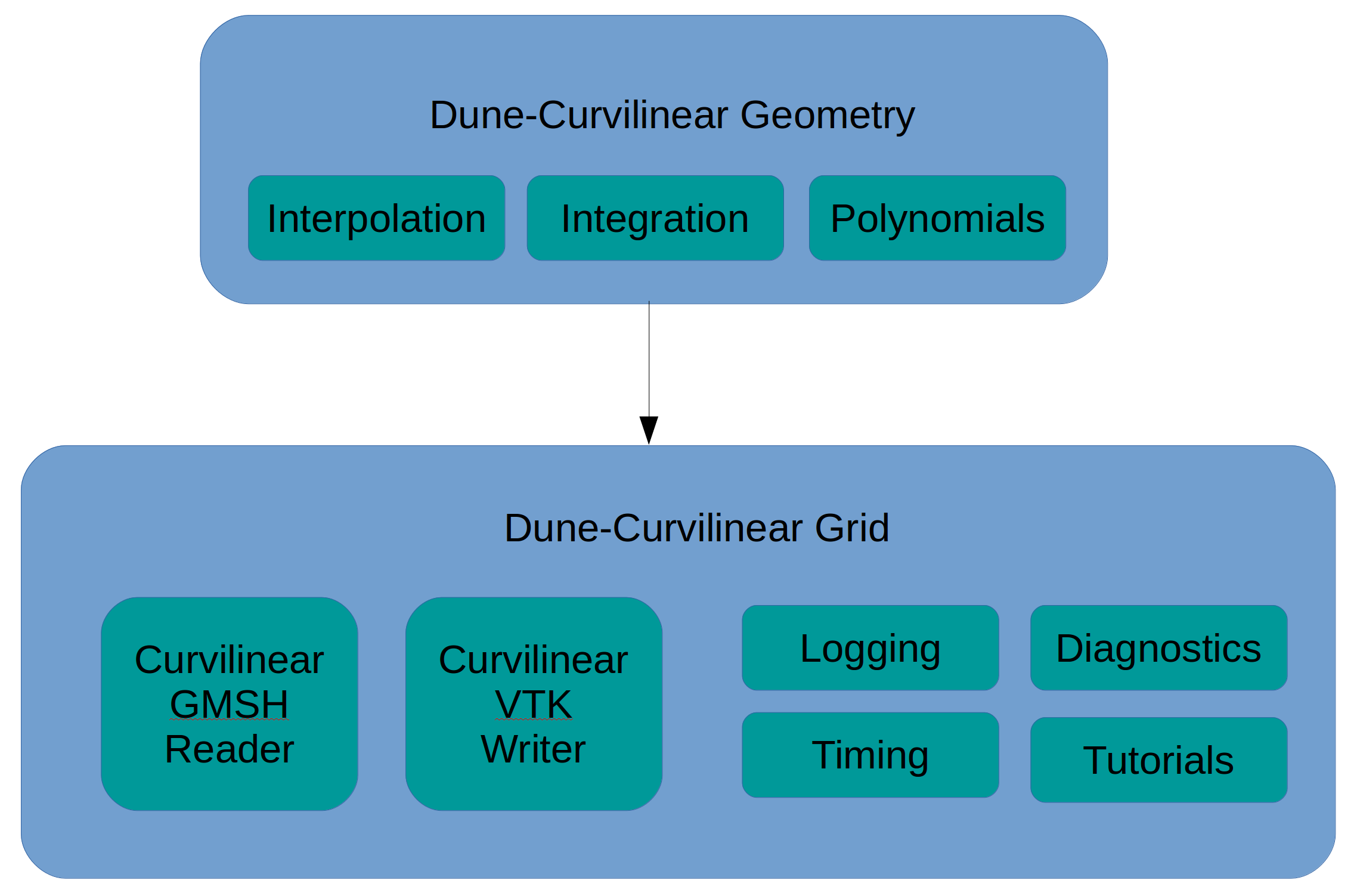}
    \caption{The structure of \curvgrid{}}
    \label{fig:introduction:curvgrid:structure}
\end{figure}

\begin{figure}
	\begin{subfigure}[b]{0.30\textwidth} \hspace{4mm} \includegraphics[scale=0.22]{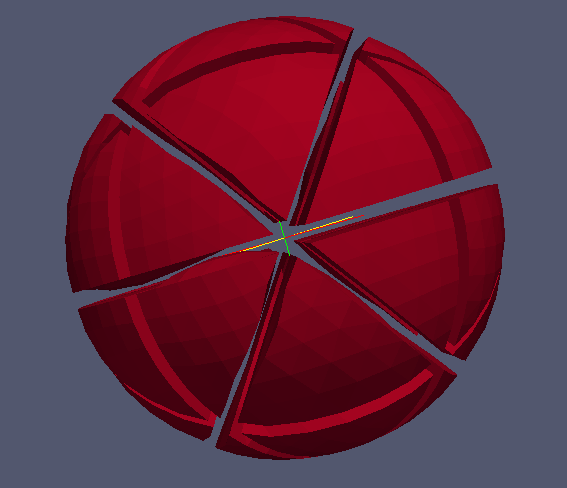} \captionsetup{width=0.8\textwidth} \caption{ Interior elements } \end{subfigure}
	\begin{subfigure}[b]{0.30\textwidth} \hspace{4mm} \includegraphics[scale=0.18]{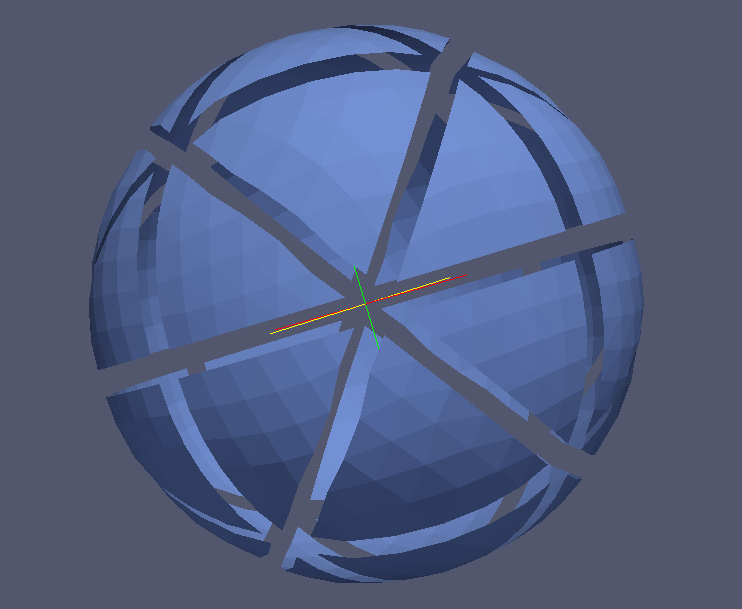}    \captionsetup{width=0.8\textwidth} \caption{ Domain Boundary surfaces} \end{subfigure}
	\begin{subfigure}[b]{0.33\textwidth} \hspace{4mm} \includegraphics[scale=0.22]{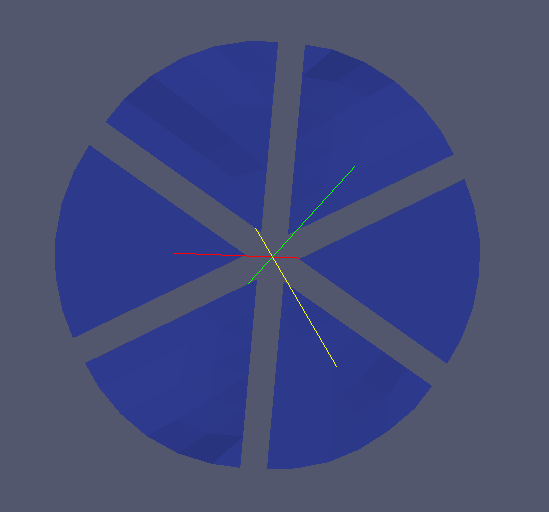}    \captionsetup{width=0.8\textwidth} \caption{ Interprocessor Boundary surfaces} \end{subfigure}
	\begin{subfigure}[b]{0.46\textwidth} \vspace{5mm} \hspace{12mm} \includegraphics[scale=0.25]{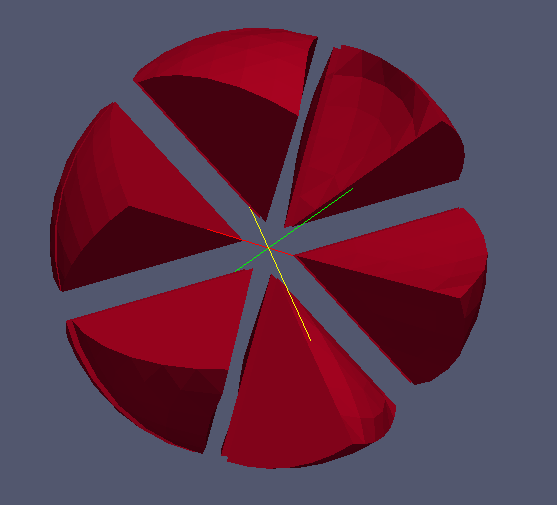} \captionsetup{width=0.6\textwidth} \caption{ Ghost elements, borrowed from neighboring processes} \end{subfigure}
	\begin{subfigure}[b]{0.46\textwidth} \vspace{5mm} \hspace{12mm} \includegraphics[scale=0.21]{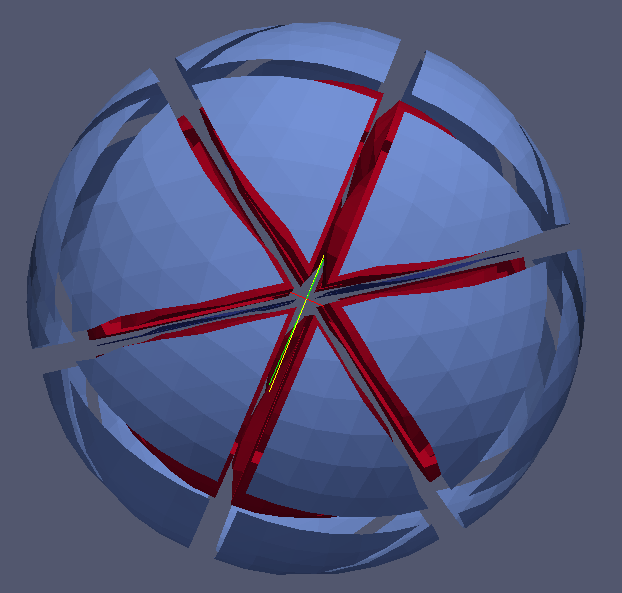}  \captionsetup{width=0.6\textwidth} \caption{ Entities of all structural types visualized at the same time } \end{subfigure}
	\caption{ Visualization of various structural (partition) types of a 32 element tetrahedral mesh, loaded in parallel on 2 cores }
	\label{fig:result:spherestruct}
\end{figure}

\begin{figure}
    \centering
	\begin{subfigure}[b]{0.45\textwidth} \hspace{3mm} \includegraphics[scale=0.20]{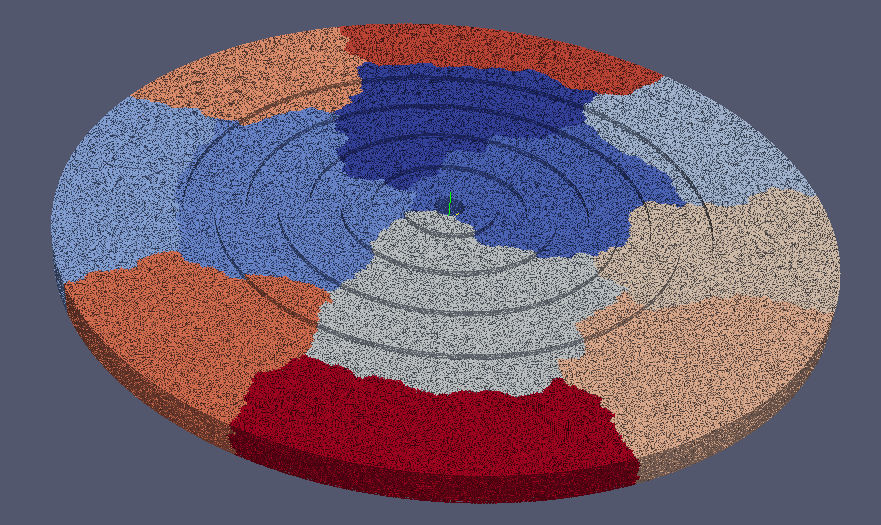}          \captionsetup{width=0.8\textwidth} \caption{Interior elements coloured by owner process rank}   \end{subfigure}
	\begin{subfigure}[b]{0.45\textwidth} \hspace{3mm} \includegraphics[scale=0.17]{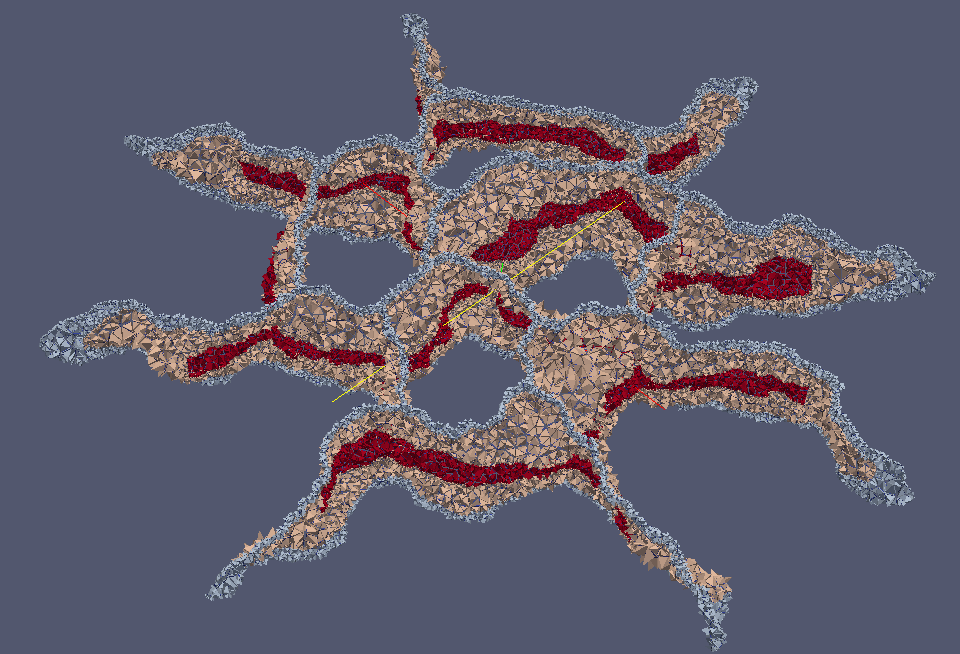}  \captionsetup{width=0.8\textwidth} \caption{Ghost elements, coloured by material tag} \end{subfigure}
	\caption{Interior and ghost elements of a 4.2 million element tetrahedral mesh of the Bullseye geometry, loaded on 12 cores}
	\label{fig:result:bullseye}
\end{figure}

\noindent
\curvgrid{} module manages the entire process of reading, manipulating, and writing of curvilinear geometries and associated fields (e.g. PDE solution). The former is accomplished by \textit{Curvilinear GMSH Reader} (\curvreader{}) class. \curvreader{} is currently capable of reading curvilinear \textit{.msh} files of orders 1 to 5, where 1 corresponds to linear meshes. \curvreader{} is fully parallel and scalable for large parallel architectures. Each process only reads the necessary local part of the mesh, distributing the memory equally among all processes. It must be noted that earlier implementation of \textit{GMSH Reader} in the \dunegrid{} module suffered from serial reading of the mesh on the master process, which is no longer a bottleneck in our implementation. \curvreader{} has the option to partition the mesh using \ParMETIS{} during the reading procedure before reading the curvature vertices, further decreasing the file access time. \curvreader{} also reads material elementary and boundary tags provided by \gmsh{}. It extends the standard \textit{Grid Factory} interface, providing tag information, as well as curvilinear order. The grid output is accomplished by \textit{Curvilinear VTK Writer} (\curvwriter{}) module, supporting \textit{VTK}, \textit{VTU} and \textit{PVTU} file formats. \curvwriter{} can either write the entire grid automatically, or write a set of individual entities, one at a time. When writing the entire grid, each element is supplied by fields denoting its rank, partition type (\cref{fig:result:spherestruct}) and physical tag, which can be used to visually inspect the parallel connectivity of the domain. The scalability of the grid assembly and visualization has been tested on parallel architectures containing from 12 to 128 cores. By the time of writing, the \curvgrid has been successfully run on several dozen different meshes, the largest being the 4.4 million element tetrahedral mesh \cref{fig:result:bullseye}. The user has full flexibility to define the codimensions of the entities that will be written, the choice to write interior, domain, process boundaries and/or ghost elements, as well as the order of virtual refinement of curvilinear entities. The output mesh can be supplied with an arbitrary number of vector and scalar fields representing, for example, the solution(s) of a PDE. We have tested the visualization capabilities of \curvgrid{} using \ParaView{} \cite{johnson+2005} and \visit{} \cite{childs+2012} end user software. \\

\noindent
The core of \curvgrid{} provides the essential indexing and communication capabilities. The global and local indices are provided for entities of all codimensions. Interprocessor communication is performed via the \dune{} standard \textit{DataHandle} interface for provided \textit{Ghost} elements entities of all codimensions. As an extension to the \dunegrid{} interface, it is possible to directly address the core curvilinear geometry of each entity, as well as the associated physical tags. \curvgrid{} is also equipped with a set of useful utilities:
\begin{itemize}
    \item \textit{Timing mechanisms}: parallel timing of separate parts of the code with statistics output over all processes
    \item \textit{Logging mechanisms}: real time logging of the current state of the code, as well as the current memory consumption on each core of a machine, allowing for the real-time diagnostics of memory bottlenecks of the code.
    \item \textit{Nearest-neighbor communication} - wrapper for the implementation of \textit{MPI\_Neighbor\_alltoall} for vector communication with neighboring processes. This functionality is available as of the MPI-2 standard \cite{MPI-3.1}
    \item \textit{Global boundary container} - interior/domain boundary all-to-all communication, useful for dense PDE solvers, such as the Boundary Integral method. \cite{kern+2009}
    \item \textit{Grid diagnostics} - collects statistics on entity volumes, qualities and their distribution among processes
\end{itemize}

\section{Theory}

%%%%%%%%%%%%%%%%%%%%%%%%%%%%%%%%%%%%%%%
% Theory for Lagrange Polynomials
%%%%%%%%%%%%%%%%%%%%%%%%%%%%%%%%%%%%%%%
\subsection{Lagrange Polynomial Interpolation}
\label{theory-lagrange}

Below we present the theory of interpolation using \textit{Lagrange} polynomials, applied to simplex geometries, This section is inspired by \cite{koshiba+2000, ilic+2003, berrut+2004}, and is a summary of well-known results. The goal of \textit{Lagrange} interpolation is to construct a mapping $\vec{x} = \vec{p}(\vec{r})$ from local coordinates of an entity to global coordinates of the domain. In its own local coordinates, the entity will be denoted as a reference element \citeDune{}. A simplex reference element $\Delta_d$ of dimension $d$ is given by the following local coordinates:
\begin{table}[H]
\centering
\begin{tabular}{l l l}
\hline
  Label & Dimension & Coordinates \\ \hline
  $\Delta_0$ & 0 & $\{ 0 \}$ \\
  $\Delta_1$ & 1 & $\{ 0\}, \{ 1\}$ \\
  $\Delta_2$ & 2 & $\{ 0, 0 \}, \{ 1, 0 \}, \{ 0, 1 \}$ \\
  $\Delta_3$ & 3 & $\{ 0, 0, 0 \}, \{ 1, 0, 0 \}, \{ 0, 1, 0 \}, \{ 0, 0, 1 \}$
\end{tabular}
\caption{Reference element local coordinates}
\label{table:lagrange:refelement}
\end{table}
\noindent
Local simplex geometries can be parametrized using the local coordinate vector $\vec{r}$:
\begin{table}[H]
\centering
\begin{tabular}{l l l}
\hline
  Entity      & Parametrization    & Range \\ \hline
  Edge        & $\vec{r}=(u)$      & $u \in [0,1]$ \\
  Triangle    & $\vec{r}=(u,v)$    & $u \in [0,1]$ and $v \in [0, 1-u]$ \\
  Tetrahedron & $\vec{r}=(u,v,w)$  & $u \in [0,1]$, $v \in [0, 1-u]$ and $w \in [0, 1-u-v]$
\end{tabular}
\caption{Reference element parametrization in local coordinates}
\label{table:lagrange:parametrization}
\end{table}

%\paragraph{Interpolatory Vertices}
%\label{theory-lagrange-vertices}
\noindent
\textbf{Interpolatory Vertices}

\noindent
In order to define the curvilinear geometry, a set of global coordinates $\vec{x}_i = \vec{p}_i(\vec{r}_i)$, known as interpolatory vertices, is provided. By convention, the interpolatory vertices correspond to a sorted structured grid on a reference simplex, namely
\[\vec{r}_{i,j,k} = \frac{(k,j,i)}{Ord}, \;\;\; i=[0..Ord], \;\;\; j=[0..Ord-i], \;\;\; k=[0..Ord-i-j]\]
where $Ord$ is the interpolation order of the entity. It is useful to construct a bijective map from a structured local grid to the provided global coordinates. It is the job of the meshing software to ensure that the global geometry of an entity does is not self-intersecting, non-singular, and that its curvature is optimized for PDE convergence \cite{lenoir1986}. In general, a non-uniform local interpolatory grid should be used in order to minimize the effect of \textit{Runge} phenomenon \cite{runge1901}. It is not an issue for lower polynomial orders, and is the standard currently provided by the available meshing software, so we shall restrict our attention only to uniform interpolation grids. The number of interpolatory points on a uniform grid over the reference simplex is described by triangular/tetrahedral numbers \cref{table:lagrange:nvertex}. These numbers are conveniently also the numbers describing the total number $N_{Ord}$ of polynomially-complete monomials up to a given order:
\begin{table}[H]
\centering
\begin{tabular}{l l l l l l l}
\hline
  Entity \textbackslash Order & 1 & 2  & 3  & 4  & 5 & general \\ \hline
  Edge                        & 2 & 3  & 4  & 5  & 6 & $Ord+1$\\
  Triangle                    & 3 & 6  & 10 & 15 & 21 & $(Ord+1)(Ord+2)/2$\\
  Tetrahedron                 & 4 & 10 & 20 & 35 & 56 & $(Ord+1)(Ord+2)(Ord+3)/6$
\end{tabular}
\caption{Number of vertices in a uniform interpolatory grid over the reference simplex}
\label{table:lagrange:nvertex}
\end{table}

% \paragraph{Interpolatory Polynomials}
% \label{theory-lagrange-polynomials}
\noindent
\textbf{Interpolatory Polynomials}

\noindent
The number of interpolatory vertices $N_{Ord}$ given in \cref{table:lagrange:nvertex} exactly matches the total number of monomials necessary to construct a complete polynomial of order $Ord$ or less. It can be observed that the uniform simplex discretization exactly matches the binomial/trinomial simplex, also known as the Pascal's triangle, commonly used to visualize the complete monomial basis. We define the function $z^{(dim, i)}(\vec{u})$ as the set of all monomials of dimension $dim$ and order less than or equal to $i$. The first few sets for 1D, 2D and 3D are as follows:

\begin{flushleft}
\begin{table}[H]
\centering
\small
\begin{tabular}{l l l}
\hline
  Edge & Triangle & Tetrahedron \\ \hline

  \begin{minipage}[l]{0.34 \textwidth}
  \begin{eqnarray*}
	z^{(1,1)}(u) &=& \{1, u\}, \\
	z^{(1,2)}(u) &=& \{1, u, u^2\}, \\
	z^{(1,3)}(u) &=& \{1, u, u^2, u^3\}, \\
	z^{(1,4)}(u) &=& \{1, u, u^2, u^3, u^4\}, \\
	z^{(1,5)}(u) &=& \{1, u, u^2, u^3, u^4, u^5\},
  \end{eqnarray*}
  \; etc.
  \end{minipage} & 

  \begin{minipage}[l]{0.27 \textwidth}
  \begin{eqnarray*}
	z^{(2,1)}(u,v)	&=& \{1, u, v\}, \\
	z^{(2,2)}(u,v) &=& \{1, u, v, \\
	& & u^2, uv, v^2\},
  \end{eqnarray*}
  \; etc.
  \end{minipage} & 
  
  \begin{minipage}[l]{0.27 \textwidth}
  \begin{eqnarray*}
	z^{(3,1)}(u,v,w) &=& \{1, u, v, w\}, \\ 
	z^{(3,2)}(u,v,w) &=& \{1, u, v, w, \\
	& & u^2, uv, v^2, \\
	& & wu, wv, w^2\},
  \end{eqnarray*}
  \; etc.
  \end{minipage}
\end{tabular}
\normalsize
\caption{First few orders of the complete monomial basis for simplex entities }
\label{table:lagrange:monomial:function}
\end{table}
\end{flushleft}

\noindent
The mapping $\vec{p}(\vec{r})$ is chosen to exactly fit all the interpolatory vertices $\vec{x}_i$. Since the numbers of interpolatory vertices and monomials is the same, the interpolatory vertices will have a \textit{unique} associated \textit{complete} polynomial basis. This is not the same for entities of other geometry types. For example, for hexahedra, the above numbers do not match. Therefore, one either has to use a structured local grid with incomplete polynomial order basis, or choose a more sophisticated local discretization. Volakis et al \cite{volakis+2006} adopt the former approach, interpolating a 9 node 2\textsuperscript{nd} order rectangle with 4\textsuperscript{th} order incomplete polynomial basis that has a convenient separable tensor product form. \\

\noindent
One way to formulate the local-to-global coordinate mapping is
\begin{equation}
	\vec{p}(\vec{r}) = \sum_j L_j(\vec{r})\vec{x}_j 
\end{equation}
\noindent
where $\vec{p}_j $ are the fixed interpolatory vertices, and $L_j$ are the \textit{Lagrange} polynomials, defined by their interpolatory property
\begin{equation}
	\label{equation-lagrangepol-interpolatory-property}
	L_j(\vec{r}_i) = \delta_{ij}
\end{equation}
\noindent
for all local interpolatory vertices $\vec{r}_i$. The advantage of this formulation is that the \textit{Lagrange} polynomials are independent of the interpolatory vertices $\vec{x}_i$, and thus can be pre-computed and reused for all entities of a given order. It remains to determine the exact form of \textit{Lagrange} polynomials. We will present a short proof that \cref{equation-lagrangepol-basis-link} holds
\begin{equation}
	\label{equation-lagrangepol-basis-link}
	z_i(\vec{r}) = \sum_j L_j(\vec{r}) z_i (\vec{r}_j) 
\end{equation}
\noindent
Here, $\{z\}$ is a vector of monomials as defined in \cref{table:lagrange:monomial:function}. Given a fixed dimension $dim$, \cref{equation-lagrangepol-basis-link} should hold for all polynomial orders $Ord$. Both sides of \cref{equation-lagrangepol-basis-link} are polynomials of order at most $Ord$, which means that they have at most $N_{Ord}$ free parameters. Therefore, to prove that \cref{equation-lagrangepol-basis-link} holds in general, it is sufficient to show that it holds for $N_{Ord}$ different arguments. Thus, it is enough to show that it holds for all $\vec{r} = \vec{r}_k$, which in turn is true due to \cref{equation-lagrangepol-interpolatory-property}. Finally, we can combine all monomials and \textit{Lagrange} polynomials into corresponding vectors
\begin{equation}
	\vec{z} (\vec{r}) = V \vec{L} (\vec{r})
\end{equation}
\noindent
where $V_{ij} = z_i (\vec{r}_j)$, and find the \textit{Lagrange} polynomial coefficients by solving the linear system
\begin{equation}
	\label{equation-lagrange-linear-system}
	\vec{L} (\vec{r}) = V^{-1} \vec{z} (\vec{r})
\end{equation}

\noindent
It is important to note that the resulting interpolated geometry in global coordinates is not completely defined by the shape of its boundary, as the entire volume of the geometry inside the entity undergoes this polynomial transformation. \\

% \paragraph{Implementation for Simplices}
% \label{subsection-simplexgrid}
\noindent
\textbf{Implementation for Simplices}

\noindent
In this section we discuss how to efficiently enumerate the simplex interpolatory points and to construct the reference simplex grid. \\

\noindent
Let us define a simplex $\Delta^{\dim}_{n}$ of integer side length $n$, and place a set of points $\vec{\eta} \in \mathbb{Z}^{\dim}$ at unit intervals. This can be done by using nested loops
\begin{itemize}
	\item $\Delta^{1}_n = \{(i)\}$, for $i = [1$ to $n]$
	\item $\Delta^{2}_n = \{(j,i)\}$, for $i = [1$ to $n]$, $j = [1$ to $n - i]$
	\item $\Delta^{3}_n = \{(k,j,i)\}$, for $i = [1$ to $n]$, $j = [1$ to $n - i]$, $k = [1$ to $n - i - j]$
\end{itemize}

\begin{figure}[hp]
    \centering
    \includegraphics[scale=2.0]{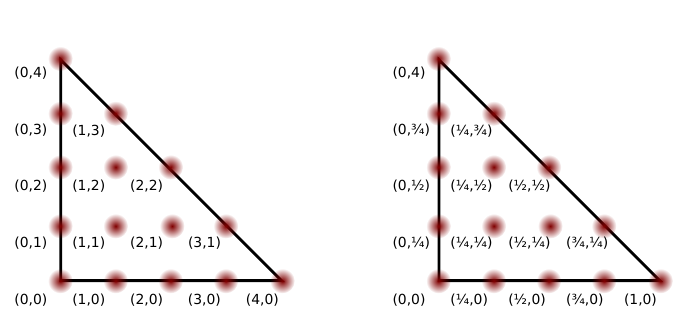}
    \caption{Construction of the uniform grid interpolatory points (right) from the \textit{Cartesian} coordinate enumeration}
    \label{fig:lagrange:enumerationconstruction}
\end{figure}

\noindent
Then, each integer vertex $(\Delta^{d}_n)_i$ corresponds exactly to the power of $u,v,w$ in the expansion of
\[ (1 + u)^n = \sum_{i=0}^n C^{(\Delta^{1}_n)_i}_n u^{(\Delta^{1}_n)_{i,1}} \]
\[ (1 + u + v)^n = \sum_{i=0}^n C^{(\Delta^{1}_n)_i}_n u^{(\Delta^{1}_n)_{i,1}} v^{(\Delta^{1}_n)_{i,2}} \]
\[ (1 + u + v + w)^n = \sum_{i=0}^n C^{(\Delta^{1}_n)_i}_n u^{(\Delta^{1}_n)_{i,1}} v^{(\Delta^{1}_n)_{i,2}} w^{(\Delta^{1}_n)_{i,3}} \]

\noindent
where $C^{i}_n, C^{i,j}_n$ and $C^{i,j,k}_n$ are the binomial, trinomial and quatrinomial coefficients. The powers of the parameters given in the above expansion correspond to the complete monomial basis for a polynomial of order $d$. The local coordinates of the uniform grid over the reference simplex can then be written as $r_i = \frac{1}{n}(\Delta^{d}_n)_i$ (see \cref{fig:lagrange:enumerationconstruction})

\noindent
After the monomials and the parametric interpolation points have been constructed, it remains to construct the interpolation matrix $V$ by evaluating the monomials at the interpolation points and to solve the linear system \cref{equation-lagrange-linear-system}, obtaining the \textit{Lagrange} polynomials. This has been implemented both explicitly, calculating and hard-coding all the simplex \textit{Lagrange} interpolatory polynomials, and implicitly, implementing symbolic polynomial arithmetic. The latter has the advantage of unrestricted polynomial order, as well as the freedom of further analytic manipulation using of symbolic arithmetic and explicit differential operators, but comes at the cost of slight computational overhead. \\

\noindent
The optimization of \textit{Lagrange} polynomial evaluation is of crucial importance, since they are typically evaluated a significant amount of times, especially during the integration and minimization procedures. Our implementation of \textit{Lagrange} polynomials benefits from the following observations:
\begin{itemize}
  \item Each \textit{Lagrange} polynomial of a given order uses the same monomial summands. It is thus of an advantage to evaluate all the \textit{Lagrange} polynomials at the same time, first evaluating all the necessary monomials, and then re-using the evaluated monomials to compute the polynomials.
  \item Along the same lines, evaluating all monomials of a given order at the same time is cheaper than evaluating them separately. Lower order monomials can be used to construct higher order monomials by doing a single multiplication per monomial.
\end{itemize}

%%%%%%%%%%%%%%%%%%%%%%%%%%%%%%%%%%%%%%%
% Theory for global and local mappings
%%%%%%%%%%%%%%%%%%%%%%%%%%%%%%%%%%%%%%%
\subsection{Coordinate transformation}
\label{sec:theory:coordinatetransform}

In order to calculate the coordinate transformation properties, one requires the knowledge of the local-to-global map $\vec{p}(\vec{r})$ and its first partial derivatives. Currently, \curvgeom{} only provides \textit{Lagrange} polynomials themselves as hard-coded expressions. Their derivatives are not yet available as hard-coded quantities, and thus are constructed by differentiating the symbolic polynomial map. This is, naturally, a little slower than having hard-coded derivatives. The advantage of analytical formulation is that the user can further apply algebraic and differential operators to the symbolic map to obtain, for example, a \textit{Hessian} matrix of the transformation.  \\

\noindent
\textbf{Local-to-Global map}
Local-to-global map $\vec{p}(\vec{r})$ is computed numerically using hard-coded \textit{Lagrange} polynomials when the order is below or equal to 5, and through analytic procedures otherwise. \\

\noindent
\textbf{\textit{Jacobian} and Inverse \textit{Jacobian}}
The local-to-global mapping is represented by a vector of symbolic polynomials, further computing \textit{Jacobian} matrix $J_{ij}(\vec{r}_0) = \partial_{r_i} p_j (\vec{R}) |_{\vec{r}_0}$ using exact partial differentiation provided by the polynomial class. This results in a matrix of polynomials, which can be evaluated for the desired local coordinates. The \textit{Jacobian} inverse and the integration element are then computed numerically using the \dune{} provided linear algebra routines, the matrix inverse $J^{-1}$ and pseudo-determinant $dI = \sqrt{\det(JJ^T)}$ respectively (see \cref{appendix:integrationelements:proof}). \\

\noindent
\textbf{Global-to-Local map}
For polynomial elements, global-to-local map is the inverse of a polynomial map. Given the matching world and entity dimension, the method searches for the exact coordinate local to the element, that corresponds to the provided coordinate. Further, this method is extended to elements with $(dim_{elem} \leq dim_{world})$ by converting it to an optimization problem
\begin{equation}
  \label{eq-theory-mapping-optimization}
  \vec{r} : |\vec{p}(\vec{r}) - \vec{x} |^2 \rightarrow \min
\end{equation} 
searching for the local coordinate closest to the inverse of the desired global coordinate in terms of distance in global coordinates. \\

\noindent
While this problem is always uniquely solvable in linear case, in the curvilinear case it poses several additional challenges
\begin{itemize}
	\item The polynomial interpolatory map $\vec{p}(\vec{r})$ is strictly bijective inside the reference element, which must be ensured by the mesh generator. However, this need not be the case outside it. For example, $p(r) = r^2$ is a bijective 1D local-to-global map for an edge defined on $[0,1]$. However, the map is clearly non-invertible for all $p(r) \leq 0$, and thus asking for a local coordinate corresponding to the global coordinate $-1$ has no defined answer.
	\item Curvilinear geometries have singularities, e.g. $r = 0$ in the previous example. At these points the integration element is zero, which most simple iterative methods can not handle. It is expected that the meshing software provides curvilinear entities with non-singular geometries, since this would result in infinitesimal global volumes associated with finite local volumes, destabilizing optimization methods and integration routines. There is no restriction on the singularity being in the near vicinity of the element, which may be enough to significantly damage convergence.
	\item For $(dim_{elem} \leq dim_{world})$, the optimization problem given by \cref{eq-theory-mapping-optimization} is highly nonlinear. There may be multiple solutions, even uncountably many.
\end{itemize}

\noindent
For obvious reasons we will not solve the problem directly, as searching for the roots of a system of polynomial equations in 2D and 3D is well-known to be a challenging task \cite{canny+1989}. Instead, the problem is solved by a first order \textit{Gauss-Newton} method \cite{bjoerck+1996}, extending the implementation from \texttt{dune-multilineargeometry}. \\

\noindent
Based on an exchange with the \dune{} user community, we have realized that in order to satisfy all use cases we need to implement two distinct methods
\begin{itemize}
  \item Restrictive method, useful to those who want to find the element containing the global coordinate, as well as the local coordinate inside that element. If the provided global coordinate is inside the element, the method will return a success and a valid local coordinate. Otherwise, the method will return a fail and no coordinate at all, meaning that the global coordinate is not inside the element. This method also extends to lower dimension entities, finding the local coordinate within the element (!), which minimizes the distance to the provided global coordinate. Given a well-defined map (non-singular in the vicinity of the element), this method is guaranteed to converge.
  \item Non-restrictive method, useful to those who wish to extrapolate the global-to-local map beyond the reference element. This method searches for the inverse (or the distance minimizer) over the entire local domain. This is a best effort method - due to the above mentioned difficulties, it is expected to fail to converge for some maps and global coordinates. In this case, an exception is thrown.
\end{itemize}

\noindent
Below we outline the algorithm of the restrictive method:

\begin{mybox}
\begin{enumerate}
	\item Let $\vec{x}_0$ be the requested global coordinate
	\item Start with a local point $\vec{r}_0$ guaranteed to be inside the element (e.g. its center),
	\item Iteratively choose better approximations for local coordinate using \[\vec{r}_{n+1} = \vec{r}_n + \vec{d}(\vec{r}_n)\] where $\vec{d}(\vec{r}_n)$ is the least squares solution of
	        \[ J(\vec{r}_n) \vec{d}(\vec{r}_n) = \vec{p}(\vec{r}_n) \] and $J(\vec{r})$ is the \textit{Jacobian} matrix.
	\item The iterative process is finished when the global coordinate distance converges to a given tolerance level $\epsilon$ in terms of the two-norm
	        \[ \epsilon_n = |\vec{p}(\vec{r}_n) - \vec{x}_0 |^2 \leq \epsilon \]
	\item The iteration is terminated prematurely if there is enough evidence that the optimal vertex is outside the element. For this, two criteria are used: the running estimate being far outside the element \[|\vec{p}_0 - \vec{p}_i|_2 > 4 R_{elem}\] and the convergence speed being significantly slower than quadratic.
\end{enumerate}
\end{mybox}

\noindent
We are still looking to improve this method. It correctly predicts the global coordinates being inside and outside the element for most of our tests, but fails to identify the boundary points inside the element for certain cases.

%%%%%%%%%%%%%%%%%%%%%%%%%%%%%%%%%%%%%%%
% Theory for integration over curvilinear entities
%%%%%%%%%%%%%%%%%%%%%%%%%%%%%%%%%%%%%%%
\subsection{Integration}
\label{sec:theory:integration}

In this section we discuss the computation of scalar and vector integrals. Any global integral can be subdivided into elementary integrals, and then parametrized using the local elementary coordinates
\[
	\int_{\Omega} f(\vec{x}) d^{\dim} x =
	\sum_e \int_e f(\vec{x}) d^{\dim} x =
	\sum_e \int_e f(\vec{r}_e) \mu_e(\vec{r}_e) d^{\dim} r_e \]
where $\vec{x}$ are the global coordinates, $\vec{r}_e$ are local coordinates of element $e$, $f(\vec{x})$ an arbitrary function defined over the domain, and the function $\mu_e(\vec{r}_e)$, associated with the change of coordinates, is called the integration element. In the remainder of the section we focus on elementary integrals, so we drop the element index $e$. \\

\noindent
In the original \dunegeom{} paradigm, the geometry class does not explicitly compute integrals, only the integration element$\mu(\vec{r})$. The user can compute the integral over the reference element by using a quadrature rule \cite{abramowitz+1970} provided in \dunegeom{}, or another external integration module. Any numerical integral, in particular the numerical quadrature, can written as a weighted sum
\[ \int f(\vec{r}) \mu(\vec{r}) d^{\dim} r = \sum_i f(\vec{r}_i) \mu(\vec{r}_i) w_i  \]
where the $r_i$ and $w_i$ are the quadrature (sampling) points and associated weights. The sampling points and weights are a property of the particular quadrature rule, and can be reused for all integrands over the reference element. Given a polynomial order $p$, one can construct a finite numerical quadrature, which will be exact for polynomial functions of order $p$ and below, and thus well approximate integral over any function, that is well-approximated by a polynomial. \\

\noindent
Numerical quadrature methods in practice are considerably faster than any other known method for geometry dimensions $\dim \leq 3$ \cite{schurer2003}, but they also have disadvantages. Many smooth functions, for example, sharply peaked, nearly singular, or those given by fractional polynomial order are known to have very slow Taylor series convergence, and hence may require very high quadrature order to achieve the desired accuracy. Also, numerical quadratures for non-trivial domains (e.g. simplices) have so far only been calculated to moderate orders. For example, Zhang et al \cite{zhang+2008} present order 21 triangular quadrature. Finding a numerical quadrature reduces to finding polynomial roots, which is known to be a hard problem in more than 1 dimension due to progressively higher decimal precision necessary to distinguish the roots from one another. One way to overcome these difficulties is to transform the integration domain to a cuboid using a \textit{Duffy} transform (for full derivation, see abstract \ref{section-abstract-duffy-transform})

\[ \int_0^1 \int_0^{1-x} f(x,y) dx dy = \int_0^1 \int_0^1 f(x, (1-x)t ) dx dt  \]

\noindent
The advantage of the cuboid geometry is that a quadrature rule can be constructed from a tensor product of 1D quadratures, which are readily available at relatively high orders. Quadrature rules created in this way have more points per order than the optimal rules, created specifically for the desired (e.g. simplex) geometries. At the time of writing, \dunegeom{} provides 1D quadratures up to order 61 and specific 2D and 3D simplex quadratures up to order 13. We are aware of the existence of advanced methods to improve performance of quadrature integration, such as sparse grids \cite{petras2000}, but they are beyond the scope of this paper. \\

\noindent
The integration functionality in \curvgeom{} addresses two additional problems:
\begin{itemize}
	\item Integrating polynomials of arbitrary order
	\item Integrating smooth functions with no a priori knowledge of the optimal polynomial approximation order.
\end{itemize}

\vspace{6pt}

\noindent
\textbf{Symbolic Integration} \\
\curvgeom{} implements a symbolic polynomial class, which is stored as a sum of monomials of a given order. Integrals over monomials of any given order can be computed analytically \cref{appendix-proof-simplexintegral}, and so can the integral over any arbitrary polynomial, which is a sum of monomial integrals. \\

\begin{table}[h]
\centering
\begin{tabular}{l | l}
\hline
Cuboid Integrals &
\begin{tabular}{@{}c@{}}
$ \int_0^1 x^i dx = \frac{1}{i+1} $ \\
$ \int_0^1 \int_0^1 x^i y^j dx dy = \frac{1}{(i+1)(j+1)} $ \\
$ \int_0^1 \int_0^1 \int_0^1 x^i y^j z^k dx dy dz = \frac{1}{(i+1)(j+1)(k+1)} $ \\
\end{tabular} \\ \hline
Simplex Integrals &
\begin{tabular}{@{}c@{}}
$ \int_0^1 x^i dx = \frac{1}{i+1} $ \\
$ \int_0^1 \int_0^{1-x} x^i y^j dx dy = \frac{i! j!}{(i + j + 2)!} $ \\
$ \int_0^1 \int_0^{1-x} \int_0^{1-x-y} x^i y^j z^k dx dy dz = \frac{i! j! k!}{(i + j + k + 3)!} $ \\
\end{tabular} \\
\end{tabular} \\
\captionsetup{width=0.8\textwidth}
\caption{Monomial integrals over cuboid and simplex reference elements. For derivation see \cref{appendix-proof-simplexintegral}}
\label{table:integration:monomialintegral}
\end{table}

\noindent
The \curvgeom{} polynomial class provides the exact integration functionality. \\

\noindent
\textbf{Adaptive Integration} \\
In its general form, a scalar integral over an element can be written as

\[\int f(\vec{x}) d^{\dim} x = \int f(\vec{r}) \mu(\vec{r}) d^{\dim} r,\]

\noindent
where the integration element is given by \[\mu(\vec{r}) = \sqrt{\det(J J^T)} \] and $J$ is the \textit{Jacobian} matrix (see \cref{appendix:integrationelements:proof}). \\

\noindent
In the case of matching element and space dimension, e.g. volume in 3D, or area in 2D, the integration element simplifies to $\mu(\vec{r}) = |\det J(\vec{r})|$. Even though the absolute value function is not polynomial, it can be observed that $\det J$ is not allowed to change sign inside the element, as that would result in self-intersection. The change of sign implies that the global geometry contains both "positive" and "negative" volume, which happens due to twisting the global coordinates inside out at the singular point $\det J = 0$. Also, the singular point $\det J = 0$ should not be present within the element, as it leads to zero volumes in global coordinates. Modern curvilinear meshing tools take account of these constraints when constructing meshes. Thus, in the case of well-conditioned elements, it remains to evaluate the integration element it anywhere inside the element and discard the minus sign if it happens to be negative. Then, given a polynomial integrand $f(\vec{x})$, the integral can be computed exactly using the quadrature rule of appropriate order. \\

\noindent
In the case of mismatching dimensions, e.g. area in 3D, or length in 2D and 3D, $\mu(\vec{r})$ cannot be simplified. It is a square root a polynomial that itself is not a square of another. Such integrals, in general, do not possess a closed form solution and have to be evaluated numerically. To address this problem, \curvgeom{} provides a recursive integrator class, which iteratively increases the quadrature order until the estimated integration error converges to a desired tolerance level. This method can consume several milliseconds for calculation of the surface area of near-singular geometries, but for well-conditioned geometries it converges much faster. The method accepts integrands in terms of functors overloading a skeleton class, and the \curvgeom{} uses it internally to provide volumes and surfaces of curvilinear entities, only requiring the user to additionally specify the desired tolerance level. \\

\noindent
In addition, the integrator class supports simultaneous integration of vector and matrix integrands via $Dune::DynamicVector$ and $Dune::DynamicMatrix$, as well as $std::vector$. The motivation of this implementation is due to the fact that, frequently, the simultaneous evaluation of a vector or a matrix is considerably cheaper than the individual evaluation of each of its components. The method provides several matrix and vector convergence error estimates, such as 1 and 2-norm, which can be selected by user to adapt to the problem at hand.\\

\noindent
According to \cite{schurer2003}, best results in low-dimensional numerical integration are achieved by the adaptive quadrature of high degree, whereas \textit{Monte-Carlo} methods perform better for high-dimensional integrals. Using an external adaptive library, for example the GSL extension due to Steven G. Johnson (\url{http://ab-initio.mit.edu/wiki/index.php/Cubature}) could be of advantage. This library is based on \textit{Clenshaw-Curtis} quadrature, which has the advantage of being hierarchical. This means that the computation of next quadrature order reuses all previously-computed quadrature points, decreasing the computational effort.

%%%\subsection{Integration Element - Vector}
%%%
%%%When integrating vector functions we are mostly interested in the integrals over boundary surfaces and edges, namely $\int_{\partial V} \vec{f}(\vec{r}) \cdot \vec{n}(\vec{r}) d(\partial V)$. For an edge in 2D the following expression for the tangential and normal integration elements (up to a sign convention) can be found:
%%%\[ d\vec{l}_{\parallel} = (\partial_u p_x, \partial_u p_y)du \; \; \; \; \; d\vec{l}_{\perp} = (\partial_u p_y, -\partial_u p_x)du  \]
%%%
%%%\noindent
%%%For a vector in 3D the tangential integration element is not defined, but the normal integration element is
%%%\[ d\vec{S} = (\partial_u \vec{p} \times \partial_v \vec{p})du \; dv  \]
%%%
%%%\noindent
%%%Thus, given polynomial vector basis functions $\vec{f}$ and polynomial interpolation, the scalar (and, if necessary, vector) products $\vec{f}(u) \cdot d\vec{l}(u)$ and $\vec{f}(u,v) \cdot d\vec{S}(u,v)$ are also polynomial, and can be integrated exactly using analytic polynomial integration code.

%%%%%%%%%%%%%%%%%%%%%%%%%%%%%%%%%%%%%%%
% Theory for point location (OCTree)
%%%%%%%%%%%%%%%%%%%%%%%%%%%%%%%%%%%%%%%
%%\input{manual-octree}

%%%%%%%%%%%%%%%%%%%%%%%%%%%%%%%%%%%%%%%%%%%%%%%%%%%%%%%%%%%%%%%%%%%%%
% Implementation Details - Curvilinear GMSH Reader
%%%%%%%%%%%%%%%%%%%%%%%%%%%%%%%%%%%%%%%%%%%%%%%%%%%%%%%%%%%%%%%%%%%%%

\pagebreak
\section{Reading Curvilinear Grid}

% \subsection{Structure of .msh files}
% 
% \begin{mybox}
% \begin{lstlisting}
% $MeshFormat
% ver f_type data_size    # This line is mostly irrelevant
% $EndMeshFormat
% $Nodes
% n_vertices
% 1 x y z
% 2 x y z
% .......
% n_vertices x y z
% $EndNodes
% $Elements
% n_elem
% 1 elem_type n_tags (process_tags) v_1 v_2 ... v_N
% 2 elem_type n_tags (process_tags) v_1 v_2 ... v_N
% .......
% n_elem elem_type n_tags (process_tags) v_1 v_2 ... v_N
% $EndElements
% \end{lstlisting}
% \end{mybox}
% 
% \noindent
% where
% \begin{itemize}
% 	\item $ver$             - version of the GMSH file
% 	\item $f\_type$          - type of file (irrelevant)
% 	\item $data\_size$       - size of file (irrelevant)
% 	\item $n\_vertices$      - number of vertices of the mesh
% 	\item $i\ x\ y\ z$         - index of the vertex and its coordinates
% 	\item $n\_elem$          - number of elements of the mesh
% 	\item $elem\_type$       - Integer which determines element type and interpolation order
% 	\item $n\_tags$          - Total number of tags. If $>2$, then have $process\_tags$
% 	\item $process\_tags$    - Tags which determine the process the vertex belongs to. Only if GMSH is told to partition the mesh
% 	\item $v\_1\ v\_2\ ...\ v\_N$ - Indices of interpolatory vertices associated with this element (includes corners)
% \end{itemize}
% \subsection{Parallel Implementation}

From the outset the parallel implementation of the Curvilinear GMSH Reader is targeted at high parallel scalability. It loads the mesh evenly on all involved processes, avoiding master process bottlenecks. The algorithm to achieve this requires reading the mesh file several times on each process:

\begin{mybox}
\begin{enumerate}
  \item VERTEX PASS 1: Skip all vertices, and place file pointer before the element section

  \item ELEMENT PASS 1: Count the total number of elements and boundary segments

  \item ELEMENT PASS 2: Read corners for all elements within the block associated to this process. Given equal splitting of elements across all processes, the process with index $rank$ should read the elements with indices \[interv(rank) = \floor[\Big]{ [rank, rank+1] \cdot N_{elem} / p_{tot} } + 1.\]
        
  \item If partitioning is enabled, partition the elements among all processes. The partitioning uses the \textit{ParMETIS\_V3\_PartMeshKway} function of \ParMETIS{} \citeParMetis{}. It produces contiguous subdomains on each process, with a roughly equal number of elements on each process. \ParMETIS{} naturally also minimizes the number of boundary connections, thus minimizing the number of interprocessor boundaries and the amount of parallel communication necessary at a later stage. We have also implemented support for \ParMETIS{} multiple constraint partitioning capabilities, but, as of time of writing \ParMETIS{} does not guarantee contiguous subdomains for multi-constraint partitioning.
        
  \item ELEMENT PASS 3: Read all data associated with elements on this process partition. Map all faces to the elements sharing them using the sorted global index of the face corners. The elements are written to the grid factory
       
  \item ELEMENT PASS 4: Read all data associated with boundary elements. Determine if the element belongs to this process by looking it up in the available face map. Separate the processed boundaries by boundary tag. Identify which of the boundary tags is associated with the domain boundary by determining the faces that have only one neighboring element across all processes, and sharing this information with all other processes. Thus each process is aware of all volume and boundary tags, even if there are no entities with this tag on the process. The boundary segments are written to the grid factory.

  \item VERTEX PASS 1: Read the coordinates of the vertices associated with the entities present on this process. The vertex coordinates are written to the grid factory.
\end{enumerate}
\end{mybox}

\noindent
The implementation has an option to directly output the processed mesh into $.VTK$ file set for debugging purposes.

% \begin{mybox}
% \noindent
% Currently using brute-force, because it is not much slower than improved for \\
% 
% \noindent
% \uline{Trivial Algorithm: (Complexity $O(12 N_{elem} N_{\beta} / p_{tot}^2)$)}\\
% \textit{Loop over all stored boundary elements $\beta_i$, and over all stored internal elements $E_j$.} \\
% \textit{ If $\beta_i \in E_j$ for some $j$ then store $\beta_i$ }\\
% 
% \noindent
% \uline{Improved Algorithm: (Complexity $O(12 N_{elem} \log_2 (N_{\beta} / p_{tot}) / p_{tot}$)}
% 
% \begin{enumerate}
% 	\item Construct map from boundary vertex index set to boundary id
% 	\item Add all boundaries to the map
% 	\item Loop over each face of all internal elements
% 	\begin{enumerate}
% 		\item If $map[face]$ is non-null, link the element and boundary
% 	\end{enumerate}
% \end{enumerate}
% 
% \end{mybox}	
% 		
% \begin{enumerate}[resume]
% 	\item Add internal elements to factory
% \end{enumerate}
% \begin{mybox}
% 	\begin{itemize}
% 		\item For debugging purposes write each element to a .vtk file using CurvilinearVTKWriter.
% 		\item Add element vertices and global element index to factory
% 		\item If creating grid with boundaries, also pass $internal\_to\_boundary\_element\_linker$. This array stores the indices of boundaries which are connected to this element (if any).
% 	\end{itemize}
% \end{mybox}

%%%%%%%%%%%%%%%%%%%%%%%%%%%%%%%%%%%%%%%%%%%%%%%%%%%%%%%%%%%%%%%%%%%%%
% Implementation Details - Curvilinear Grid Constructor
%%%%%%%%%%%%%%%%%%%%%%%%%%%%%%%%%%%%%%%%%%%%%%%%%%%%%%%%%%%%%%%%%%%%%

\pagebreak
\section{Constructing Curvilinear Grid}
\label{impl-grid-constructor}

The grid construction is called by the \textit{Grid Factory} (see \ref{interface-grid-factory}) after all the vertices, elements and boundary segments have been inserted. The construction of the \curvgrid{} is in accordance with the following plan

\begin{mybox}
\begin{enumerate}
	\item Construction of grid entities - Interior ($I$), Process Boundary ($PB$), Domain Boundary ($DB$) and Interior Boundary ($IB$) edges and faces.
	\item Construction of the global index for all entities. By default, the global index for vertices and elements is re-used from the mesh file.
	\item Construction of Ghost elements ($G$)
	\item Construction of entity subsets used for iteration over different entity partition types and codimensions
	\item Construction of communication maps, used to perform communication via the \textit{DataHandle} interface
	%\item Construction of OCTree for hierarchical element location
\end{enumerate}
\end{mybox}

\subsection{Storage}
\label{impl-grid-storage}

\noindent
The paradigm used in \curvgrid{} is to store all data corresponding to the grid in a single class, namely the \textit{CurvilinearGridStorage} class. Entities of each codimension are stored in arrays indexed by the local entity index, which is contiguous and starts from 0 on all processes. Each entity stores its global index, geometry type and partition type. For the detailed explanation on the correct assignment of partition types to entities of each codimension, the user is referred to the corresponding section of \dune{} grid usage manual, found on the website of the project. Elements and faces also store the associated material (physical) tag. Elements store the local index of all interpolatory vertices in the \textit{Sorted Cartesian} order (see \cref{impl-gmsh-numbering-convention}). The edges and faces do not store the interpolatory vertex indices, as it would be wasteful for higher curvilinear orders. Instead, each edge and face is stored as a subentity of an associated parent element - any of the elements containing it. Thus, each subentity stores the parent element local index, as well as the subentity index, by which the subentity is indexed within the reference element. Finally, each face stores its boundary type - Interior, Domain Boundary or Periodic Boundary, as well as the index of the 2nd parent element that shares it. By convention, the primary parent element must always be interior to the process storing the face. The secondary parent element may be either interior, or Ghost (\cref{fig:impl:ghostelements}). The data associated with a Ghost element is stored on the neighboring process, but can be accessed from this process as part of interprocessor communication. In case of Domain Boundaries, there is no secondary parent. By convention, any attempts to access the secondary parent of a Domain Boundary result in an exception, as the user code must take the provided partition type into account when addressing neighboring entities. \\

\begin{figure}
    \centering
	\includegraphics[scale=1.5]{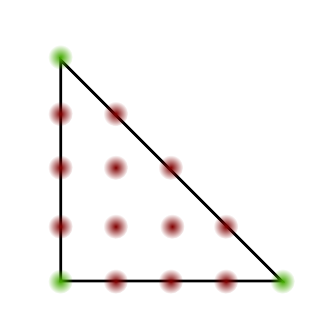}
	\caption{The interpolatory vertices of the 5th order curvilinear triangle. Corners are given in red}
	\label{fig:impl:storage:vertexvscorner}
\end{figure}

\noindent
\curvgrid{} contains several different local index sets, designed to uniquely index only the entities of a given subset. Namely, Domain Boundary segments, Interior Boundary segments, Process Boundaries, remaining Interior boundaries, as well as Ghost elements each have their own local index. \curvgrid{} provides maps between those indices and local indices of all entities of a given codimension. In addition, there is a unique index for grid corners, namely the interpolatory vertices that define the linear entities (\cref{fig:impl:storage:vertexvscorner}). This is necessary, because the \dune{} facade class operates in terms of linear elements, and thus requires a contiguous index for the corners. Importantly, among all vertices only entity corners possess unique process boundary index, since, for all interprocessor communication purposes, the mesh can be assumed to be linear without loss of generality. Finally, a map from global to local index is provided for entities of all codimensions. \\

\noindent
\curvgrid{} operates with two different types of \textit{GlobalIds}, designed to uniquely identify the entity among all processes. First of all, it is a pre-requisite that all vertices have an associated global index, which is either re-used from the mesh file, or constructed in the beginning of the grid construction procedure. Before the global indices for other codimensions are generated, these entities are uniquely identified by the sorted set of global indices of the entity corners, which can be used in a map to determine if a given communicated entity is already present on the receiving process. Later, when entities of all codimensions possess global index, the \textit{GlobalId} is simply a wrapper of the global index, such as to use minimal amount of resources necessary. It is the latter \textit{GlobalId} that is made available for the user code at the end of construction procedure. \\

\noindent
For the purposes of iterating over frequently used entity subsets, the corresponding local index subsets are provided for each codimension. Namely, subsets are provided for all entities of a given codimension, Interior entities, Process Boundary entities, Domain Boundary entities, Ghost entities, Interior + Domain Boundaries (called \textit{interior} in \dune{}), Interior + Domain Boundaries + Process Boundaries (called \textit{interior border} in \dune{}). \\

\noindent
For communication purposes, all communicating entities need to store a vector of ranks of processes with which these entities are shared. For more details on communicating entities, see \cref{impl-grid-constructor-comm}.

\subsection{Global index construction}
\label{impl-grid-constructor-globalindex}

\noindent
In this section we briefly describe the algorithm used to construct the global indices for all codimensions. The challenge in computing the global indices comes from the fact that originally the processes are not aware of their neighbors. Due to the insertion of complete boundary segments by the grid factory, each process can readily identify all of its process boundaries as the faces that have only one containing element on the process and are not already marked as domain boundaries. The algorithm has four definite stages - determining the neighboring process for each process boundary, assigning (virtual) ownership of each shared entity to only one of the processes sharing it, enumerating the global index on processes owning the entities, and communicating the global index to all other processes containing each shared entity. \\

\begin{figure}
    \centering
	\includegraphics[scale=0.5]{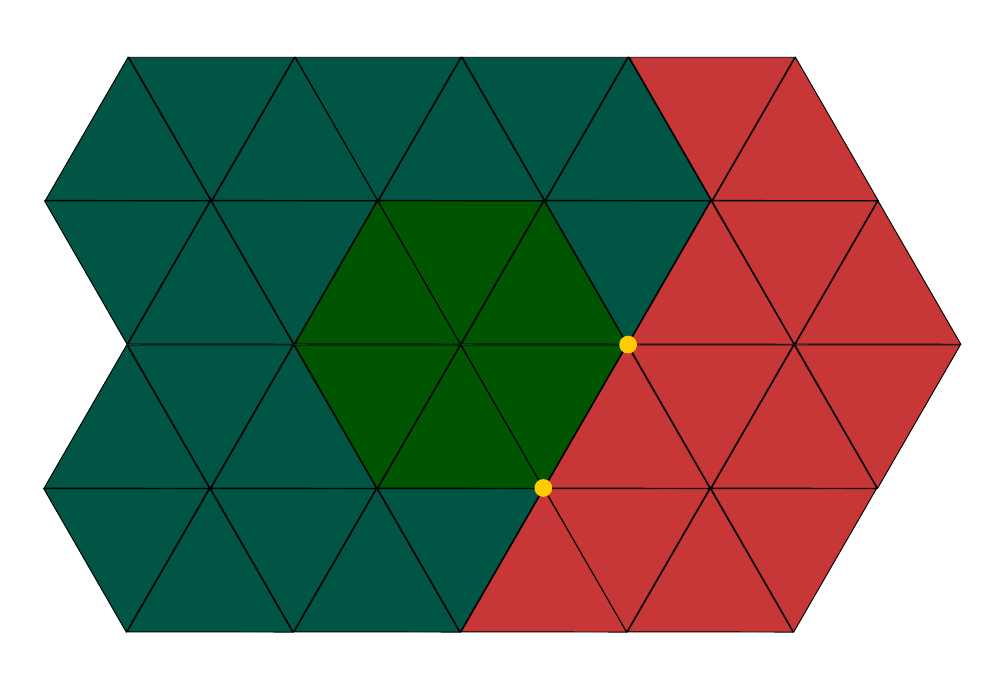}
	\captionsetup{width = 0.8\textwidth}
	\caption{The two yellow vertices are shared between three processes. The edge in-between them only exists on red and green processes, but not on the blue one}
	\label{fig:impl:globalindex:fakeedge}
\end{figure}

\noindent
The algorithm starts by determining neighbor ranks of each Process Boundary ($PB$) corner. Currently, each process communicates all process boundary corner global indices to all other processes. From the received global indices each process can deduce all other processes sharing each of its corners. Then, each process assigns provisional neighbor processes to edges and faces by considering the processes that share all entity corners. If two processes share all the corners of a given entity, it does not mean that they share the entity as well (\cref{fig:impl:globalindex:fakeedge}). The ambiguity is quite rare, because most entities are shared only by two processes, and since each process boundary entity must be shared by at least two processes, there is no need to check if the entity exists. Nevertheless, the grid must be able to handle the rare case of an entity being shared by more than two processes. In this case, the edge and face \textit{GlobalIds} (see \cref{impl-grid-storage}) are communicated to all provisional neighboring processes. Each of the provisional neighbor processes then responds, whether or not each entity corner set corresponds to an entity. The ranks that are not confirmed to share the entities are then removed from the neighbor list. \\

\noindent
For each $PB$ edge and face, an owner is determined. The ownership strategy is flexible, as long as all processes agree. Currently, a shared entity is considered to be owned the process with the lowest rank among those sharing it. Each process owns all of its non-shared entities. The number of entities owned by each process is then communicated to all other processes. \\

\noindent
Each process then locally enumerates the global index of all its entities. To begin with, each process computes the shift in global index due to the entities of each codimension enumerated by the processes with ranks lower than this process rank. All processes enumerate global indices consecutively, starting with 0 on rank 0 process. This constructs a global index individually for each codimension. A global index over all entities is also constructed, shifting the enumeration also by all the enumerated entities of higher codimension. \\

\noindent
The enumerated global indices are then communicated among all processes sharing the entities. By analyzing entity neighbors, each process can compute how many global indices it needs to send to and receive from each other process, avoiding extra communication. At the end of this process, the global-to-local index map is filled on each process.

\subsection{Ghost element construction}
\label{impl-grid-constructor-ghost}

\begin{figure}
    \centering
	\includegraphics[scale=0.5]{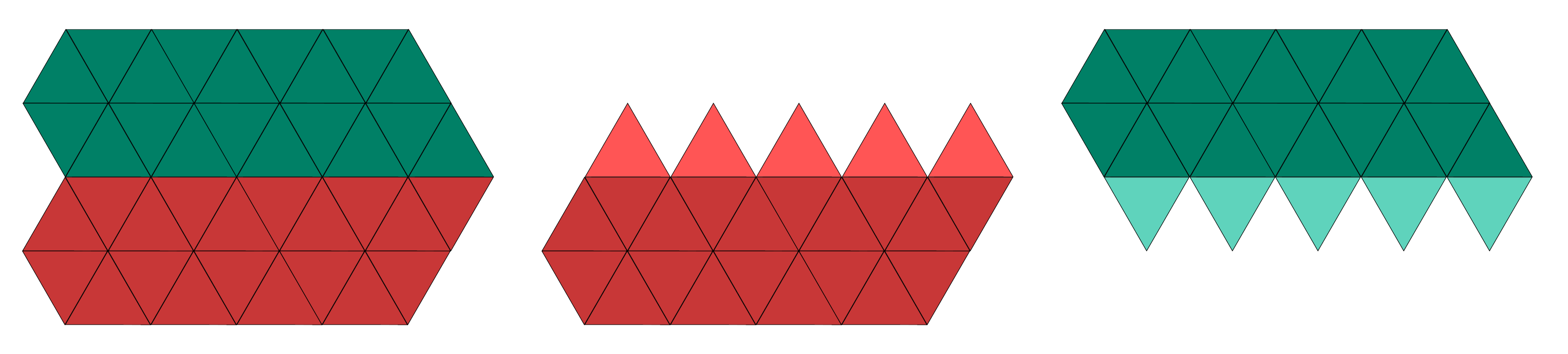}
	\caption{The first image depicts two neighboring processes without Ghost elements. The second and third images contain only the first and only the second process entities respectively, including Ghost elements borrowed from the other process. }
	\label{fig:impl:ghostelements}
\end{figure}

\noindent
Ghost entities are the subentities of the element on the other side of the process boundary face, including the element itself. The process boundary entities are not considered Ghost entities. Thus, the Ghost entities are the internal/domain boundary entities of another process, borrowed by this process. Construction of Ghost entities involves communicating all the information associated with the entities to the neighboring processes, and then incorporating the Ghost entities into the grid on the receiving side (\cref{fig:impl:ghostelements}). \\

\noindent
Since for every $PB$ face the neighboring process rank has already been determined in the previous section, and the global index is already known, it remains only to communicate the corresponding neighbor entities and correctly integrate them into the local grid. Firstly, one needs to communicate the properties of the structure to be communicated. Thus, for each interior element next to the $PB$ the interpolation order is communicated, as well as the number of $PB$ faces it shares with the receiving side. It is important to note that a Ghost element can have more than one $PB$ face associated to it, and the receiving side does not know in advance that two or three of its $PB$ faces are associated with the same Ghost element. This is also the reason it is not possible to know in advance how many ghosts will be received from a neighbor process. Afterwards, for each Ghost element the global index, physical tag, global indices of all codimension subentities and subentity indices of $PB$ faces of the element are communicated. The corresponding Ghost elements are then added to the mesh, and it is determined which vertex coordinates are missing. The vertex coordinates are not communicated immediately, because the neighbor process may already have some of the global coordinates due to narrow mesh appendices. Thus, each process communicates the number of vertex coordinates missing from each of its neighbors, and then communicates the missing coordinates.

\subsection{Communication interface construction}
\label{impl-grid-constructor-comm}

The communication paradigm of \dune{} \textit{DataHandle} interface is to communicate data between instances of the same entity on different processes. Depending on the communication protocol, only entities of specific structural types will communicate. We will slightly redefine the partition type classes in this section for compactness reasons. There are three different communicating partition types
\begin{itemize}
	\item $PB$ - Process boundary entities, that is, process boundary faces and their subentities
	\item $I$ - Interior entities, including the $DB$ but excluding the $PB$ entities.
	\item $G$ - Ghost elements and their subentities, excluding $PB$
\end{itemize}

\begin{figure}
    \centering
	\begin{subfigure}[b]{0.48\textwidth}
	  \includegraphics[scale=0.4]{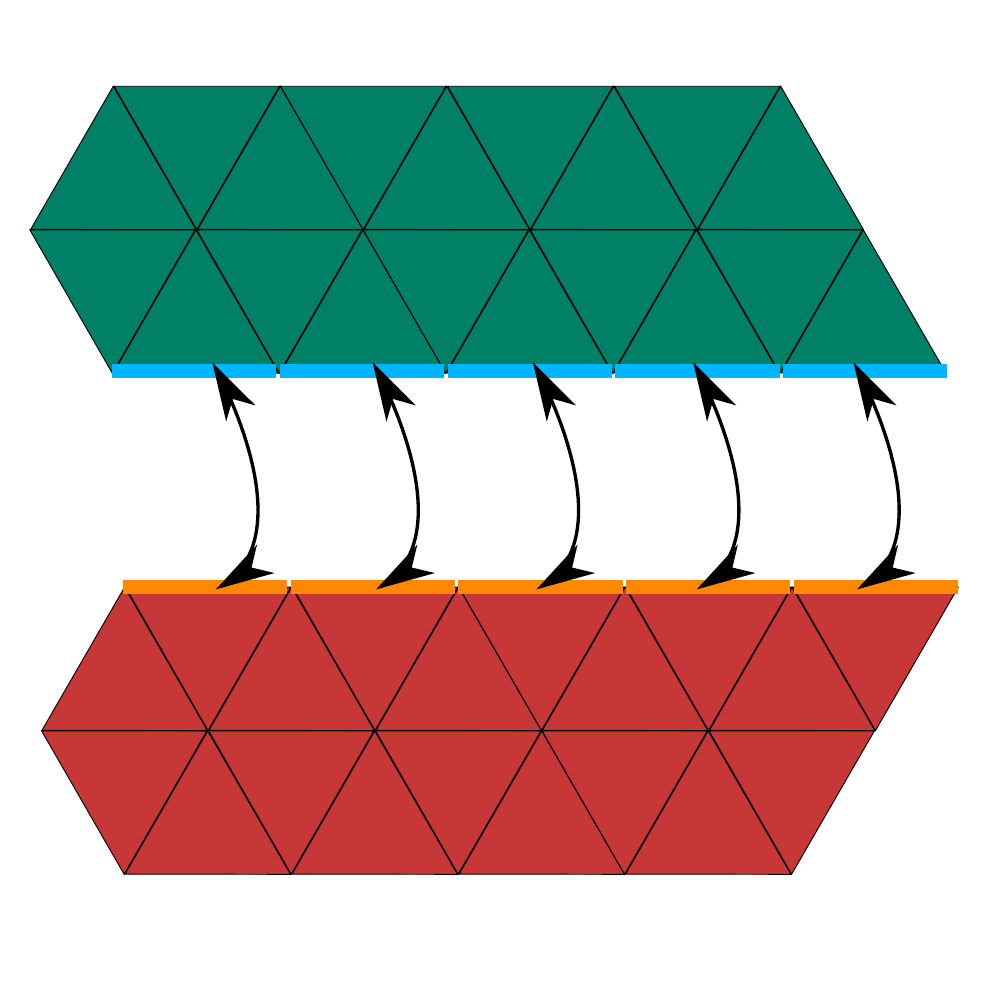}
	  \captionsetup{width=0.8\textwidth} 
	  \caption{$PB \leftrightarrow PB$. Communication of neighboring process boundaries}
	\end{subfigure}
	\begin{subfigure}[b]{0.48\textwidth}
	  \includegraphics[scale=0.4]{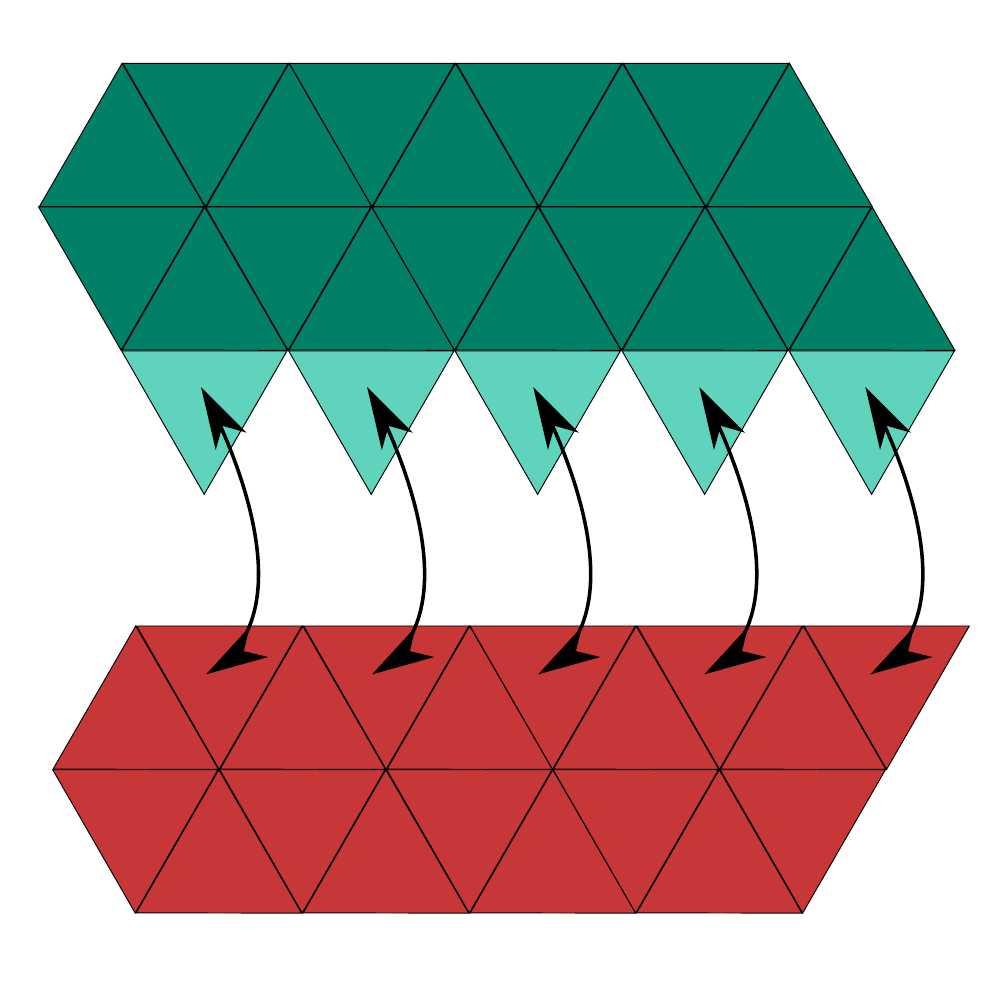}
	  \captionsetup{width=0.8\textwidth} 
	  \caption{$I \leftrightarrow G$. Communication of interior element and its Ghost on another process }
	\end{subfigure}
	\begin{subfigure}[b]{0.48\textwidth}
	  \includegraphics[scale=0.4]{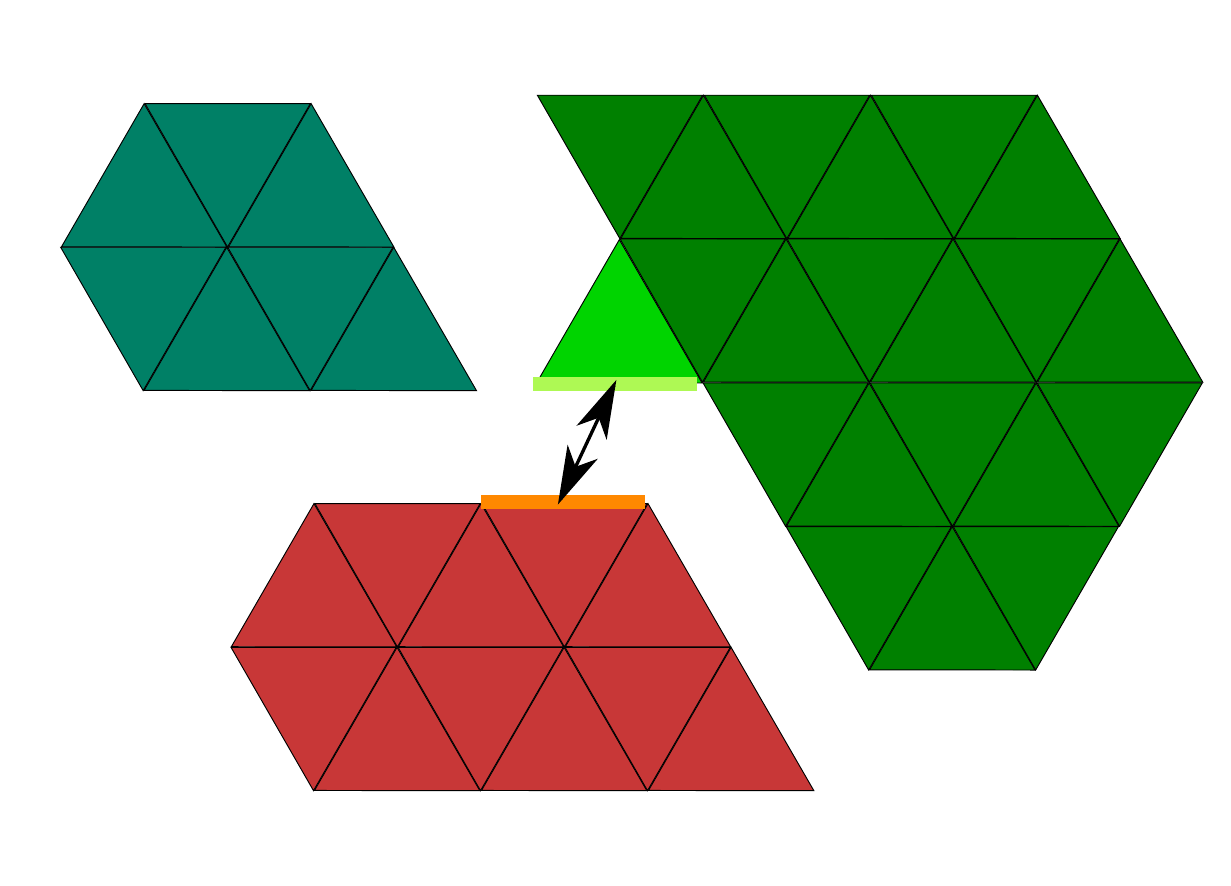}
	  \captionsetup{width=0.8\textwidth} 
	  \caption{$PB \leftrightarrow G$. Communication of a Ghost of the blue process on the green process with a process boundary on the red process}
	\end{subfigure}
	\begin{subfigure}[b]{0.48\textwidth}
	  \includegraphics[scale=0.4]{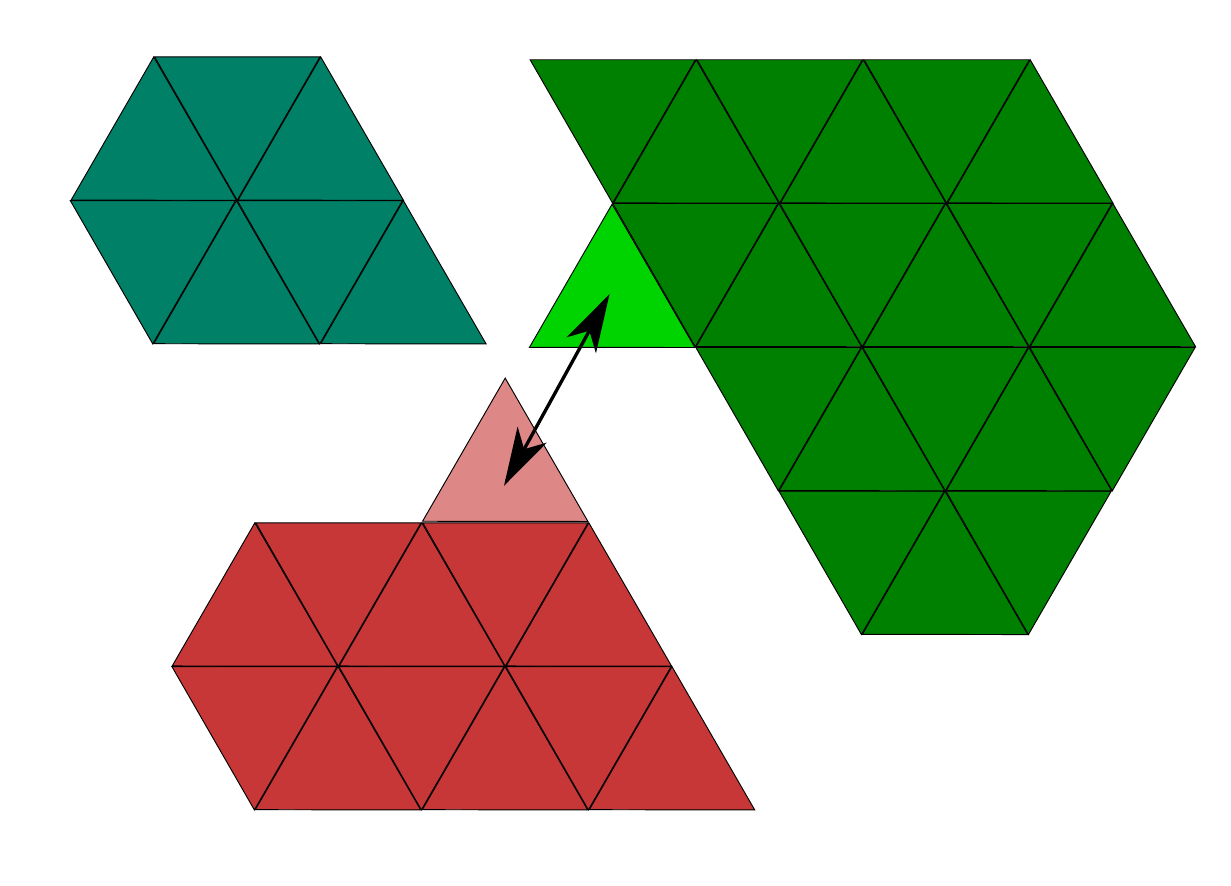}
	  \captionsetup{width=0.8\textwidth} 
	  \caption{$G \leftrightarrow G$. Communication between two ghosts of the same element of the blue process on the green and the red process}
	\end{subfigure}
	\captionsetup{width=0.8\textwidth}
	\caption{ Partition type pairs that can communicate to each other}
	\label{fig:impl:comm:partitionpairs}
\end{figure}

\noindent
We consider all partition type pairs, for which the \textit{DataHandle} communication is possible (\cref{fig:impl:comm:partitionpairs}). Note that interior entities only communicate to Ghost and vice-versa, because Process Boundaries are Process Boundaries on both neighboring processes. However, Process Boundaries can communicate to Ghosts and vice-versa, because a process boundary can be a Ghost with respect to a third process at the triple point. \\

\noindent
The \textit{DataHandle} interface foresees four communication protocols. The \cref{table:impl:datahandle:protocols} presents the communication interfaces used by Dune, and explains how they translate to communication between entities of above specified partition types \\

\begin{table}
\centering
\begin{tabular}{ | l | l | l | l | l | l | l | l | }
  \hline
  Interface & Direction &
      \small $PB \rightarrow PB$ \normalsize &
      \small $PB \rightarrow G$ \normalsize &
      \small $I \rightarrow G$ \normalsize &
      \small $G \rightarrow I$ \normalsize &
      \small $G \rightarrow PB$ \normalsize &
      \small $G \rightarrow G$ \normalsize \\ \hline 
  \begin{tabular}[x]{@{}c@{}}\lstinline|InteriorBorder_|\\ \lstinline|InteriorBorder| \end{tabular}
  & ---      & Y & N & N & N & N & N \\ \hline
  \lstinline|InteriorBorder_All|            & Forward  & Y & Y & Y & N & N & N \\ \hline
  \lstinline|InteriorBorder_All|            & Backward & Y & N & N & Y & Y & N \\ \hline
  \lstinline|All_All|                       & ---      & Y & Y & Y & Y & Y & Y \\ \hline
\end{tabular}
\caption{Communication interfaces of \textit{DataHandle}, and the associated communicating partition type pairs}
\label{table:impl:datahandle:protocols}
\end{table}

\noindent
The aim of this part of the grid constructor is to generate/reuse unique index maps for the sets $PB$, $I$ and $G$. Then, for every communicating entity, for every possible partition type pair, we require an array of ranks of the processes for which such communication is possible. Note that the map for the pair $PB\rightarrow PB$ already exists, it is constructed in the very beginning of the grid construction procedure to enable global indices and Ghost elements. The algorithm is as follows:

\begin{mybox}
\begin{enumerate}
	\item Mark the neighbor process ranks of the associated $PB$ for all $I$ and $G$ entities, whose containing elements neighbor $PB$, thus enabling the $I \rightarrow G$ and $G \rightarrow I$ communication. Note that entities of all (!!) codimensions can have more than one neighbor rank obtained this way. During the marking, elements with two or more process boundaries from different processes may be encountered. In that case, for each process boundary entity the rank of the other process boundary is marked, thus providing some information for the future construction of the $PB \rightarrow G$ communications.
	\item Then, all entities that can communicate are associated with ranks of all other processes, over which the entities are shared.% It remains to finish the $PB \rightarrow G$ communication, and calculate the remaining protocols $G \rightarrow I$, $G \rightarrow PB$ and $G \rightarrow G$.
	\item For all $PB$ entities, subtract $PB\rightarrow PB$ set from the $PB \rightarrow G$ set to ensure that the latter excludes the former. Also, mark the number of effective $PB \rightarrow G$ candidate entities of each codimension for each process
	\item For all $PB$ entities with non-empty $PB \rightarrow G$ set, communicate $G$ indices to all neighboring $PB$ entities
	\item For all $PB$, append the union of the received $G$ to the $PB \rightarrow G$ set, thus completing it %(hopefully)
	\item For all $PB$ entities with non-zero $PB \rightarrow G$, communicate self index to all $G$ of $PB \rightarrow G$ set
	\item For all $G$, append the union of the received $PB$ to the $G \rightarrow PB$ set, thus completing it %(hopefully)
	\item For all $PB$ entities with non-zero $PB \rightarrow G$, communicate to all own $G$ neighbors all the other $G$ in the set
		%\subitem Optimization - do this only if you are lowest rank among all $PB$-neighbors
		%\subitem Further optimization - do this only if you are modulus-rank among all $PB$-neighbors
	\item For all $G$, append the union of the received $G$ to the $G \rightarrow G$ set, thus completing it %(hopefully)	
\end{enumerate}
\end{mybox}

% 
% \subsection{Iteration set construction}
% \label{impl-grid-constructor-iterator}
% 
% Construction of the iterator sets involves simply iterating over all entities, and filling the sets with local indices based on the entity structural type.
% 
% 
% \subsection{OCTree construction}
% \label{impl-grid-constructor-octree}

%%%%%%%%%%%%%%%%%%%%%%%%%%%%%%%%%%%%%%%%%%%%%%%%%%%%%%%%%%%%%%%%%%%%%
% Implementation Details - Curvilinear VTK Writer
%%%%%%%%%%%%%%%%%%%%%%%%%%%%%%%%%%%%%%%%%%%%%%%%%%%%%%%%%%%%%%%%%%%%%

\pagebreak
\section{Writing Curvilinear Grid}

For the grid output, two hierarchic classes have been implemented - \textit{CurvilinearVTKWriter} and \textit{CurvilinearVTKGridWriter}. The former discretizes and accumulates entities and fields one-by-one and later writes them, and the latter uses the former to write the entities of the grid by iterating over them. The implementation of the \textit{CurvilinearVTKWriter} is best understood by considering its features.

\begin{enumerate}  
  \item \textit{nodeSet} is a vector of global coordinates of interpolatory vertices of the entity in correct order \ref{impl-gmsh-numbering-convention}
  \item \textit{tagSet} is a set of 3 scalar fields used for diagnostics, which typically accompany all written elements. Namely, they are the physical tag, the partition type, and the containing process rank in that very order. 
  \item \textit{interpolate} parameter determines how the virtual refinement of the entity is performed. If true, the regular grid of interpolatory points is re-used to subdivide the entity into smaller linear entities. If false, a new regular grid is constructed for this purpose, using the \textit{nDiscretizationPoint} parameter to determine the number of desired discretization vertices per edge. If the interpolation vertices are used the latter parameter is discarded. Note that the associated fields are sampled over the vertices of the virtual refinement grid, and thus increasing the virtual refinement is useful for better visualizing the element curvature and/or finer field resolution. Note that increasing \textit{nDiscretizationPoint} results in quadratic increase in both writing time and the resulting visualization file size, so the parameter must be used with care.
  \item \textit{explode} optional parameter shrinks all entities with respect to their mass-centers, allowing to better visualize the 3D structure of the grid. Default parameter is $0.0$, which corresponds to no shrinking, and the maximum allowed parameter is $0.99$
  \item \textit{magnify} optional parameter expands all boundary surfaces with respect to the origin, allowing to better visualize the difference between boundary segments and element faces. By default there is no magnification  
  \item \textit{writeCodim} is a vector of 4 boolean variables, one for each entity codimension. This vector allows the user to control the specific codimensions of entities to be displayed. For example, it is possible to switch on only the edge wireframe of the mesh by providing $\{ false, false, true, false \}$
\end{enumerate}

\begin{figure}
	\centering
	\includegraphics[scale=0.2]{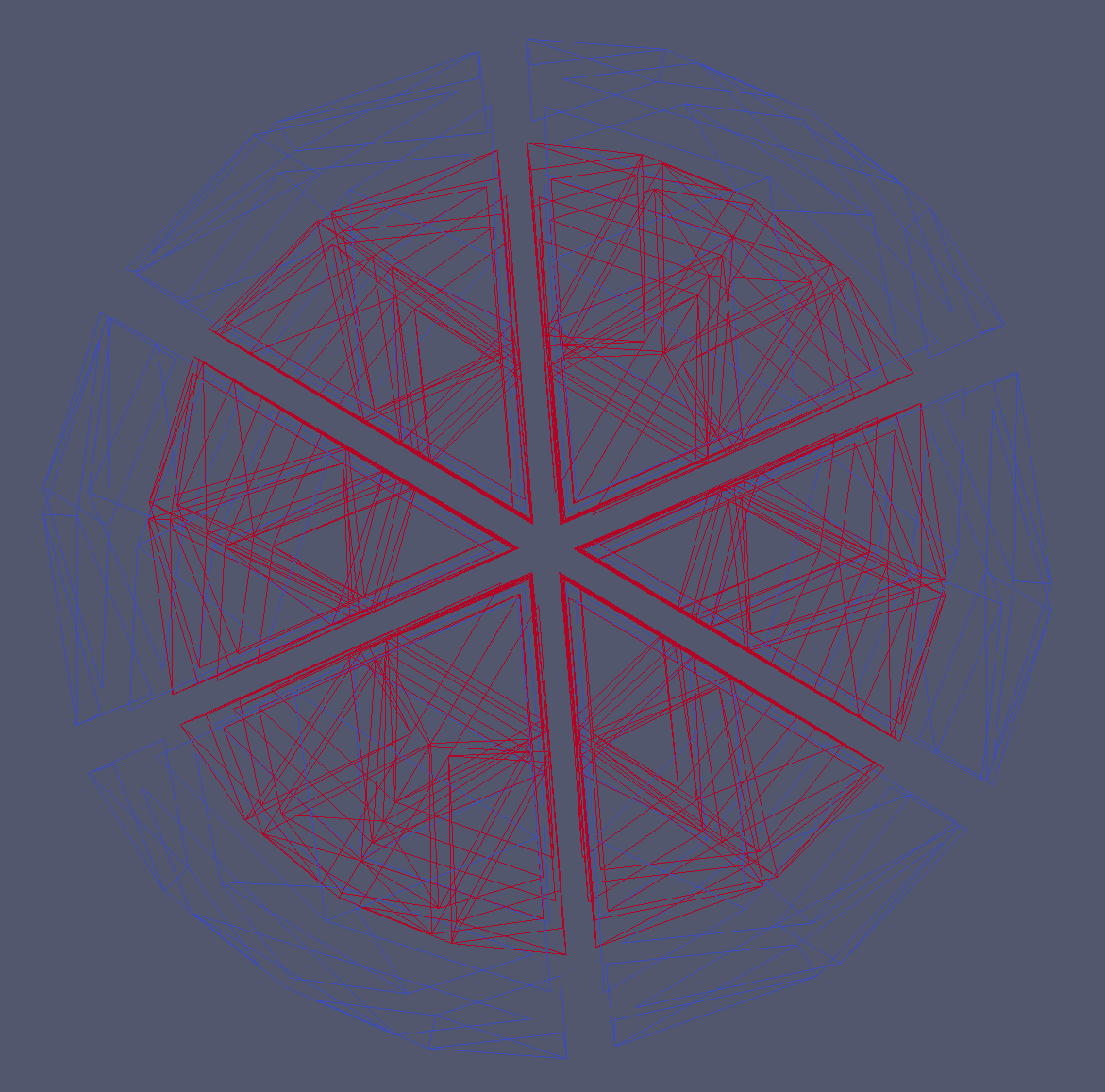}
	\caption{Visualization of an edge wireframe of a 32 element 2nd order mesh using the manual virtual refinement with  \textit{nDiscretizationPoint=3}  }
	\label{fig:gmshreader:wireframe}
\end{figure}

\pagebreak
\section{Tutorials}

%%%%%%%%%%%%%%%%%%%%%%%%%%%%%%%%%%%%%%%%%%%%%%%%%%%%%%%%%%%%%%%%%%%%%
% Curvilinear Grid Howto - Section on tutorials
%%%%%%%%%%%%%%%%%%%%%%%%%%%%%%%%%%%%%%%%%%%%%%%%%%%%%%%%%%%%%%%%%%%%%

\noindent
In order to start using the \curvgrid{} module we recommend to study the source code of the tutorials provided in \textit{curvilineargridhowto} folder. The tutorials use progressively more complicated concepts, so it is recommended to address them in the indexed order.

\subsection{Preliminaries - Creating Grid}
\label{usage-howto-tutorial-preliminaries}

All tutorials of \curvgrid{} reuse the simple auxiliary routine located in \textit{creategrid.hh}. This routine contains paths to all meshes currently used in tutorials, allowing the user to select the desired mesh by providing its index. It then proceeds to initialize the logging mechanisms
\begin{mybox}
\begin{lstlisting}
    Dune::LoggingMessage::init(mpihelper);
    Dune::LoggingTimer<Dune::LoggingMessage>::init(mpihelper);
\end{lstlisting}
\end{mybox}

\noindent
These singleton classes are essential for \curvgrid{}, as they support the parallel logging of the grid reading, construction and writing process, as well as later statistics on performance of individual routines. They can also be used by the user to write user logging output and time user functions, comparing them to the grid construction time. The next step is initializing the \textit{Curvilinear Grid Factory}

\begin{mybox}
\begin{lstlisting}
  Dune::CurvilinearGridFactory<GridType> factory(withGhostElements, withGmshElementIndex, mpihelper);
\end{lstlisting}
\end{mybox}

\noindent
where the boolean variable $withGhostElements$ determines whether the Ghost elements will be constructed and $withGmshElementIndex$ determines if the global element index from \gmsh{} is going to be reused by the grid (recommended). Otherwise, the index is automatically generated (see Tutorial 6 for discussion). After the above prerequisites, the curvilinear \gmsh{} reader is used to read the mesh file, partition it and write it to the provided grid factory.

\begin{mybox}
\begin{lstlisting}
  Dune::CurvilinearGmshReader< GridType >::read(factory, filename, mpihelper); 
\end{lstlisting}
\end{mybox}

\noindent
The \textit{Curvilinear Grid Factory} extends the interface of the standard dune grid factory, and therefore it will not work with other grids available in Dune. In order to achieve that, one must instead use the provided \textit{FactoryWrapper}. This wrapper class adheres to the standard grid factory interface, and disregards any process tag or curvilinear information provided by the curvilinear \gmsh{} reader. \\

\noindent
Finally, the grid is constructed, providing the user with the associated pointer. We note that the user must delete the grid pointer after use. %This is due to the fact that the factory used to create the grid ceases to exist before the grid itself.

\begin{mybox}
\begin{lstlisting}
  GridType *grid = factory.createGrid();
\end{lstlisting}
\end{mybox}

\subsection{Tutorial 1 - Getting started}
\label{usage-howto-tutorial-gettingstarted}

\noindent
This tutorial uses the above procedure to construct the \curvgrid{} by reading it from a \gmsh{} file. This and all other tutorials can be run both in serial and in parallel. First we define the grid

\begin{mybox}
\begin{lstlisting}
  typedef Dune::CurvilinearGrid<ctype, dim, isCached> GridType;
\end{lstlisting}
\end{mybox}

\noindent
where $dim=3$ is the dimension of the grid, $ctype=double$ is the underlying real type, and the $isCached$ is a boolean variable determining if the curvilinear geometry caching is used (recommended). Currently, only 3D tetrahedral grids are available. Then, the curvilinear grid is created by the $createGrid$ procedure described above. Finally, if we are interested in the time statistics for reading and construction of the grid, it can be obtained using the following command

\begin{mybox}
\begin{lstlisting}
  Dune::LoggingTimer<Dune::LoggingMessage>::reportParallel();
\end{lstlisting}
\end{mybox}

\subsection{Tutorial 2 - Traverse}
\label{usage-howto-tutorial-traverse}

This tutorial repeats the procedure from tutorial 1 to create the grid. It then iterates over the grid and extracts relevant information from the curvilinear entities. Currently, \curvgrid{} does not support refinement, so both leaf and level iterators will only iterate over the parent entities. As of \dune{} 2.4 revision the range-based for iterators introduced in \textit{c++11} standard are the preferred way to perform grid iteration. The below example iterator will iterate over all entities of the grid of a given codimension

\begin{mybox}
\begin{lstlisting}
  LeafGridView leafView = grid.leafGridView();
  for (auto&& elementThis : entities(leafView, Dune::Dim<dim - codim>())) {...}
\end{lstlisting}
\end{mybox}

Now, we would like to extract some relevant information from the iterator
\begin{mybox}
\begin{lstlisting}
  Dune::GeometryType gt = elementThis.type();
  LocalIndexType  localIndex
    = grid.leafIndexSet().index(elementThis);
  GlobalIndexType globalIndex
    = grid.template entityGlobalIndex<codim>(elementThis);
  PhysicalTagType physicalTag
    = grid.template entityPhysicalTag<codim>(elementThis);
  InterpolatoryOrderType interpOrder
    = grid.template entityInterpolationOrder<codim>(elementThis);
  BaseGeometry geom
    = grid.template entityBaseGeometry<codim>(elementThis);
  std::vector<GlobalCoordinate> interpVertices = geom.vertexSet()
\end{lstlisting}
\end{mybox}

\noindent
The $GeometryType$ and $LocalIndex$ are standard in \dune{}. $GlobalIndex$ provides a unique integer for each entity of a given codimension, over all processes. $PhysicalTag$ is the material tag associated with each entity, obtained from the mesh file. It can be used to relate to the material property of the entity, or to emphasize its belonging to a particular subdomain. In current \dunegrid{} standard this information can only be obtained by through the reader and requires auxiliary constructions. $InterpolatoryOrder$ denotes the integer polynomial interpolation order of the geometry of the entity. Currently, \curvgrid{} supports orders 1 to 5, the limiting factor being the curvilinear vertex indexing mapper in the reader. Finally, the $entityBaseGeometry$ gives the user direct access to the curvilinear geometry class of the entity, thus extending the interface provided by \textit{Dune::Grid::Geometry}. For example, it can be used to obtain the set of interpolatory vertices of the geometry, or an analytic polynomial matrix representing its Jacobian. \\

\subsection{Tutorial 3 - Visualization}
\label{usage-howto-tutorial-visualisation}

This tutorial demonstrates a simple way to output the curvilinear grid and associated vector fields (e.g. solutions of your PDE) to a PVTU file set using the \textit{CurvilinearVTKGridWriter} class. The writer is able to output arbitrary number of user-defined fields. The fields can be scalar or vector, and can be associated either with elements or with boundary faces. The sampled fields must overload the \textit{Dune::VTKScalarFunction} or \textit{Dune::VTKVectorFunction} standard provided by \curvgrid{}, and thus adhere to the following interface:

\begin{mybox}
\begin{lstlisting}
    // Writer initializes the functor once per each entity. This procedure can be used to pre-compute any quantities that do not change over the entity, accelerating the output
    virtual void init (const Entity & entity) {...}

    // Procedure that returns the field value as a function of local (!) coordinate
    virtual Range evaluate(const Domain & x) const  {...}

    // Procedure that returns the field name as it will appear in output file
    virtual std::string name() const  { return "localIndex2D"; }
\end{lstlisting}
\end{mybox}

\noindent
We provide 6 examples:
\begin{itemize}
  \item Element index scalar in 3D, written over all elements of the grid;
  \item Boundary segment index scalar in 2D, written over all boundary segments of the grid;
  \item Local sinusoidal vector in 3D, written over all elements of the grid;
  \item Local sinusoidal vector in 2D, written over all boundary segments of the grid;
  \item Global sinusoidal vector in 3D, written over all elements of the grid;
  \item Global sinusoidal vector in 2D, written over all boundary segments of the grid;
\end{itemize}

\noindent
These examples illustrate the important difference between local and global coordinates. Local coordinates are unique for each element, so one shall observe unique field behavior for each element. Global coordinates on the other hand are the same for all elements, thus providing a single continuous sinusoid across the entire mesh. The local fields are associated with the orientation of the element in space, which is arbitrary up to the permutation of element corners. Thus, to correctly display local fields, one must consider the orientation of the element. This can be achieved by considering the global coordinates and/or global indices of the corner vertices of the element, and of the intersection faces in between two elements. \\

\noindent
We start by constructing a curvilinear grid as previously demonstrated. We then proceed to initialize the writer, as well as fix its virtual refinement order to 15. We do this because we want to resolve the details of the sinusoid function to a very high precision, and because the mesh size is small. This operation is time-consuming, and, in general, the user should be aware of quadratic complexity of virtual refinement, and choose the refinement order that is necessary and sufficient. Should we choose to avoid specifying a fixed refinement order, the writer will calculate this order itself by considering the curvilinear order of the element. This works fine for most FEM applications, unless the basis order of the element is larger than its curvilinear order.

\begin{mybox}
\begin{lstlisting}    
  Dune::CurvilinearVTKGridWriter<GridType> writer(*grid);
    
  const int userDefinedVirtualRefinement = 15;
  writer.useFixedVirtualRefinement(userDefinedVirtualRefinement);
\end{lstlisting}
\end{mybox}

\noindent
We proceed to stack the pointers to the above field functors into 4 arrays, distinguished by vector/scalar and 2D/3D fields. We have decided to use dynamic polymorphism to implement this functionality. While this may be less efficient than other implementations, it allows writing fields produced by different functors by the very same compact routine.

\begin{mybox}
\begin{lstlisting}    
  std::vector<BaseVTKScalarFunction2D *> vtkFuncScalarSet2D_;
  std::vector<BaseVTKScalarFunction3D *> vtkFuncScalarSet3D_;
  std::vector<BaseVTKVectorFunction2D *> vtkFuncVectorSet2D_;
  std::vector<BaseVTKVectorFunction3D *> vtkFuncVectorSet3D_;

  vtkFuncScalarSet2D_.push_back(
    new VTKFunctionBoundarySegmentIndex<GridType>(*grid));
  vtkFuncScalarSet3D_.push_back(
    new VTKFunctionElementIndex<GridType>(*grid));

  vtkFuncVectorSet2D_.push_back(new VTKFunctionLocalSinusoidFace());
  vtkFuncVectorSet2D_.push_back(new VTKFunctionGlobalSinusoidFace());
  vtkFuncVectorSet3D_.push_back(new VTKFunctionLocalSinusoidElem());
  vtkFuncVectorSet3D_.push_back(new VTKFunctionGlobalSinusoidElem());

  writer.addFieldSet(vtkFuncScalarSet2D_);
  writer.addFieldSet(vtkFuncScalarSet3D_);
  writer.addFieldSet(vtkFuncVectorSet2D_);
  writer.addFieldSet(vtkFuncVectorSet3D_);
  writer.write("./", "tutorial-3-output"););
\end{lstlisting}
\end{mybox}

\begin{figure}[H]
	\begin{subfigure}[b]{0.45\textwidth} \hspace{4mm} \includegraphics[scale=0.14]{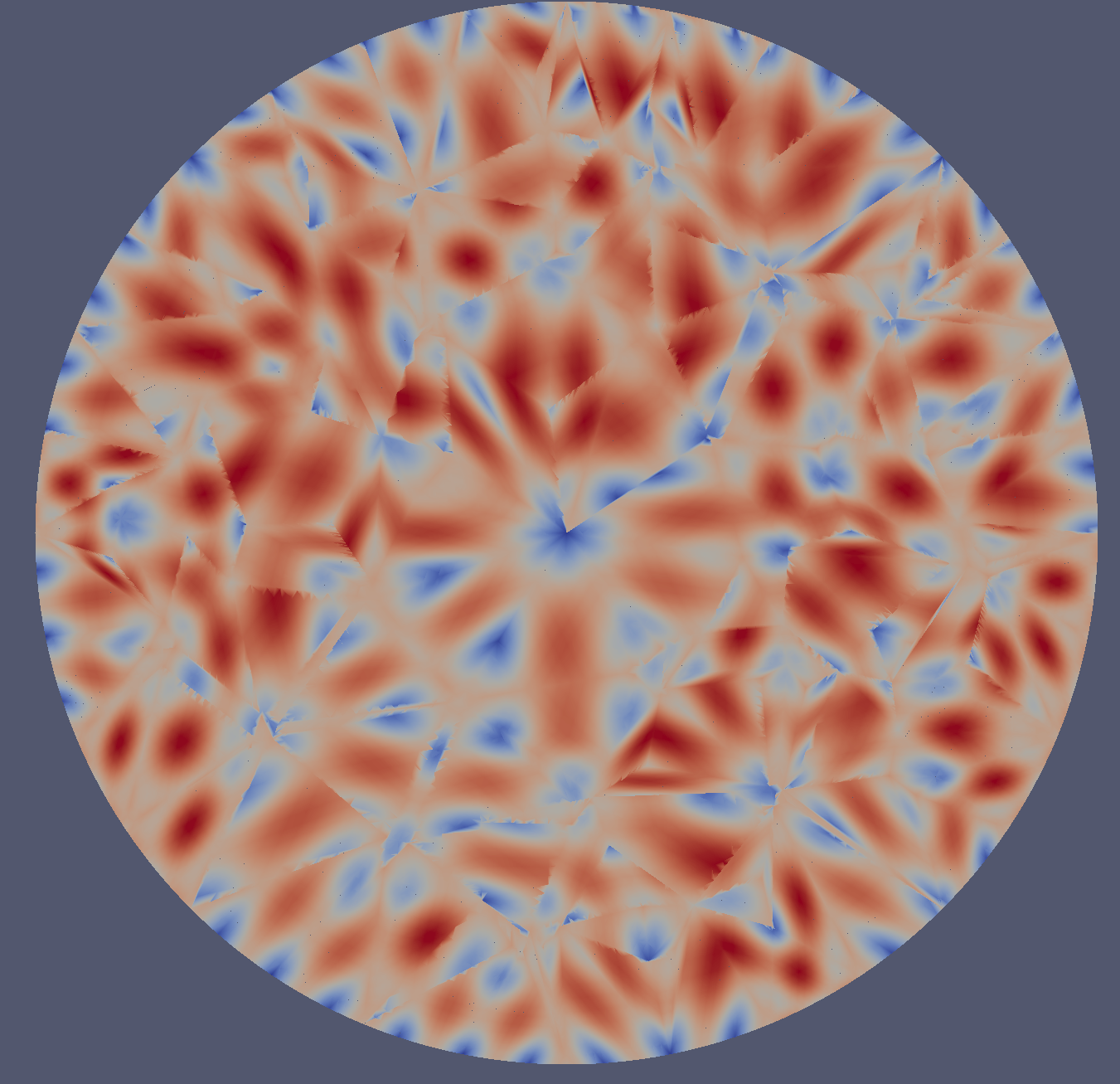} \end{subfigure}
	\begin{subfigure}[b]{0.45\textwidth} \hspace{4mm} \includegraphics[scale=0.16]{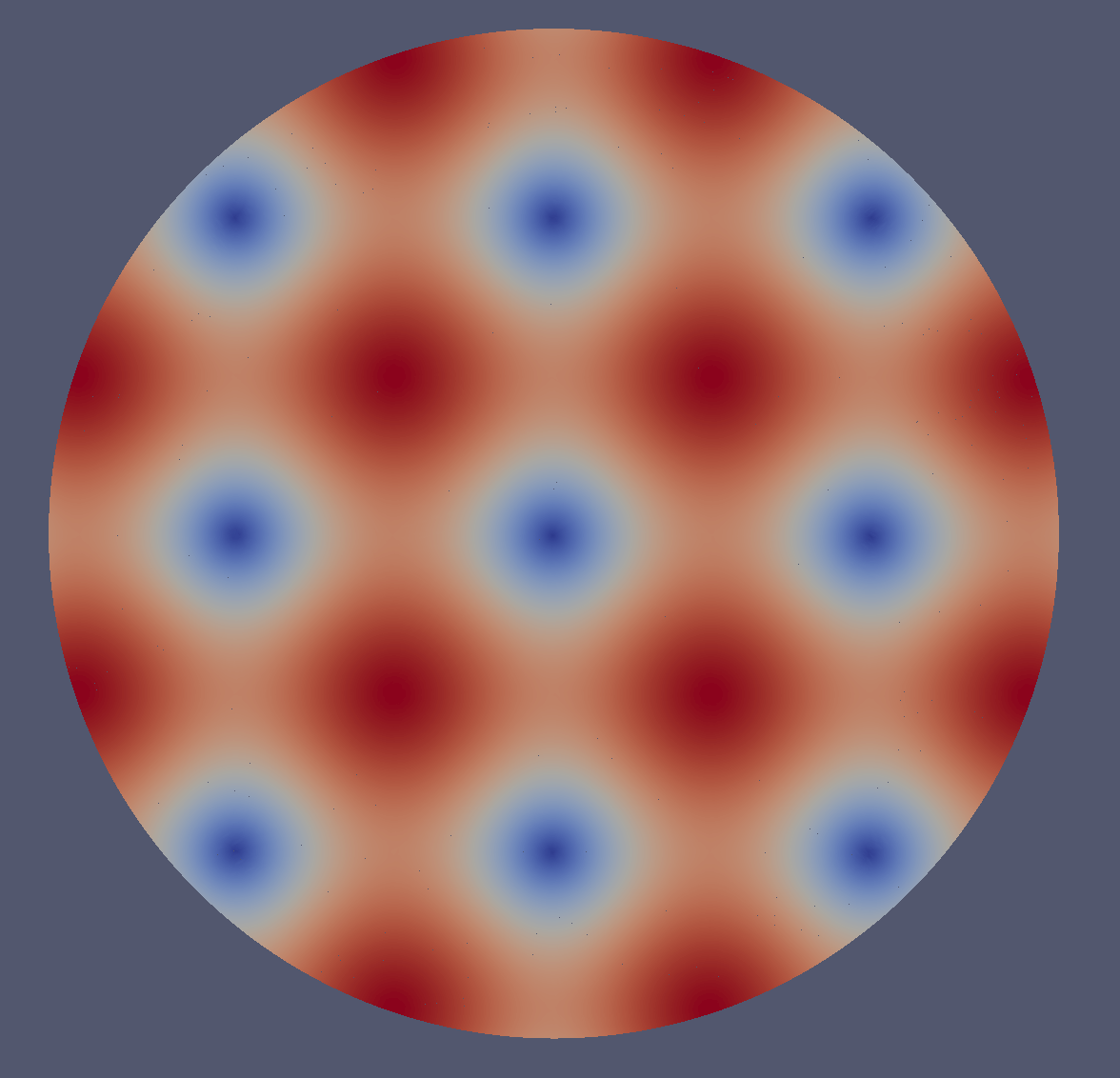} \end{subfigure}
	\caption{ Visualization of tutorial 3. A sinusoid as function of local and global coordinates. This example emphasizes that there is no a priori orientation of the local coordinates. It is the task of the user to ensure that the local field is correctly oriented by considering the global indices of intersecting entities.}
	\label{fig:tutorial3:sinusoid}
\end{figure}

\subsection{Tutorial 4 - Quadrature Integration Tutorials}
\label{usage-howto-tutorial-integration-quadrature}

The following two tutorials demonstrate the capability of \curvgrid{} to address mathematical and physical problems requiring integration of certain quantities over the grid domain. \\

\noindent
\textbf{Scalar Surface Integral - \textit{Gauss} Law} \\
\noindent
In this example we verify \textit{Gauss} law numerically by computing the surface integral of the electric field produced by a unit point charge across the domain boundary of the mesh enclosing that charge. The tutorial will demonstrate that changing the curvature of the domain boundary or the position of the charge inside the domain does not affect the result that is $4 \pi$. More precisely, we compute the integral
\[\int_{\delta \Omega} \vec{E}(\vec{x}) \cdot d\vec{S} = \int_{\delta \Omega} \vec{E}(\vec{x}(\vec{u})) \cdot \vec{n}(\vec{u}) I(\vec{u}) d^2 u \]
\noindent
where $\vec{x}$ is the global coordinate, $\vec{u}$ is the coordinate local to the surface finite element,
\[\vec{E}(\vec{x}) = \frac{\vec{x} - \vec{x}_0}{|\vec{x} - \vec{x}_0|^{-3}}\]
is the electric field of a unit charge located at $\vec{x}_0$, $\vec{n}(\vec{u})$ is the surface outer normal in global coordinates as a function of local coordinates and
\[I(\vec{u}) = \sqrt{\det [ J^T(\vec{u}) J(\vec{u}) ]}\]
is the generalized integration element due to conversion of the integral from global to local coordinates (see \cref{appendix:integrationelements:proof}). \\

\noindent
In order to use the \curvgeom{} integration capabilities, the user must provide the integrand in the form of a functor class. The functor need not be overloaded, but must implement the $()$ operator

\begin{mybox}
\begin{lstlisting}
  ResultType operator()(const LocalCoordinate & x) const {...}
\end{lstlisting}
\end{mybox}

\noindent
as a function of the coordinate local to the entity it is evaluated over, in this case, a 2D coordinate local to a face. The code then iterates over all domain boundaries of the grid, calculating integrals of the functor over each boundary segment and adding up the results. In order to integrate a functor, the \textit{QuadratureIntegrator} class needs to be provided with the entity geometry, the functor to be integrated, the relative and absolute tolerances to be used for integration, as well as the norm type to be used for the case of multidimensional integration. 
\begin{mybox}
\begin{lstlisting}
  typedef Dune::QuadratureIntegrator<ct, DIM2D>  Integrator2DScalar;
  typedef typename Integrator2DScalar::template Traits<Integrand2D>::StatInfo  StatInfo;	
  StatInfo thisIntegralG = Integrator2DScalar::template integrateRecursive<FaceGeometry, Integrand2D, NORM_TYPE>(geometry, gaussf, RELATIVE_TOLERANCE, ACCURACY_GOAL);
\end{lstlisting}
\end{mybox}

\noindent
\textit{StatInfo} is the return data type of the \textit{QuadratureIntegrator}, which is a pair of the integral result and the quadrature order at which the desired relative accuracy was achieved. Note that the absolute accuracy \textit{ACCURACY\_GOAL} is used to determine if the integral is close enough to 0, as relative accuracy can not be used for this purpose due to division by 0. \\

\noindent
\textbf{Surface Vector Integral - Normal Integral} \\
\noindent
This test demonstrates the capability of \textit{QuadratureIntegrator} to integrate vector quantities. It tests the well-known identity, stating that the integral over the unit outer normal over a closed bounded domain is 0.

\begin{equation}
	\oiint_{\partial \Omega} n_x dS = \oiint_{\partial \Omega} \vec{e}_x \cdot \vec{n} dS = \iiint_{\Omega} \nabla \cdot \vec{e}_x dV = 0
\end{equation}

\noindent
The implementation of this test is almost identical to the \textit{Gauss} integral test above. The only difference is that \textit{Dune::FieldVector} is used as a functor output, and internally, instead of taking the dot product between the outer normal and the electric field, the normal itself is returned

%\subsection{Tutorial 4 - Recursive Numerical Integration}
%\label{usage-howto-tutorial-integration-recursive}

\subsection{Tutorial 5 - Communication via the DataHandle Interface}
\label{usage-howto-tutorial-communication}

This tutorial, consisting of two parts, serves as a simple use-case of interprocessor communication through grid entities, which is achieved via the \textit{DataHandle} interface. \\

\noindent
In the first tutorial we explicitly select all entities of the grid that are supposed to communicate for each available communication protocol, and communicate a dummy constant. The goal is to confirm that all the expected entities were communicated over, and all others were not. In the second tutorial, the global index of each entity is sent according to the specified communication protocol, and compared to the global index on the receiving side. It is demonstrated that these indices match for each communicating pair, and an exception is thrown should this not be the case. \\

% \noindent
% Briefly, the procedure communicateConstant iterates over all process process boundary intersections, and marks its neighbour entities, if they will be sending or receiving within the provided interface. Then the custom DataHandleConst is used to perform the communication. This data handle communicates one and the same constant to all entities it is requested. The important thing is that whenever gather or scatter methods are called, it checks if the requested entity was marked for sending or receiving, and throws an error if it is not. Finally, the main procedure checks that the total number of entities that have received the constant is equal to the number of entities marked for receiving.

\noindent
\curvgrid{} implements the \textit{DataHandle} communicators in accordance to the standard \dune{} interface, and its functionality does not exceed that envisioned by the standard interface. Since these tutorials are rather involved, and the standard mechanism is well-documented within the standard \dune{} reference, the detailed explanation of these two tutorials is beyond the scope of this paper. For further information, the user is referred to the \textit{Dune Grid Howto} documentation, found in \url{www.dune-project.org}

\subsection{Tutorial 6 - Parallel Data Output}

This tutorial demonstrates the use of a small utility \textit{ParallelDataWriter} designed for sorted parallel output of vectors. It explores the benefits of using the global element index provided by the mesh, for debugging of parallel numerical codes. The tutorial demonstrate that a quantity sampled over all elements of the grid and sorted by the element global index is independent of the number of processes used. Thus, the user can directly compare the output file, generated by runs with varying process count. \\

\noindent
\textit{Important Note: This tutorial only works if the mesh-provided global element index is re-used by the curvilinear \gmsh{} reader (default). If this is not the case, the automatically generated global index will depend on the number of processes and destroy the above symmetry.}

\noindent
This tutorial samples the volume of the elements. However, the interface of the \textit{ParallelDataWriter} extends to writing vectors of data for each element as well. One must first define the class

\begin{mybox}
\begin{lstlisting}
  <class Grid, class IndexType, class DataType>
  class ParallelDataWriter
\end{lstlisting}
\end{mybox}

\noindent
where \textit{IndexType} and \textit{DataType} are the data types of index and data arrays respectively. They can be any of the Plain Old Datatype (POD) classes, as long as their size is fixed over all processes for the communication purposes. One would then proceed to call the writer routine

\begin{mybox}
\begin{lstlisting}
  static void writeParallelData2File(std::string filename, std::vector<IndexType> & interiorElementGlobalIndex, std::vector<int> & interiorElementNDof, std::vector<DataType> & data, const Grid & grid)
\end{lstlisting}
\end{mybox}

\noindent
where \textit{interiorElementNDof} denotes the number of data entries per global index, and \textit{data} stores all data entries for all global indices in a contiguous 1D vector. \\

\subsection{Tutorial 7 - Global Boundary Communicator}

\noindent
This tutorial demonstrates the capabilities of \curvgrid{} to handle dense boundary communication problems, such as the Boundary Integral (BI) method. In the BI method, the global matrix or part of it correspond to pairwise coupling of every two faces on a given closed surface, providing a fully populated, i.e. dense, matrix. Each processor needs to obtain complete information of all faces on a given surface, collected over all processes.

\begin{mybox}
\begin{lstlisting}
  typedef Dune::CurvGrid::GlobalBoundaryContainer<GridType> BoundaryContainer;
  BoundaryContainer container(*grid, isDomainBoundary, volumeTag, surfaceTag);
\end{lstlisting}
\end{mybox}

\noindent
In case \textit{isDomainBoundary} is set to true, the \textit{BoundaryContainer} does not require the last two parameters. Otherwise, one must specify the surface tag of the interior surface, as well as the volume tag of elements either on the interior or the exterior of the surface, all one or the other. The unit outer normal for each face of the surface is determined as the unit outer normal of the associated element. Thus, if one provides the volume tag of the elements on the outside of the interior surface, one will always receive the inner unit normal at a later stage, and will have to multiply all of them by -1 in order to obtain the unit outer normal. The \textit{BoundaryContainer} does not contain the surfaces already located on this process, in order to save space.  \\

\noindent
In order to iterate over the \textit{BoundaryContainer}, we implement the accompanying \textit{BoundaryIterator}

\begin{mybox}
\begin{lstlisting}
  BoundaryIterator iter(container);
  while (!container.end()) {...}
\end{lstlisting}
\end{mybox}

\noindent
The boundary iterator re-implements most of the functionality of the standard Dune iterator, such as \textit{geometry()}, \textit{unitOuterNormal()}, \textit{indexInInside()}, \textit{geometryInInside()}, as well as some additional functionality compactified into the same iterator to save on communication complexity, such as 

\begin{mybox}
\begin{lstlisting}
  template <int codim>
  UInt globalIndex(UInt subIndexInFace) const {...}

  template <int codim>
  UInt globalIndexInParent(UInt subIndexInElem) const 

  UInt order() const { }

  BaseGeometryEdge geometryEdge(UInt subIndex) const {..}
\end{lstlisting}
\end{mybox}

\noindent
The tutorial performs several tests of the communicated boundary, such as the normal integral from Tutorial 4, calculation of total number of edges, as well as the complementarity of the global index of boundary surfaces located on this process, and on the \textit{BoundaryContainer}. \\

\subsection{Tutorial 8 - Interior Boundary}

The last tutorial extends the \textit{Gaussian} integral tutorial to interior boundaries, performing integrals for a set of charges inside and outside of the boundary. This is a simple test to verify if the interior boundary of the mesh forms a closed surface.

%%\subsection{Tutorial 5 - Polynomial Manipulation and Integration}
%%\label{usage-howto-tutorial-polynomial}
%%
%%
%%\subsection{Tutorial 7 - Point Location - OCTree}
%%\label{usage-howto-tutorial-octree}

%%%%%%%%%%%%%%%%%%%%%%%%%%%%%%%%%%%%%%%%%%%%%%%%%%%%%%%%%%%%%%%%%%%%%
% Curvilinear Grid Diagnostics - Section on mesh statistics and visualisation
%%%%%%%%%%%%%%%%%%%%%%%%%%%%%%%%%%%%%%%%%%%%%%%%%%%%%%%%%%%%%%%%%%%%%

\section{Diagnostics tools}
\label{section-diagnostics}

\begin{figure}
\centering
	\begin{subfigure}[b]{0.48\textwidth} \includegraphics[scale=0.22]{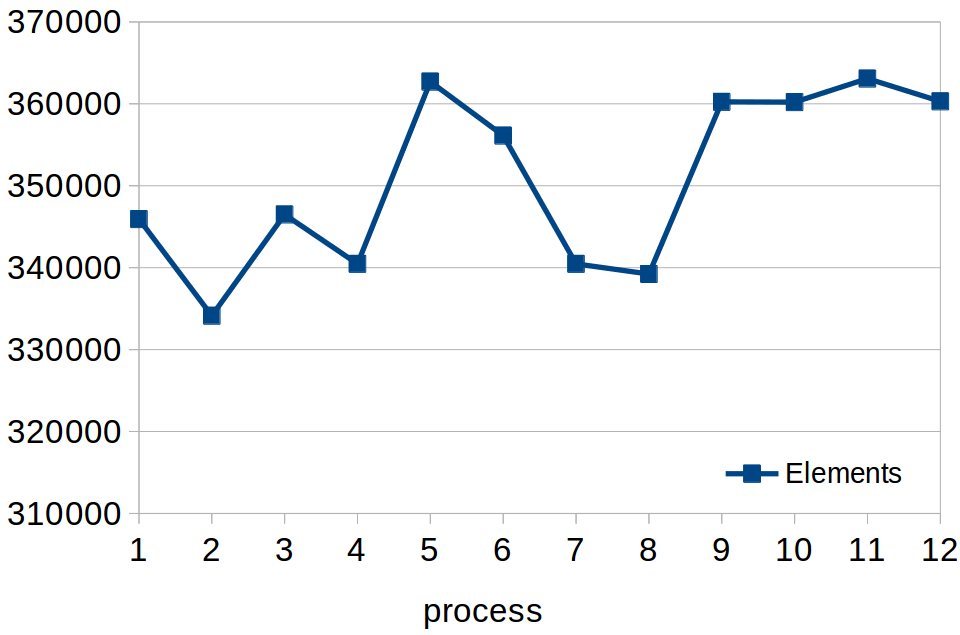} \end{subfigure}
	\begin{subfigure}[b]{0.48\textwidth} \includegraphics[scale=0.24]{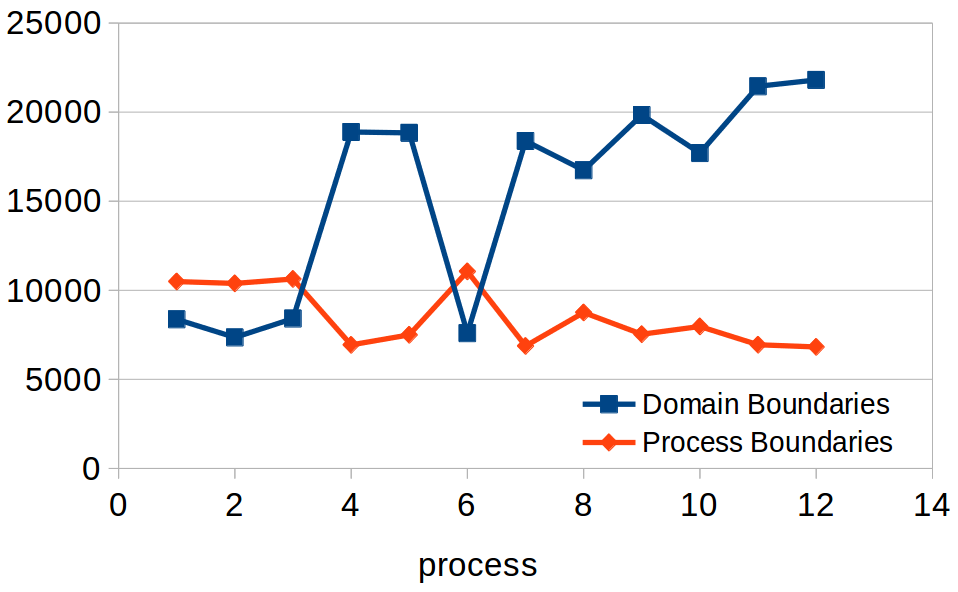} \end{subfigure}
\caption{Distribution of the Bullseye grid entities among 12 processes}
\label{fig:impl:storage:bullseyeelemdistr}
\end{figure}

\begin{figure}
\centering
	 \includegraphics[scale=0.15]{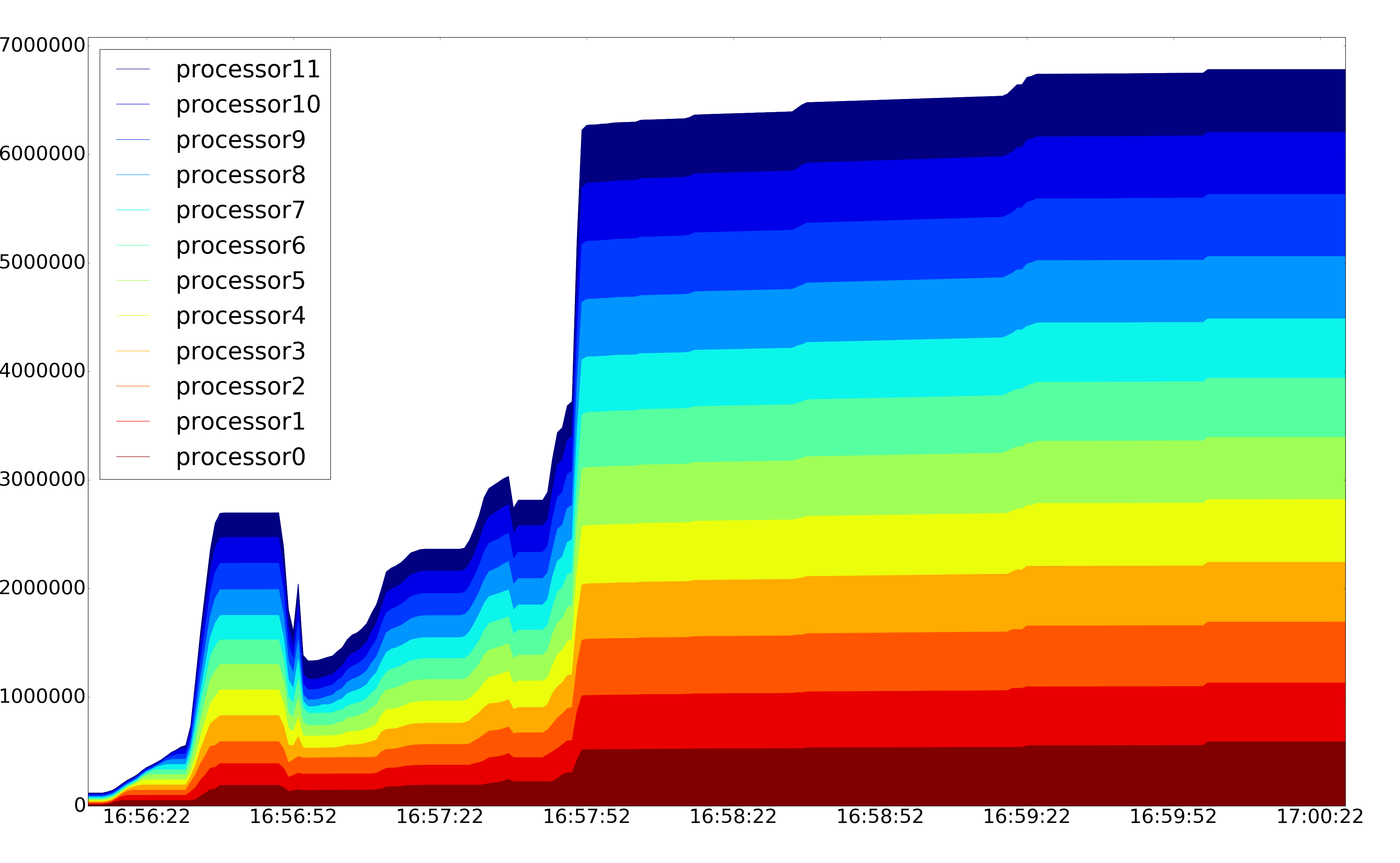}
\caption{Memory consumption during the mesh reading, grid construction and writing. Memory is given in KB, and the curves are progressively added, meaning that the memory taken by each process corresponds to the area, not the absolute value. The x-axis corresponds to system time.}
\label{fig:impl:storage:realtimelog}
\end{figure}

\curvgrid{} provides a diagnostics routine \textit{grid-diagnostics}, designed to analyze the grid construction procedure. As an example, consider running the routine on a 12 core machine, using a 4.2 million element first order \textit{Bullseye} mesh \cref{fig:result:bullseye}. The diagnostics routine uses \curvreader{}, \texttt{gridconstructor} and \curvwriter{} to read the mesh, construct the grid, and write the resulting grid into a VTU/PVTU file set. Afterwards, it runs a number of statistics tests, evaluating the element distribution, average size and curvature. \cref{fig:impl:storage:bullseyeelemdistr} presents the distribution of elements, process boundaries and domain boundaries among the 12 processes. At any point in time, \curvgrid{} can report parallel timing, that is, minimal and maximal time taken by each process for each timed stage of the grid (\cref{table:impl:storage:timelog}), using the \textit{LoggingTimer} utility. This utility can also be used to time user code. Finally, the \textit{RealTimeLog} utility, provided in \curvgrid{}, can be used to perform logging of the system memory real time by analyzing the corresponding system file. \cref{fig:impl:storage:realtimelog} presents the plot of the memory logging file, as it is output after the mesh has been written to file. By comparing the plot with the timing, one can discern that a large amount of extra memory is taken by auxiliary structures during the grid reading and writing. The grid construction is finished at 16:56:51, and takes $\approx$ 3.6 GB of total memory.

\begin{table}
\centering
\begin{tabular}{l l l}
\hline
	Min time [s]	& Max time [s]	& Action \\ \hline
	0.38411		& 0.384378	& CurvilinearGMSHReader: Skipping vertices \\
	3.86178		& 4.19604	& CurvilinearGMSHReader: Counting elements \\
	40.4335		& 40.7665	& CurvilinearGMSHReader: Reading and partitioning linear elements \\
	20.3694		& 23.1265	& CurvilinearGMSHReader: Reading complete element data \\
	11.071		& 13.6252	& CurvilinearGMSHReader: Reading boundary segment data \\
	0.758839	& 1.08354	& CurvilinearGMSHReader: Reading vertex coordinates \\
	3.50761		& 7.87511	& CurvilinearGMSHReader: Inserting entities into the factory \\
	8.29851		& 10.6347	& CurvilinearGridConstructor: Entity generation \\
	0.610859	& 2.95748	& CurvilinearGridConstructor: Global index generation \\
	0.228916	& 0.346936	& CurvilinearGridConstructor: Ghost element generation \\
	2.07785		& 2.81369	& CurvilinearGridConstructor: Index set generation \\
	0.300707	& 0.641833	& CurvilinearGridConstructor: Communication map generation \\
	0.0501765	& 0.685088	& CurvilinearGridConstructor: Communication of neighbor ranks \\
	151.23861	& 154.15874	& CurvilinearVTKWriter: Writing grid
\end{tabular}
\caption{\curvgrid{} parallel timing output for Bullseye grid diagnostics. Time is given in seconds. Minimum/maximum is taken over all processes doing the task}
\label{table:impl:storage:timelog}
\end{table}

% \subsection{Mesh statistics}
% \label{section-diagnostics-statistics}
% 
% 
% \subsection{Visualisation}
% \label{section-diagnostics-visualisation}

\pagebreak
\section{Testing}

\subsection{Curvilinear Geometry}
\label{sec:tests:curvgeom}

\noindent
\textbf{test-polynomial}. This test performs arithmetic operations, differentiation and integration of basic polynomials in 1D, 2D and 3D. \\

\noindent
\textbf{test-polynomialvector}. This test generates random polynomial vectors in 3D and checks that basic vector calculus identities $\nabla \times \nabla f(\vec{x}) = 0$ and $\nabla \cdot (\nabla \times \vec{g}(\vec{x})) = 0$ hold. \\

\noindent
\textbf{test-quadratureintegration}. This test performs recursive integration on a set of functions \cref{appendix:tests:quadintegrator:performance} (given in  \cref{appendix-geometry-tests-integral}) and reports the order at which the integral converges. \\

\noindent
\textbf{test-quadratureintegration-matrix}. This test constructs random polynomial matrices, integrates them both recursively and analytically and compares the results. \\

\noindent
\textbf{test-lagrangeinterpolation}. This test uses explicit polynomial maps given by functors and interpolates them using progressively higher order. It then evaluates the interpolated mapping, first for the interpolation points, then for a set of random points within the entity, and compares to the exact mapping. For the interpolatory points, the analytical and interpolated maps should always match due to the interpolatory property of \textit{Lagrange} polynomials, whereas for all other points within the entity the maps would only match if the polynomial interpolation order is greater or equal to the polynomial order of the original map. \\
%It is intended to also test the \textit{SubentityInterpolator} method in the future. This can be done by evaluating the local-to-global map on the subentity and on the parent entity, restricting the sampling points to the subentity in question, and verifying that they match. \\

\noindent
\textbf{test-curvilineargeometry}. This set of tests is performed for all 3 dimensions and for interpolation orders 1 to 5, first constructing the curvilinear geometries and cached geometries for a set of analytical functions \cref{appendix:tests:curvgeom:integrand} in \cref{appendix-geometry-tests-integral}.
\begin{itemize}
	\item \textbf{Test 1}. Evaluate the \textit{global()} mapping for all corners of the entity and compare to the analytical map.
	\item \textbf{Test 2}. Evaluate the \textit{global()} mapping for a random set of local coordinates within the entity, and compare the results to the analytical map. The test is omitted if the interpolation order is smaller than the order of the mapping.
	\item \textbf{Test 3}. Evaluate the \textit{local()} mapping for all global interpolation points of the entity and compare to the interpolatory reference grid. Also, check if all these points are reported to be inside the element. As described in \cref{sec:theory:coordinatetransform}, the current \textit{local()} method is imperfect:
		\subitem -It fails to converge to a boundary point if the geometry has a zero derivative on the corner or on the whole boundary. This is the expected behavior for singular entities, thus such entities should be avoided at the meshing stage.
		\subitem -It occasionally fails to correctly identify the point to be interior to the entity, if the point is close to the boundary.
	\item \textbf{Test 4}. It is verified if $\vec{p} \approx local(global(\vec{p}))$ for a set of random coordinates $\vec{p}$ within the entity. It is also checked if all the sample points are reported to be inside the entity as they should be.
	\item \textbf{Test 5}. It is verified that the global coordinates on the immediate exterior of the entity are correctly identified to be outside it, at the same time checking the functionality of the subentity normal functionality. For this, unit outer normals are constructed for a set of points across the boundary of the entity at regular intervals. The sample exterior points are then defined to be $\vec{p} = \vec{g} + \alpha \vec{n}$, where $\vec{g}$ is the global coordinate of the boundary point, $\vec{n}$ the normal at that point, and $\alpha = 0.01 L$ a small displacement, where $L$ is the length scale of all entities (see \cref{fig:geometry:test:normal}).
	\item \textbf{Test 6}. The scalar basis functions given in \cref{appendix:tests:curvgeom:integrand} are integrated over the reference geometry, and results are compared to the exact values given in \cref{appendix:tests:curvgeom:mappings}
	\item \textbf{Test 7}. The dot product surface integrals of vector basis functions are integrated over the reference geometry, and compared to  the exact values. Integrands, mappings and exact results are given in \cref{appendix:tests:curvgeom:dotproductintegral}
\end{itemize}

\begin{figure}
    \centering
    \includegraphics[scale=1.0]{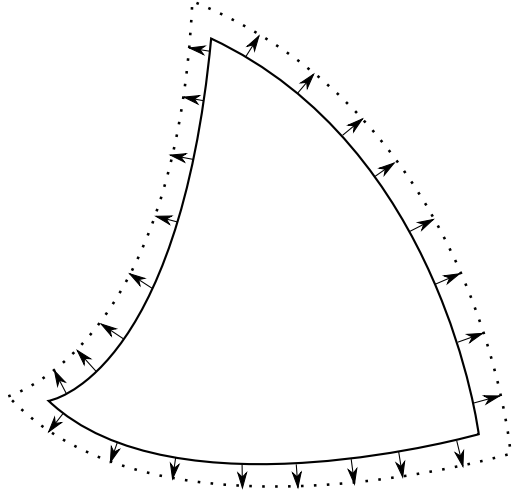}
	\captionsetup{width=0.8\textwidth} 
	\caption{A combined accuracy test to find the outer normal and capability to accurately determine whether a given global coordinate is exterior. Face normals are sampled on a regular grid over the boundary, and are used to produce global coordinates barely exterior to the entity. }
	\label{fig:geometry:test:normal}
\end{figure}

\subsection{Curvilinear Grid}
\label{sec:tests:curvgrid}

At present, \curvgrid{} does not employ explicit testing procedures. The confidence in its accuracy is based on extensive testing of \curvgeom{}, the accuracy of the provided tutorials, and the fact that our 3D FEM code based on the \curvgrid{} successfully reproduces reproduces standard benchmarks and experimental results. For discussion on future development of \curvgrid{} testing procedures, please refer to \cref{sec:conclusion:futurework}

\pagebreak
%\section{Conclusion}
\section{Further work}
\label{sec:conclusion:futurework}

% \subsection{Further work}
% \label{sec:conclusion:futurework}

\noindent
In this section we discuss the further work to be undertaken in the development of \curvgrid{}. \cref{table:conclusion:todolist} presents a list of directions for further work in terms of functionality, performance and scalability. In addition, it is fully intended to implement an automatic \curvgrid{} testing procedure. The essential part of automatic testing is the integration of the standard \dunegrid{} testing procedures, applicable to all grids that adhere to the facade class interface. From one side, \curvgrid{} extends the standard \dunegrid{} interface, and it is not yet decided if it is best to extend the standard interface to include the additional curvilinear functionality, or if it is preferable to perform the additional curvilinear-only tests in a separate routine. From the other side, several tests for the current standard interface (for example, global-to-local mapping) are hard-wired to linear geometries and do not foresee the difference in expected behavior of such tests between linear and curvilinear grids. The integration of curvilinear standard functionality is an ongoing discussion within the \dune{} community. \\

\noindent
We would also like to elaborate on the improvement of scalability of \curvgrid{} assembly. Currently, the assembly of grid connectivity requires an all-to-all communication step to determine the neighboring processes for each shared vertex. An optimal algorithm to determine process neighbors should make use of the following observations:
\begin{itemize}
 \item Most interprocessor boundary vertices are shared by two processes only, all other cases are progressively more rare
 \item If an interprocessor boundary vertex is shared between two processes, the surrounding vertices are likely to be shared by the same processes
\end{itemize}

\noindent
Another convention responsible for the scalability of the construction process, albeit not as much as the all-to-all communication, is the shared entity ownership. Currently the shared entities are owned by the lowest rank containing them. This results in progressively higher workload of global index enumeration for lower rank processes. This paradigm can and should be replaced by a more balanced one. Another desired property is avoiding parallel communication in entity ownership determination. Thus, the ownership must be uniquely determined from the sharing process ranks and the shared entity corner global indices, and must result in the same ordering on all processes. An example of such convention is a $XOR$ operation between the entity corner global indices and the containing process ranks. This quantity is the same over all processes computing it, and is more or less random, resulting in a much better workload distribution than the lower rank convention. That said, the global index enumeration part is a $O(n)$ algorithm, and thus is one of the faster parts of the constructor even without optimization.

\begin{flushleft}
\begin{table}
\begin{tabularx}{\textwidth}{@{}| l X |@{}}
\hline
Functionality & 1D and 2D curvilinear simplex grids \\
Functionality & Arbitrary polynomial order curvilinear simplex meshes. Requires a generalized mapper from \gmsh{} to \textit{Sorted Cartesian} node indexing. \\
Functionality & Non-uniform polynomial order curvilinear meshes \\
Functionality & Additional geometry types (e.g. hexahedral, prismatic) \\
Functionality & Non-conformal curvilinear meshes (with hanging nodes) \\
Functionality & Global and local refinement of curvilinear grids, including adaptive refinement \\
Functionality & Mixed element grids \\
Functionality & Usage of \gmsh{} partition tags to read pre-partitioned meshes \\
Functionality & Multi-constraint grid partition, for example, for simultaneous element and boundary segment load balancing \\
Functionality & Dynamic load balancing \\
Functionality & Front/Overlap partition types \\
Functionality & Identification and management of periodic boundaries directly from boundary tags \\
Performance & Symbolic polynomial vector and matrix classes that share pre-computed monomials \\
Performance & Adaptive quadrature (e.g. \textit{Clenshaw-Curtis}), sparse grids \cite{petras2000} \\
Performance & Efficient location of the curvilinear element containing a given global coordinate (for example, via Octant Tree) \\
Performance & \textit{BoundaryContainer} interior surface outer normal computation instead of communicating it  \\
Performance & Optimization of \curvwriter{} performance, memory footprint, and resulting file size	 \\
Scalability & Complete switch of \curvgrid{} to neighbor MPI communication \\
Scalability & Improved load balance of shared entity ownership during grid construction \\
Scalability & \textit{ParallelDataWriter} scalability improvement using the \textit{MPI-3} parallel file output  \\
Scalability & \textit{BoundaryContainer} boundary surface communication using blocks that fit process memory \\
\hline
\end{tabularx}
\caption{To do - list of \curvgrid{}}
\label{table:conclusion:todolist}
\end{table}
\end{flushleft}

\pagebreak
\appendix

\section{Proofs and Concepts}

%%%%%%%%%%%%%%%%%%%%%%%%%%%%%%%%%%%%%%%
% Proof of explicit expression for the integration elements
%%%%%%%%%%%%%%%%%%%%%%%%%%%%%%%%%%%%%%%
\subsection{Integration Elements}
\label{appendix:integrationelements:proof}

When transforming an integral over an entity from global to local coordinates, the resulting integral acquires an additional prefactor called Integration Element

\[ \int_{V} dx dy dz = \int_{V'} I(u,v,w) du dv dw \]

\noindent
\textbf{Edge(1D Local Coordinates)}:
\[\vec{dl} = \vec{p}(u + du) - \vec{p}(u) = \frac{\partial \vec{p}}{\partial u} du \]
\[dl = |\vec{dl}| = |\frac{\partial \vec{p}}{\partial u}| du = I(u)du \]
For example, if global coordinates are 3D, then
\[I(u) = \sqrt{\bigl (\frac{\partial x}{\partial u})^2 + \bigl (\frac{\partial y}{\partial u})^2 + \bigl (\frac{\partial z}{\partial u})^2} \]

\noindent
\textbf{Face(2D Local Coordinates)}:
\[\vec{dA} = (\vec{p}(u + du, v) - \vec{p}(u, v)) \times (\vec{p}(u, v + dv) - \vec{p}(u, v)) = \frac{\partial \vec{p}}{\partial u} \times \frac{\partial \vec{p}}{\partial v} du dv \]
\[dA = |\vec{dA}| = \bigl | \frac{\partial \vec{p}}{\partial u} \times \frac{\partial \vec{p}}{\partial v} \bigr | du dv \]
For example, if global coordinates are 3D, then
\[I(u,v) = \biggl | \begin{pmatrix}
  \partial_u y \partial_v z - \partial_u z \partial_v y \\
  \partial_u z \partial_v x - \partial_u x \partial_v z \\
  \partial_u x \partial_v y - \partial_u y \partial_v x
\end{pmatrix} \biggr | du dv \]

\noindent
\textbf{Element(3D Local Coordinates)}:
\begin{eqnarray*}
  dV & = & ((\vec{p}(u + du, v, w) - \vec{p}(u, v, w)) \times (\vec{p}(u, v + dv, w) - \vec{p}(u, v, w))) \\
     && \cdot  (\vec{p}(u, v + dv, w) - \vec{p}(u, v, w)) du dv dw  \\
     & = & (\partial_u \vec{p} \times \partial_v \vec{p}) \cdot \partial_w \vec{p} \; du dv dw
\end{eqnarray*}
\[ I(u,v,w) = (\partial_u \vec{p} \times \partial_v \vec{p}) \cdot \partial_w \vec{p} \]

\noindent
It can be shown that the three above cases can all be rewritten as
\[ I(\vec{r}) = \sqrt{\det(J(\vec{r})^T J(\vec{r}))} \]
where $\vec{r}$ are the local coordinates, and $J(\vec{r})$ is the Jacobian transformation defined as
\[J(\vec{r}) = \frac{\partial \vec{p}}{\partial \vec{r}}\]
which, for 3D global coordinates can be written as
\[J_{Edge}(u) = \bigl(\frac{\partial x}{\partial u}, \frac{\partial y}{\partial u}, \frac{\partial z}{\partial u} \bigr)  \]
\[J_{Face}(u,v) = 
\begin{pmatrix}
   \partial_u x &&    \partial_u y  &&    \partial_u z \\
   \partial_v x &&    \partial_v y  &&    \partial_v z
\end{pmatrix}
\]
\[J_{Elem}(u,v,w) = 
\begin{pmatrix}
   \partial_u x &&    \partial_u y  &&    \partial_u z \\
   \partial_v x &&    \partial_v y  &&    \partial_v z \\
   \partial_w x &&    \partial_w y  &&    \partial_w z
\end{pmatrix}
\]

\noindent
Finally, in case where local and global dimension matches (e.g. edge in 1D, face in 2D, element in 3D), the Jacobian is a square matrix. If the geometry is strictly non-singular ($\det{J} > 0$ everywhere within the entity), the general expression for the integration element can be simplified to
\[ I(\vec{r}) = \det{J(\vec{r})} \]

%%%%%%%%%%%%%%%%%%%%%%%%%%%%%%%%%%%%%%%
% Proof of polynomial integral over simplex
%%%%%%%%%%%%%%%%%%%%%%%%%%%%%%%%%%%%%%%
\subsection{Duffy Transform}
\label{section-abstract-duffy-transform}

The \textit{Duffy} transform is a bijective map between the local coordinate of a hypercube and another entity, in our case, the reference simplex. One can use the \textit{Duffy} transform to directly apply the hypercubic integration routines (e.g. tensor product quadrature rules) to simplex domains. Effectively, the \textit{Duffy} transform claims that
\[\int_{\triangle} f(\vec{x}) d\vec{x} = \int_{\Box} g(\vec{\tau}) f(\vec{x}(\vec{\tau})) d\vec{\tau},\]
giving explicit expressions for $g(\vec{\tau})$ and $\vec{x}(\vec{\tau})$. \\

\noindent
We consider simplices of all dimensions up to 3: \\

\noindent
\textbf{1D}: The reference hypercube and simplex are both an edge defined on $[0, 1]$, so they have exactly the same parameter. \\

\noindent
\textbf{2D}: Here we map from reference triangle to reference square. We transform the integral of interest, namely

\[ \int_0^1 \int_0^{1-x} f \biggl(
\begin{matrix}  x \\ y  \end{matrix} \biggr)
 dx dy = \int_0^1 \int_0^1 (1-x) 
f \biggl( \begin{matrix} x \\ (1-x)t \end{matrix} \biggr)
dx dt \]

\noindent
using a substitution $t = y / (1 - x)$, such that $t \in [0, 1]$. \\

\noindent
\textbf{3D}: Here we map from the reference tetrahedron to the reference cube. We transform the integral of interest, namely

\[ \int_0^1 \int_0^{1-x} \int_0^{1-x-y} f \scalebox{1.2}{\Bigg(}
\begin{matrix}  x \\ y \\ z  \end{matrix} \scalebox{1.2}{\Bigg)}
dx dy = \int_0^1 \int_0^1 \int_0^1 (1-x)^2 (1-t)
f \scalebox{1.2}{\Bigg(} \begin{matrix}  x \\ (1-x)t \\ (1-x)(1-t)\tau  \end{matrix} \scalebox{1.2}{\Bigg)}
dx dt d\tau \]

\noindent
using a substitution $t = \frac{y}{1 - x}, \tau = \frac{z}{(1 - x)(1 - t)}$, such that $t,\tau \in [0, 1]$. \\

%%%%%%%%%%%%%%%%%%%%%%%%%%%%%%%%%%%%%%%
% Method of dealing with first order singular integrals
%%%%%%%%%%%%%%%%%%%%%%%%%%%%%%%%%%%%%%%
%% \input{manual-appendix-geometry-integral-singularityelimination}

%%%%%%%%%%%%%%%%%%%%%%%%%%%%%%%%%%%%%%%
% Proof of polynomial integral over simplex
%%%%%%%%%%%%%%%%%%%%%%%%%%%%%%%%%%%%%%%
\subsection{Proof for polynomial summand integrals}
\label{appendix-proof-simplexintegral}

\noindent
Note that the series for beta-function can be written as
\[B(a+1,b+1) = \int_0^1 x^a (1-x)^b dx = \int_0^1 x^a \sum_{i=0}^b C_b^i (-1)^i x^i = \sum_{i=0}^b \frac{(-1)^i C_b^i}{a+1+i}\]

\noindent
where $C_b^i = \frac{b!}{i!(b-i)!}$ is the binomial coefficient. \\

\noindent
\textbf{1D:} \\

\begin{equation}
	I_{1D} = \int_0^1 x^{\alpha} dx = \frac{x^{a + 1}}{a + 1} \biggr |_0^1 = \frac{1}{a + 1}
\end{equation}

\noindent
\textbf{2D:} \\

\begin{eqnarray*}
	I_{2D} = \int_0^1 \int_0^{1-x} x^{a} y^{b} dx dy
	& = & \frac{1}{b+1} \int_0^1 x^{a} (1-x)^{b+1} dx = \frac{1}{b+1} \beta(a + 1, b + 2) \\
	& = & \frac{1}{b+1} \frac{a!(b+1)!}{(a+b+2)!} = \frac{a!b!}{(a+b+2)!}
\end{eqnarray*}

\noindent
\textbf{3D:} \\

\begin{eqnarray*}
	I_{3D}
	& = & \int_0^1 \int_0^{1-x} \int_0^{1-x-y} x^a y^b z^c dx dy dz \\
	& = & \frac{1}{c+1} \int_0^1 \int_0^{1-x} x^a y^b (1-x-y)^{c+1} dx dy \\
	& = & \frac{1}{c+1} \int_0^1 \int_0^{1-x} x^a y^b \sum_{i=0}^{c+1} C_{c+1}^i (-1)^i y^i (1-x)^{c+1-i} dx dy \\
	& = & \sum_{i=0}^{c+1} \frac{C_{c+1}^i (-1)^i}{c+1} \int_0^1 \ x^a (1-x)^{c+1-i} \int_0^{1-x} y^{b+i} dx dy \\
	& = & \sum_{i=0}^{c+1} \frac{C_{c+1}^i (-1)^i}{(c+1)(b+1+i)} \int_0^1 \ x^a (1-x)^{c+1-i} (1-x)^{b+1+i} dx \\
	& = & \sum_{i=0}^{c+1} \frac{C_{c+1}^i (-1)^i}{(c+1)(b+1+i)} \int_0^1 \ x^a (1-x)^{b+c+2} dx \\
	& = & \sum_{i=0}^{c+1} \frac{C_{c+1}^i (-1)^i}{(c+1)(b+1+i)} \beta(a+1, b+c+3) \\
	& = & \frac{1}{c+1} \beta(b+1,c+2) \beta(a+1, b+c+3) \\
	& = & \frac{1}{c+1} \frac{b!(c+1)!}{(b+c+2)!} \frac{a! (b+c+2)!}{(a+b+c+3)!} \\
	& = &  \frac{a! b! c!}{(a+b+c+3)!}
\end{eqnarray*}

%%%%%%%%%%%%%%%%%%%%%%%%%%%%%%%%%%%%%%%
% Conversion from GMSH to DUNE notation
%%%%%%%%%%%%%%%%%%%%%%%%%%%%%%%%%%%%%%%
\subsection{Convention for numbering interpolatory vertices}
\label{impl-gmsh-numbering-convention}

This section discusses the different conventions for numbering interpolatory vertices for \textit{Lagrange}-based curvilinear entities. The convention used in \textit{Lagrange} interpolation is placing of the interpolatory vertices on a regular grid over the entity, the number of vertices on each edge corresponding to the interpolatory order minus 1, i.e. 2 points to define linear edge, 3 to define quadratic and so on. The convention used in \textit{} can be called \textit{Sorted Cartesian} indexing, and is described by the following vertex enumeration algorithm.

\begin{mybox}
\begin{lstlisting}
for (z=0 to 1, y=0 to 1-z, x=0 to 1-z-y) { vertex(x,y,z); }
\end{lstlisting}
\end{mybox}

\noindent
Instead, the \gmsh{} community uses the a recursive convention

\begin{enumerate}
	\item First number all corners, then all edges, then all faces, then the vertices interior to the element element
	\item Inside an edge, vertices are numbered sequentially from starting to finishing corner
	\item Inside a face, a new triangle is defined recursively from the outer-most interior vertices, with the same order as the original triangle
	\item Inside an element, a new element is defined recursively from the outer-most interior vertices, with the same order as the original triangle
	\item For a triangle, the order of edges is $(0,1)$, $(1,2)$, $(2,0)$. (in 2D)
	\item For a tetrahedron, the order of edges is $(0,1)$, $(1,2)$, $(2,0)$, $(3,0)$, $(3,2)$, $(3,1)$.
	\item For a tetrahedron, the order of faces is $(0, 2, 1)$, $(0, 1, 3)$, $(0, 3, 2)$, $(3, 1, 2)$, including orientation
\end{enumerate}

\noindent
We are still looking for a nice algorithm to analytically map between both conventions for arbitrary order entities. For now, we hard code the  \textit{GMSH} to  \textit{Dune} map for simplex geometries up to order 5. The following table contains the Dune-notation vertex indices corresponding to ascending \textit{GMSH} vertex index

\begin{figure}[H]
    \centering
    \includegraphics[scale=1]{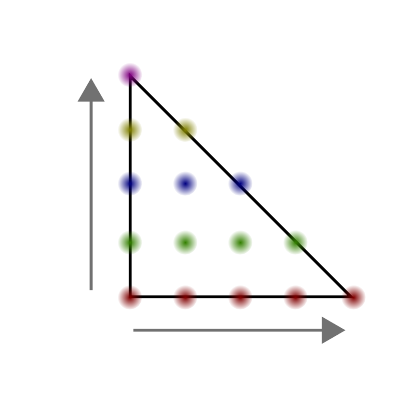}
    \caption{Sorted Cartesian interpolatory vertex enumeration}
    \label{fig:appendix:gmsh:enumeration:dune}
\end{figure}

\begin{figure}[H]
    \centering
    \includegraphics[scale=1]{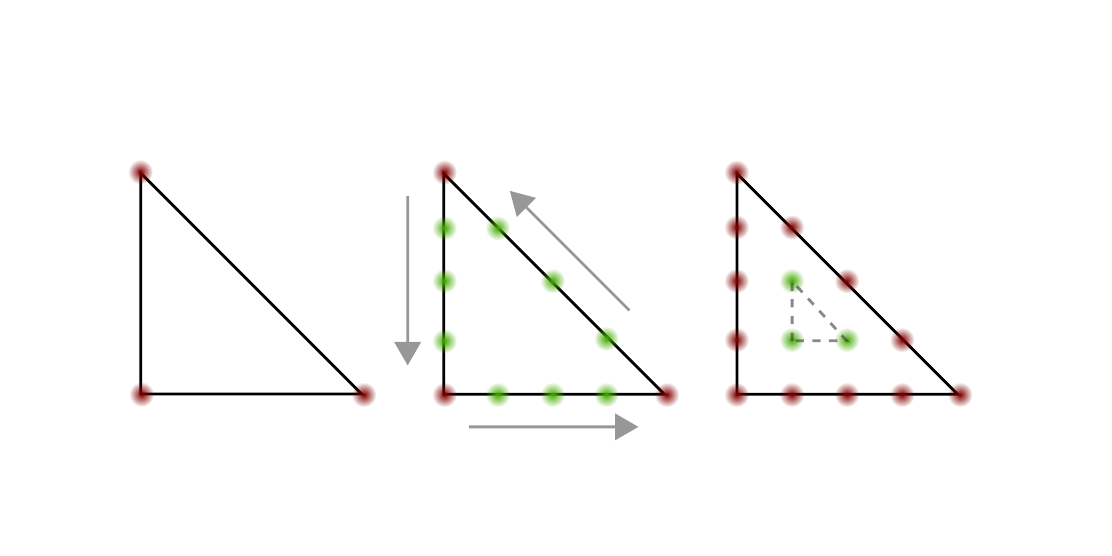}
    \caption{ \textit{GMSH} recursive interpolatory vertex enumeration}
    \label{fig:appendix:gmsh:enumeration:gmsh}
\end{figure}

\begin{mybox}
\begin{itemize}
	\item Triangle Order 1: \{0, 1, 2\}
	\item Triangle Order 2: \{0, 3, 1, 5, 4, 2\}
	\item Triangle Order 3: \{0, 3, 4, 1, 8, 9, 5, 7, 6, 2\}
	\item Triangle Order 4: \{0, 3, 4, 5, 1, 11, 12, 13, 6, 10, 14, 7, 9, 8, 2\}
	\item Triangle Order 5: \{0, 3, 4, 5, 6, 1, 14, 15, 18, 16, 7, 13, 20, 19, 8, 12, 17, 9, 11, 10, 2\}
	
	\item Tetrahedron Order 1: \{0, 3, 1, 2\}
	\item Tetrahedron Order 2: \{0, 7, 3, 4, 9, 1, 6, 8, 5, 2\}
	\item Tetrahedron Order 3: \{0, 11, 10, 3, 4, 17, 14, 5, 15, 1, 9, 18, 12, 16, 19, 6, 8, 13, 7, 2\}
	\item Tetrahedron Order 4: \{0, 15, 14, 13, 3, 4, 25, 27, 19, 5, 26, 20, 6, 21, 1, 12, 28, 29, 16, 22, 34, 31, 24, 32, 7, 11, 30, 17, 23, 33, 8, 10, 18, 9, 2\}
	\item Tetrahedron Order 5: \{0, 19, 18, 17, 16, 3, 4, 34, 39, 36, 24, 5, 37, 38, 25, 6, 35, 26, 7, 27, 1, 15, 40, 43, 41, 20, 28, 52, 55, 46, 33, 53, 49, 30, 47, 8, 14, 45, 44, 21, 31, 54, 51, 32, 50, 9, 13, 42, 22, 29, 48, 10, 12, 23, 11, 2\}
\end{itemize}
\end{mybox}

\section{Interfaces}

%%%%%%%%%%%%%%%%%%%%%%%%%%%%%%%%%%%%%%%
% Implementation of Polynomial Class
%%%%%%%%%%%%%%%%%%%%%%%%%%%%%%%%%%%%%%%
\subsection{Polynomial Class}
\label{interface-geometry-polynomial}

\noindent
An arbitrary polynomial of order $n$ with $d$ parameters can be represented in its expanded form as
\[ p(\vec{u}) = \sum_i A_i \prod_{j = 0}^d u_j^{\mathrm{pow}_{i,j}},  \]
where ${pow}_{i,j}$ is the power $j$\textsuperscript{th} dimension of $i$\textsuperscript{th} summand. For example, in 3D this can be written as
\[ p(\vec{u}) = \sum_i A_i u^{pow_{u,i}} v^{pow_{v,i}} w^{pow_{w, i}},  \]

\noindent
We define a Monomial class, which stores a constant multiplier $A$ and vector of powers $pow$.

\begin{mybox}
\begin{lstlisting}
  PolynomialTraits::Monomial(double prefNew, std::vector<int> powerNew)
\end{lstlisting}
\end{mybox}

\noindent
A polynomial can be constructed either empty, from a single monomial or from another polynomial.

\begin{mybox}
\begin{lstlisting}
  Polynomial()
  Polynomial(Monomial M)
  Polynomial(const Polynomial & other)
\end{lstlisting}
\end{mybox}

\noindent
The below interface provides methods to perform basic algebraic operations with polynomials and scalars. The method $axpy$ is the scaled addition, equivalent to $this += other * a$

\begin{mybox}
\begin{lstlisting}
  LocalPolynomial & operator+=(const Monomial & otherM)
  LocalPolynomial & operator+=(const LocalPolynomial & other)
  LocalPolynomial & operator*=(const double c)  
  LocalPolynomial & operator*=(const LocalPolynomial & other)  
  void axpy(LocalPolynomial other, double c)
  LocalPolynomial operator+(const LocalPolynomial & other)
  LocalPolynomial operator+(const ctype a)  
  LocalPolynomial operator-(const LocalPolynomial & other)  
  LocalPolynomial operator-(const ctype a)  
  LocalPolynomial operator*(const ctype a)  
  LocalPolynomial operator*(const LocalPolynomial & other)
\end{lstlisting}
\end{mybox}

\noindent
We have implemented differentiation, integration and evaluation a polynomials. $derivative$ routine returns the partial derivative of a polynomial w.r.t. coordinate indexed by the parameter; $evaluate$ routine evaluates the polynomial at the provided local coordinate; $integrateRefSimplex$ routine integrates the polynomial over the reference entity of the same dimension as the polynomial, returning a scalar. An integral of a monomial over the reference simplex has an analytical expression, see \ref{appendix-proof-simplexintegral}

\begin{mybox}
\begin{lstlisting}
  LocalPolynomial derivative(int iDim)
  double evaluate(const LocalCoordinate & point)
  double integrateRefSimplex()
\end{lstlisting}
\end{mybox}

\noindent
The following auxiliary methods can be used to provide additional information about a polynomial. $order$ routine returns the largest power among all monomials, that is, the sum of powers of that monomial;
$magnitude$ routine returns the largest absolute value prefactor over all monomials. $to\_string$ routine converts the polynomial to a string for further text output

\begin{mybox}
\begin{lstlisting}
  unsigned int order()
  double magnitude()
  std::string to_string()
\end{lstlisting}
\end{mybox}

\noindent
Also, caching is implemented via the $cache()$ routine. This method can be called after the polynomial will no longer be changed, but only evaluated. Pre-computing factorials and monomial powers accelerates further evaluation and analytical integration of the polynomial.

%\noindent
%\uline{Compactify}: adds up all summands with the same power. Sorts the summands by $(x_1,y_1,z_1) < (x_2, y_2, z_2)$, where $x$ has the highest priority and $z$ has the lowest priority. Then all of the repeating powers will be consecutive. Simply loop over sorted polynomial, and to a new polynomial add the sums of all consecutive repeating polynomials.

%%\subsection{Tests}
%%
%%
%%\noindent
%%Currently the tests are only for 1, 2 and 3 dimensions. Most of the tests use intrinsic functionality like polynomial operators and derivatives to construct polynomials and print them to the screen, and request the user to to verify manually if they match the expected polynomials which are also printed. For each dimension there is one test which integrates a non-linear polynomial over simplex and prints out the result which is also compared manually. \\
%%
%%\textbf{TODO:} These tests can and should be automatized in the future using integer string comparison. The test program should throw an error if a test fails

%%%%%%%%%%%%%%%%%%%%%%%%%%%%%%%%%%%%%%%
% Implementation of Curvilinear Geometry Helper
%%%%%%%%%%%%%%%%%%%%%%%%%%%%%%%%%%%%%%%
\subsection{CurvilinearGeometryHelper}
\label{interface-geometry-helper}

\noindent
This section will discuss the interface of the \textit{CurvilinearGeometryHelper} class. This is an auxiliary class of \curvgeom{}, which provides functionality for addressing subentities of a uniform interpolatory grid, which are later used by \curvgeom{} to address the subentity geometries. \\

\noindent
Hard-coded number of curvilinear degrees of freedom as given in \ref{theory-lagrange}

\begin{mybox}
\begin{lstlisting}
  static int dofPerOrder(Dune::GeometryType geomType, int order)
\end{lstlisting}
\end{mybox}

\noindent
Hard-coded map between a corner internal index and vertex internal index.

\begin{mybox}
\begin{lstlisting}
  static InternalIndexType cornerIndex(Dune::GeometryType geomType, int order, InternalIndexType i)
\end{lstlisting}
\end{mybox}

\noindent
A procedure to extract corner indices from a vertex index set % using the $cornerIndex$ command

\begin{mybox}
\begin{lstlisting}
  template<class ct, int mydim>
  static std::vector<int> entityVertexCornerSubset(Dune::GeometryType gt, const std::vector<int> & vertexIndexSet, InterpolatoryOrderType order)
\end{lstlisting}
\end{mybox}

\noindent
A procedure to find the coordinate of a corner of a reference element based on its index. %Should be replaced by $ref.position()$ together with other methods of this class.

\begin{mybox}
\begin{lstlisting}
  template <typename ctype, int cdim>
  static Dune::FieldVector<ctype, cdim> cornerInternalCoordinate(GeometryType gt, InternalIndexType subInd)
\end{lstlisting}
\end{mybox}

\noindent
A procedure to generate a set of integer coordinates for the uniform interpolatory simplex grid (see \cref{fig:lagrange:enumerationconstruction})

\begin{mybox}
\begin{lstlisting}
  template <int mydim>
  static IntegerCoordinateVector simplexGridEnumerate(int n)
\end{lstlisting}
\end{mybox}

\noindent
A procedure to generate the local vertex set for a uniform interpolatory simplex grid (see \cref{fig:lagrange:enumerationconstruction}). Can re-use the already existing \textit{simplexGridEnumerate} array for speedup

\begin{mybox}
\begin{lstlisting}
  template <class ct, int mydim>
  static std::vector<Dune::FieldVector<ct, mydim> > simplexGridCoordinateSet(int n)
  template <class ct, int mydim>
  static std::vector<Dune::FieldVector<ct, mydim> > simplexGridCoordinateSet(IntegerCoordinateVector integerGrid, int n)
\end{lstlisting}
\end{mybox}

\noindent
A procedure to extract the local interpolatory vertex grid of a subentity of a given entity. \textit{subentityCodim} determines the codimension of the subentity, \textit{subentityIndex} determines the internal index of the subentity within the parent entity.

\begin{mybox}
\begin{lstlisting}
    template <class ct, int cdim>
    static std::vector<InternalIndexType> subentityInternalCoordinateSet(Dune::GeometryType entityGeometry, int order, int subentityCodim, int subentityIndex)
\end{lstlisting}
\end{mybox}

%%%%%%%%%%%%%%%%%%%%%%%%%%%%%%%%%%%%%%%
% Interface of Curvilinear Grid Factory
%%%%%%%%%%%%%%%%%%%%%%%%%%%%%%%%%%%%%%%
\subsection{Curvilinear Grid Factory}
\label{interface-grid-factory}

This section will discuss the information that needs to be provided in order to construct a curvilinear grid.

\begin{mybox}
\begin{lstlisting}
  Dune::CurvilinearGridFactory<GridType> factory(withGhostElements, withGmshElementIndex, mpihelper);
\end{lstlisting}
\end{mybox}

\noindent
In the above constructor, \textit{withGhostElements} determines if Ghost elements will be constructed, \textit{withGmshElementIndex} determines if the global element index will be re-used from the \gmsh{} file, or constructed from scratch, and \textit{mpihelper} is the MPI Helper class provided by \dune{}. \\

\noindent
A vertex must be inserted using its global coordinate and global index. At the moment, \curvgrid{} construction procedure requires \textit{a priori} knowledge of the vertex global index. All vertices belonging to each process must be inserted this way.

\begin{mybox}
\begin{lstlisting}
  insertVertex ( const VertexCoordinate &pos, const GlobalIndexType globalIndex )
\end{lstlisting}
\end{mybox}

\noindent
A curvilinear element must be inserted using its geometry type, interpolatory vertex local index STL vector, interpolatory order and physical tag. Currently, only 3D simplex elements are supported. All elements present on each process must be inserted. One must not insert elements not present on this process. The local index of an interpolatory vertex corresponds to the order the vertices were inserted into the grid. The order in which the vertex indices appear within the element is in accordance with the dune convention, discussed in \cref{impl-gmsh-numbering-convention}. Currently the available interpolation orders are 1-5. The interpolation order must correspond to the number of interpolatory vertices. Currently, physical tag is an integer, corresponding to the material property of the entity or otherwise.

\begin{mybox}
\begin{lstlisting}
  void insertElement(GeometryType &geometry, const std::vector< LocalIndexType > &vertexIndexSet, const InterpolatoryOrderType elemOrder, const PhysicalTagType physicalTag)
\end{lstlisting}
\end{mybox}

\noindent
A curvilinear boundary segment must be inserted using its geometry type, interpolatory vertex local index vector, interpolatory order, physical tag, and boundary association. Currently, only 2D simplex boundary segments are supported. All boundary segments present on this process must be inserted. One must not insert boundary segments not present on this process. Domain boundary segments must always be present in the mesh file. If interior boundaries are present in the geometry, they are also inserted using this method, setting $isDomainBoundary = false$.

\begin{mybox}
\begin{lstlisting}
  void insertBoundarySegment(GeometryType &geometry, const std::vector< LocalIndexType > &vertexIndexSet, const InterpolatoryOrderType elemOrder, const PhysicalTagType physicalTag, bool isDomainBoundary)
\end{lstlisting}
\end{mybox}

\noindent
Same as the facade \textit{Grid Factory} class, after the grid construction a pointer to that grid is returned. It is the duty of the user to delete the grid before the end of the program.

\begin{mybox}
\begin{lstlisting}
  GridType * createGrid()
\end{lstlisting}
\end{mybox}

%%%%%%%%%%%%%%%%%%%%%%%%%%%%%%%%%%%%%%%
% Interface of AllCommunication 
%%%%%%%%%%%%%%%%%%%%%%%%%%%%%%%%%%%%%%%
\subsection{AllCommunicate}
\label{interface-allcommunicate}

This section will discuss the templated the interface of our wrappers for MPI all-to-all communication and the nearest neighbor communication. \\

\noindent
A wrapper for $MPI\_Alltoallv$ method allows arrays of arbitrary type $T$, as long as its size is fixed and can be determined at compile-time (Plain Old Datatype, POD). This communication protocol is not scalable for very large architectures, since the number of communications performed by each process grows linearly with the process count. Its optimal use case is completely dense communication - every two processes exchange some information. The user needs to provide the input and output arrays, as well as the integer arrays denoting how many entries will be sent to and received from each process. Note that $out$ and $lengthOut$ need not be known \textit{a priori}, but need to have sufficient memory reserved for the output to be written.
\begin{mybox}
\begin{lstlisting}
  template <typename T>
  void communicate(const T * in, const int * lengthIn, T * out, int * lengthOut)
\end{lstlisting}
\end{mybox}
\noindent
A more comfortable interface for the above communication uses STL vectors. The meaning of the arguments is the same, however, the memory is automatically reserved for the output vectors, so there is no need to compute the required memory \textit{a priori}.
\begin{mybox}
\begin{lstlisting}
  template <typename T>
  void communicate(const std::vector<T> & in, const std::vector<int> & lengthIn, std::vector<T> & out, std::vector<int> & lengthOut)
\end{lstlisting}
\end{mybox}

\noindent
It is frequently the case that several processes need to communicate to several others, but most of the processes do not communicate to each other. Typically in finite difference or finite element implementations, each node communicates with the neighboring nodes only, and the communication per process stays constant with increasing process count. In this scenario, it is impractical to use all-to-all communication, and implementation of pairwise communication may be tedious. Starting from MPI-2 \cite{MPI-3.1}, the standard includes the nearest neighbor communication paradigm \textit{MPI\_Neighbor\_alltoallv}, designed especially for this purpose. We provide wrappers for this function. In the following protocol, $in$ and $out$ concatenate all the data sent to and received from neighbor processes only. $nNeighborIn$ and $nNeighborOut$ specify the number of send-to-neighbors and receive-from-neighbors. $ranksIn$ and $ranksOut$ specify the ranks of all neighbor processes. Same as in the first protocol, all output variables need not be known \textit{a priori}, but must have sufficient memory reserved.
\begin{mybox}
\begin{lstlisting}
  template <typename T>
  void communicate_neighbors(const T * in, int nNeighborIn, const int * ranksIn, const int * lengthIn, T * out, int & nNeighborOut, int * ranksOut, int * lengthOut)
\end{lstlisting}
\end{mybox}
\noindent
We also present an STL vector version of the above, which automatically reserves memory for output vectors
\begin{mybox}
\begin{lstlisting}
  template <typename T>
  void communicate_neighbors(const std::vector<T> & in, const std::vector<int> & ranksIn, const std::vector<int> & lengthIn, std::vector<T> & out, std::vector<int> & ranksOut, std::vector<int> & lengthOut)
\end{lstlisting}
\end{mybox}

\section{Explicit tests and solutions}

\subsection{Curvilinear Geometry Integral Tests}
\label{appendix-geometry-tests-integral}
This section presents explicit polynomial maps, integrands and exact results for integrals used in \curvgeom{} testing procedures (see \cref{sec:tests:curvgeom})

\begin{table} [H]
\centering
\begin{tabular}{| l | l | l | l | l |}
\hline
Ref. Element & Function       	& Analytic result   & Convergence order & Time to solution (ms) \\ \hline
1D Simplex & $1.0$               	& 1                 & 4                 & 0.06 \\ \hline
1D Simplex & $x$                  	& 0.5               & 4                 & 0.06 \\ \hline
1D Simplex & $x^3 - 3x + 3$	& 1.75              & 6                 & 0.07 \\ \hline
1D Simplex & $\sqrt{x}$		& 0.666666667       & 28                & 0.07 \\ \hline

2D Simplex & $1$	                			& 0.5               & 3                 & 0.06 \\ \hline
2D Simplex & $1 + x$                  			& 0.666666667       & 3                 & 0.07 \\ \hline
2D Simplex & $1 + x^2 + y^2$        	  	& 0.666666667       & 4                 & 0.07 \\ \hline
2D Simplex & $xyy$              			& 0.016666667       & 5                 & 0.08 \\ \hline
2D Simplex & $\sqrt{xy}$              			& 0.130899694       & 30                & 0.29 \\ \hline
2D Simplex & $2000 x^3 y^3$   	   	& 1.785714286       & 9                 & 0.07 \\ \hline
2D Simplex & $3628800 x^7 y^{10}$		& 0.545584447       & 18                & 0.19 \\ \hline
2D Simplex & $\sqrt{x^7 y^{10} + 0.5}$	& 0.353554             & 3                 & 0.07 \\ \hline

3D Simplex & $1$                    			& 1/6               		& 3                 & 0.07 \\ \hline
3D Simplex & $\sqrt{xyz}$            			& 0.013297747       & 39                & 10.3 \\ \hline
3D Simplex & $\sqrt{x^2 + y^2 + z^2}$	& 0.0877136        	& 13                & 0.36 \\ \hline
\end{tabular}
\caption{Performance of recursive integration routine with relative accuracy $\epsilon = 10^{-5}$}
\label{appendix:tests:quadintegrator:performance}
\end{table}

\begin{table} [H]
\centering
\begin{tabular}{ | l | l | l |}
  \hline
  Ord & Dim & Scalar Basis Function \\ \hline
  0 & 1 & $1$ \\ \hline
  1 & 1 & $1 + 2x$ \\ \hline
  2 & 1 & $1 + 2x + 3x^2$ \\ \hline
  3 & 1 & $1 + 2x + 3x^2 + 4x^3$ \\ \hline
  4 & 1 & $1 + 2x + 3x^2 + 4x^3 + 5x^4$ \\ \hline
  5 & 1 & $1 + 2x + 3x^2 + 4x^3 + 5x^4 + 6x^5$ \\ \hline
  0 & 2 & $1$ \\ \hline
  1 & 2 & $1 + 2(x + y)$ \\ \hline
  2 & 2 & $1 + 2(x + y) + 3(x^2 + y^2) + xy$ \\ \hline
  3 & 2 & $1 + 2(x + y) + 3(x^2 + y^2) + xy + 4(x^3 + y^3) + xy^2$ \\ \hline
  4 & 2 & $\begin{array}{lcl} 1 & + & 2(x + y) + 3(x^2 + y^2) + xy + 4(x^3 + y^3) + xy^2 \\ & + & 5(x^4 + y^4) + xy^3 \end{array}$ \\ \hline
  5 & 2 & $\begin{array}{lcl} 1 & + & 2(x + y) + 3(x^2 + y^2) + xy + 4(x^3 + y^3) + xy^2 \\ & + & 5(x^4 + y^4) + xy^3 + 6(x^5 + y^5) + xy^4 \end{array}$ \\ \hline
  0 & 3 & $1$ \\ \hline
  1 & 3 & $1 + 2(x + y + z)$ \\ \hline
  2 & 3 & $1 + 2(x + y + z) + 3(x^2 + y^2 + z^2) + xy$ \\ \hline
  3 & 3 & $1 + 2(x + y + z) + 3(x^2 + y^2 + z^2) + xy + 4(x^3 + y^3 + z^3) + xyz$ \\ \hline
  4 & 3 & $\begin{array}{lcl} 1 & + & 2(x + y + z) + 3(x^2 + y^2 + z^2) + xy + 4(x^3 + y^3 + z^3) + xyz \\ & + & 5(x^4 + y^4 + z^4) + xyz^2 \end{array}$ \\ \hline
  5 & 3 & $\begin{array}{lcl} 1 & + & 2(x + y + z) + 3(x^2 + y^2 + z^2) + xy + 4(x^3 + y^3 + z^3) + xyz \\ & + & 5(x^4 + y^4 + z^4) + xyz^2 + 6(x^5 + y^5 + z^5) + xyz^3 \end{array}$ \\ \hline
\end{tabular}
%\vfill
\caption{Scalar basis functions used in \curvgeom{} test. There is one function for each polynomial order and each geometry dimension}
\label{appendix:tests:curvgeom:integrand}
\end{table}

\begin{landscape}

\begin{table} [H]
\centering
\begin{tabular}{ | l | l | l | l | l | l | l | l | l |}
  \hline
  $d_e$ & Map & $\mu(\vec{r})$ & $I_0$ & $I_1$ & $I_2$ & $I_3$ & $I_4$ & $I_5$ \\ \hline
  1 & $(x)$                & $1$ & $1.0$ & $2.0$ & $3.0$ & $4.0$ & $5.0$ & $6.0$ \\ \hline
  1 & $(x,0)$              & $1$ & $1.0$ & $2.0$ & $3.0$ & $4.0$ & $5.0$ & $6.0$ \\ \hline
  1 & $(x,0,0)$            & $1$ & $1.0$ & $2.0$ & $3.0$ & $4.0$ & $5.0$ & $6.0$ \\ \hline
  1 & $(1+2x)$             & $2$ & $2.0$ & $4.0$ & $6.0$ & $8.0$ & $10.0$ & $12.0$ \\ \hline
  1 & $(2x,3x)$            & $\sqrt{13}$ & $\sqrt{13}$ & $2\sqrt{13}$ & $3\sqrt{13}$ & $4\sqrt{13}$ & $5\sqrt{13}$ & $6\sqrt{13}$ \\ \hline
  1 & $(2x,0.5+3x,5x)$     & $\sqrt{38}$ & $1\sqrt{38}$ & $2\sqrt{38}$ & $3\sqrt{38}$ & $4\sqrt{38}$ & $5\sqrt{38}$ & $6\sqrt{38}$ \\ \hline
  1 & $(x^2)$              & $2x$ & $1.0$ & $7/3$ & $23/6$ & $163/30$ & $71/10$ & $617/70$ \\ \hline
  1 & $(x,x^2)$            & $\sqrt{1 + 4x^2}$ & $1.47894286$ & $3.175666172$ & $4.994678155$ & $6.89140143$ & $8.84167808$ & $10.83102449$ \\ \hline
  1 & $(x,x^2,2)$          & $\sqrt{1 + 4x^2}$ & $1.47894286$ & $3.175666172$ & $4.994678155$ & $6.89140143$ & $8.84167808$ & $10.83102449$ \\ \hline
  2 & $(x,y)$              & $1$ & $1/2$ & $7/6$ & $41/24$ & $17/8$ & $37/15$ & $2.75714$ \\ \hline
  2 & $(x,y,0)$            & $1$ & $1/2$ & $7/6$ & $41/24$ & $17/8$ & $37/15$ & $2.75714$ \\ \hline
  2 & $(1+x,x+y)$          & $1$ & $1/2$ & $7/6$ & $41/24$ & $17/8$ & $37/15$ & $2.75714$ \\ \hline
  2 & $(y,3x,x+y)$         & $\sqrt{19}$ & $\sqrt{19}/2$ & $7\sqrt{19}/6$ & $41\sqrt{19}/24$ & $17\sqrt{19}/8$ & $37\sqrt{19}/15$ & $2.75714 \sqrt{19}$ \\ \hline
  2 & $(x^2,y^2)$          & $4xy$ & $1/6$ & $13/30$ & $59/90$ & $103/126$ & $0.94127$ & $1.03915$ \\ \hline
  2 & $(x^2,y^2,xy)$       & $2\sqrt{x^4+y^4+4x^2 y^2}$ & $0.360858$ & $0.938231$ & $1.47326$ & $1.93004$ & $2.33506$ & $2.70079$ \\ \hline
  3 & $(x,y,z)$            & $1$ & $1.0/6$ & $5.0/12$ & $23.0/40$ & $0.676389$ & $0.748214$ & $0.801935$ \\ \hline
  3 & $(x+y,y+z,x+z)$      & $2$ & $1.0/3$ & $5.0/6$ & $23.0/20$ & $2\cdot 0.676389$ & $2\cdot 0.748214$ & $2\cdot 0.801935$ \\ \hline
  3 & $(x^2,y^2,z^2)$      & $8xyz$ & $1.0/90$ & $0.0301587$ & $0.0416667$ & $0.0481922$ & $0.0522134$ & $0.05483$ \\ \hline
\end{tabular} %\vfill
\caption{Curvilinear mappings used in \curvgeom{} test, their integration elements, and integral values for integrals $I_{Ord}$ from  \cref{appendix:tests:curvgeom:integrand}. Integrals are indexed by the associated polynomial order $Ord$, and dimension chosen to correspond to the dimension of the mapping.}
\label{appendix:tests:curvgeom:mappings}
\end{table}

\end{landscape}

\begin{table}
\centering
\begin{tabular}{ | l | l | l | l | l | l | }
  \hline
  mydim & cdim & map               & Normal                  & Integrand                  & Result     \\ \hline
  1     & 2    & $(x,0)$           & $(0,-1)$                & $-x$                       & $-1/2$     \\ \hline
  1     & 2    & $(2x,3x)$         & $(3,-2)$                & $x$                        & $1/2$      \\ \hline
  1     & 2    & $(x,x^2)$         & $(2x,-1)$               & $2x^2-x$                   & $1/6$      \\ \hline
  2     & 3    & $(x,y,0)$         & $(0,0,-1)$              & $-xy$                      & $-1/24$    \\ \hline
  2     & 3    & $(y,3x,x+y)$      & $(-3,-1,3)$             & $-3x-y+3xy$                & $-13/24$   \\ \hline
  2     & 3    & $(y^2,x^2,xy)$    & $(-2y^2,-2x^2,4xy)$     & $-2x^3-2y^3+4x^2y^2$      & $-17/180$   \\ \hline
\end{tabular}
%\\
\caption{Table 3. Dot product integrals of vector basis functions over curved element faces}
\label{appendix:tests:curvgeom:dotproductintegral}
\end{table}

%%%%%%%%%%%%%%%%%%%%%%%%%%%%%%%%%%%%%%%
% Bibliography
%%%%%%%%%%%%%%%%%%%%%%%%%%%%%%%%%%%%%%%

\newpage
\bibliographystyle{plainnat}
\bibliography{bib/oswald,bib/numerical_libraries,bib/finite_element_method,bib/electromagnetic_antenna,bib/electromagnetic_methods,bib/misc,bib/material_properties,bib/interpolation,bib/integration,bib/molecular_plasmonics,bib/numerical_algebra}

%\printindex

\end{document}